\documentclass[12pt,a4paper]{article}
\pdfoutput=1
\usepackage{jheppub, amsthm,amsbsy,amsfonts,mathrsfs,enumerate,float,wrapfig,amsmath,amssymb,mathtools, tikz}

\allowdisplaybreaks

\title{Seiberg-Witten curves with O7$^\pm$-planes}
\author[a]{Hirotaka Hayashi,}
\author[b]{Sung-Soo Kim,}
\author[c]{Kimyeong Lee,}
\author[d]{Futoshi Yagi}
\affiliation[a]{Department of Physics, School of Science, Tokai University, 4-1-1 Kitakaname, \\ Hiratsuka-shi, Kanagawa 259-1292, Japan}
\affiliation[b]{School of Physics, University of Electronic Science and Technology of China,\\
 No. 2006 Xiyuan Ave, West Hi-Tech Zone, Chengdu, Sichuan 611731, China}
 \affiliation[c]{School of Physics, Korea Institute for Advanced Study, 85 Hoegi-ro Dongdaemun-gu, Seoul 02455, Korea}
 \affiliation[d]{School of Mathematics, Southwest Jiaotong University,
West zone, High-tech district, Chengdu, Sichuan 611756, China}

 \emailAdd{h.hayashi@tokai.ac.jp}
 \emailAdd{sungsoo497@gmail.com}
 \emailAdd{klee@kias.re.kr}
 \emailAdd{futoshi\_yagi@swjtu.edu.cn}

 \abstract{We construct Seiberg-Witten curves for 5d $\mathcal{N}=1$ gauge theories whose Type IIB 5-brane configuration involves an O7-plane and discuss an intriguing relation between theories with an O7$^+$-plane and those with an O7$^-$-plane and 8 D7-branes. We claim that 5-brane configurations with an O7$^+$-plane can be effectively understood as 5-brane configurations with a set of an O7$^-$-plane and eight D7-branes with some special tuning of their masses such that the D7-branes are frozen at the O7$^-$-plane. We check this equivalence between SU($N$) gauge theory with a symmetric hypermultiplet and SU($N$) gauge theory with an antisymmetric with 8 fundamentals, and also between SO($2N$) gauge theory and Sp($N$) gauge theory with eight fundamentals. We also compute the Seiberg-Witten curves for non-Lagrangian theories with a symmetric hypermultiplet, which includes the local $\mathbb{P}^2$ theory with an adjoint. 
\\
\vspace{2cm}

\begin{center}
{\it 
Dedicated to the memory of Lars Brink}	
\end{center}
}
\begin{document}
 \preprint{KIAS-P23025}
\maketitle
\allowdisplaybreaks
\section{Introduction}\label{sec:intro}

There has been much progress on supersymmetric theories of eight supercharges in five and six dimensions (5d/6d), and String theory and M-theory provide useful tools for studying them. For instance, $(p,q)$ 5-brane webs in Type IIB string theory \cite{Aharony:1997bh} and M-theory compactified on a Calabi-Yau threefold \cite{Morrison:1996xf,Douglas:1996xp,Intriligator:1997pq} have shed light on uncovering rich non-perturbative aspects of higher dimensional supersymmetric theories. Different methods of computing BPS partitions on the Omega background are developed \cite{Nekrasov:2002qd,Nekrasov:2003rj, Aganagic:2003db,Iqbal:2007ii,Awata:2008ed,Nakajima:2003pg,Nakajima:2005fg, Gottsche:2006bm,Keller:2012da,Kim:2012gu,
Huang:2017mis,Kim:2019uqw,Kim:2020hhh}, and various limits of the partition functions give rise to interesting physical observable capture vacuum structure of higher dimensional supersymmetry theories. Of particular interest is the Seiberg-Witten curves \cite{Seiberg:1994rs,Seiberg:1994aj} of higher dimensional theories compactified to four dimensions on a circle $S^1$ or a torus $T^2$, which captures M5-brane configuration in $\mathbb{R}^2\times T^2$ for the theories. Various ways of obtaining 5d/6d Seiberg-Witten curves are developed \cite{Sakai:2017ihc,Closset:2020afy,Haouzi:2020yxy,Closset:2021lhd,Jia:2021ikh,Brini:2021wrm,DelMonte:2022kxh, Sakai:2023fnp} which includes thermodynamic limit \cite{Haghighat:2016jjf, Haghighat:2018dwe, Li:2021rqr}, Nekrasov-Shatashvili limit of the partition function leading to quantum curves \cite{Chen:2020jla,Chen:2021ivd,Chen:2021rek,Chen:2023aet} as well as 5-brane webs \cite{Bao:2013pwa, Kim:2014nqa,Hayashi:2017btw, HKSWY:2023}.

From Type IIB 5-brane webs, in particular, one can systematically compute the corresponding Seiberg-Witten curves even for non-toric cases which involve gauge groups of SU-type with a sufficiently large number of hypermultiplets in the fundamental representation \cite{Bao:2013pwa, Kim:2014nqa}. The dual (non-)toric diagrams \cite{Hayashi:2017btw}, also known as generalized toric diagrams, are represented by black dots and white dots \cite{Benini:2009gi} which lead to the characteristic equation that is eventually identified as the Seiberg-Witten curve by associating the coefficients of the characteristic equation with physical parameters. The white dots of the dual diagram, in particular, represent more than one 5-brane bound to a single 7-brane, and hence yield a degenerated polynomial. There are also intrinsically non-toric cases that are of Sp/SO types of gauge groups and their 5-brane webs require an orientifold plane, for instance, an O5-plane (or its S-dual ON-plane) or an O7-plane. The construction of such theories involving an O5-plane is discussed in \cite{Hanany:2000fq, Zafrir:2015ftn} and one needs to consider the covering space which includes projected 5-brane configurations due to an O5-plane. The corresponding characteristic equation or Seiberg-Witten curve hence has a $\mathbb{Z}_2$ symmetry arising from the identification of reflected mirror images \cite{Hayashi:2017btw}.

Study of the Seiberg-Witten curves for theories based on 5-brane webs with an O7-plane is, however, limited to the cases where an O7$^-$-plane is resolved into a pair of two 7-branes \cite{Sen:1996vd}, leading to the curves for dual SU-gauge theories \cite{Bergman:2015dpa,Hayashi:2015fsa}. The curves for theories whose brane configuration involves an O7$^+$-plane are still not explored much. As 5-brane webs with an O7$^\pm$-plane enrich our understanding of SO/Sp gauge theories as well as SU gauge theories with hypermultiplet in the second rank (antisymmetric/symmetric) tensor representations, we study the construction of the Seiberg-Witten curves based on 5-brane webs with an O7$^\pm$-plane in this work.

The focus of this paper is two-fold: First, we provide a systematic way of computing the Seiberg-Witten curve for 5d theories whose  5-brane webs involve an O7-plane. Previously, these authors have constructed Seiberg-Witten curves for 5d theories based on 5-brane webs with or without an O5-plane, by imposing boundary conditions for an O5-plane satisfies. This construction is applicable for Sp($N$), SO($N$), and $G_2$ gauge theories with fundamental hypermultiplets (flavors) \cite{Hayashi:2018bkd}, and also SU($N$) gauge theories at the high Chern-Simons level $\kappa$ possibly with flavors. Other orientifold planes in 5d are  ON, O7$^-$, and O7$^+$ planes. Note that an ON-plane can be understood as an S-dual configuration of O5-plane \cite{Sen:1998rg,Sen:1998ii,Kapustin:1998fa, Hanany:1999sj} and hence constructing the Seiberg-Witten curve is not much different from that for an O5-plane. Also, recall that O7$^-$-plane can be resolved into a pair of 7-branes whose monodromy is the same as that of an O7$^-$-plane. This means that those theories which involve an O7$^-$-plane thus can be treated as 5-brane webs where an O7$^-$-plane is resolved. 5-brane configurations with an O7$^+$-plane are, however, not considered before in the study of the Seiberg-Witten curve. Recently, new theories, including non-Lagrangian theories, whose 5-brane configuration has an O7-plane have been suggested. Local $\mathbb{P}^2$ with adjoint matter is a noticeable example of non-Lagrangian theories that involve an O7$^+$-plane. As its brane configuration is known, a study based on its 5-brane webs would be a fruitful direction for better understanding the theory. One of which is Seiberg-Witten curves based on the corresponding 5-brane web.  In this paper, we discuss a systematic way of constructing Seiberg-Witten curves for those theories involving an O7$^+$-plane as well as an O7$^-$plane, which provides Seiberg-Witten curves for non-Lagrangian theories and also may complete the program that we have pursued.

The second is a proposal that an O7$^+$-plane can be effectively understood as the combination of an O7$^-$-plane with 8 fundamental hypermultiplets ($\mathrm{O7}^-+8\mathrm{D7}$'s) of specially tuned masses for computing some physical quantities. As being associated with frozen singularities, an O7$^+$-plane is, of course, not the same as an O7$^-$-plane with eight flavors. There are, however, many properties that suggest that two orientifolds are closely related.  The monodromy of an O7$^+$-plane $\begin{pmatrix}
-1 & -4 \\
0 & -1
\end{pmatrix}
$ is equivalent to that of $\mathrm{O7}^-+8\mathrm{D7}$'s, which reflects a similarity between two 5-brane configurations: one is with an O7$^+$-plane and the other is with 8 massless fundamental hypermultiplets that are frozen at the position of the O7$^+$-plane where half of them having the opposite phase. For SU($N$) gauge theories, the Chern-Simons shift by decoupling a hypermultiplet in the symmetric representation is equivalent to that by decoupling an antisymmetric hyper and eight fundamental flavors together. The contribution of a symmetric hypermultiplet to the cubic prepotential can be shown to be equivalent to the contribution from an antisymmetric and eight massless fundamental hypermultiplets. We study the relation between theories that can be constructed from an O7$^+$-plane and  $\mathrm{O7}^-+8 \mathrm{D7}$'s, 
\begin{align*}
 \mathrm{O7}^+ ~~\longleftrightarrow ~~
 \mathrm{O7}^- + 8 \, \mathrm{D7}\big|_{\rm frozen}\ ,
\end{align*}
and discuss equivalence between observables of two theories when eight flavors are stuck at an O7$^-$-plane in a particular way. We use the Seiberg-Witten curves that we discuss in this paper to demonstrate that the Seiberg-Witten curve for 5d theories whose brane configuration involves an O7$^+$-plane is equivalent to that for 5d theories involving an O7$^-$-plane with 8 D7 branes.

The organization of the paper is as follows: In section \ref{sec:prepotential}, we consider the cubic prepotentials of 5d gauge theories whose 5-brane webs can be described by an O7-plane, and demonstrate the relationship between an O7$^+$ and  $\mathrm{O7}^-+8\mathrm{D7}$'s. In section \ref{sec:O7-construction}, we discuss a systematic way of computing Seiberg-Witten curves for SU($N$) gauge theories with a symmetric hypermultiplet and SO($2N$) gauge theories, based on 5-brane webs of an O7$^+$-plane. We then extend our construction to  Seiberg-Witten curves for SU($N$) gauge theories with an antisymmetric hypermultiplet and Sp($N$) gauge theories, based on 5-brane webs of an O7$^-$-plane, in section \ref{sec:O7-construction}. In connection with the relation between O7$^+$ and $\mathrm{O7}^-+8\mathrm{D7}$'s, we compare two Seiberg-Witten curves and give an account for the relation from 5-brane webs in section \ref{sec:comparison}. We summarize and discuss future directions in the conclusion. In Appendices, we discuss a possible extension of our construction to a certain class of quiver theories whose 5-brane web has an O7$^+$-plane and also list some details of the computations for equivalence between SU(3) gauge theory with a hypermultiplet in the antisymmetric representation and that with a hypermultiplet in the fundamental representation. 
\vspace{.5cm}

\bigskip
\section{Cubic prepotentials}\label{sec:prepotential}
A 5d $\mathcal{N}$ = 1 supersymmetric gauge theory of a gauge group $G$ has a Coulomb branch which is parameterized by the real scalar field $\phi$ in the vector multiplet. On the Coulomb branch, gauge group $G$ is broken to U(1)$^{\mathrm{rank}(G)}$, and the corresponding low-energy Abelian gauge theory is governed by the cubic prepotential given as follows  \cite{Seiberg:1996bd,Morrison:1996xf, Intriligator:1997pq}:
\begin{align}
\mathcal{F}(\phi) = \frac{1}{2g_0^2}h_{ij}\phi_i\phi_j + \frac{\kappa}{6}d_{ijk}\phi_i\phi_j\phi_k + \!\frac{1}{12}\bigg(\sum_{r\in\Delta}\left|r\cdot \phi\right|^3 - \!\sum_f\!\sum_{w \in R_f}\!\left|w\cdot \phi +m_f\right|^3\!\bigg).  \label{eq:prepot}
\end{align}
Here, $g_0$ is the gauge coupling, $\kappa$ is the classical Chern-Simons
level, and $m_f$ is mass parameter for the matter $f$. $h_{ij} = \mathrm{Tr}(T_iT_j)$ where $T_i$ are the Cartan generators of $\mathfrak{g}$, and $d_{ijk} = \frac12\mathrm{Tr}(T_i\{T_j,T_k\})$ $d_{ijk}$ which is only non-zero for  $G=\mathrm{SU}(N\ge3)$. The last term of the prepotential $\mathcal{F}(\phi)$ comes from one-loop contribution where $\Delta$ is the root system of Lie algebra $\mathfrak{g}$ associated to $G$ and $w$ is a weight of the representation $R_f$ of the Lie algebra $\mathfrak{g}$.

\begin{figure}
    \centering
    \includegraphics[width=12cm]{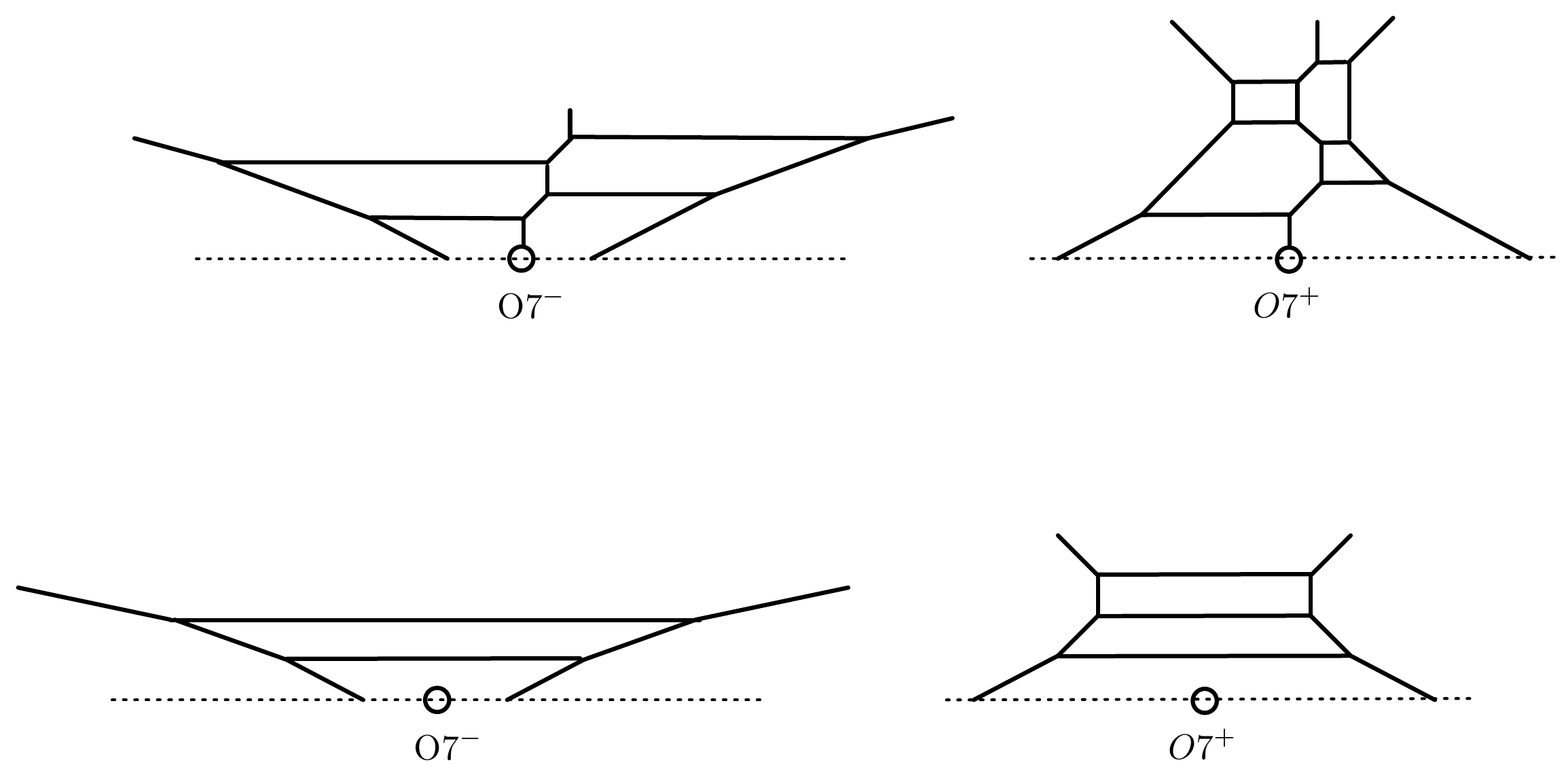}
    \caption{Examples of 5-branes web with an O7$^\pm$-plane: From the top left, they represent the following gauge theories: SU($4)+1\mathbf{AS}$, SU($6)+1\mathbf{Sym}$, Sp($2)$, and SO($6$).}
    \label{fig:webO7+-}
\end{figure}
In this section, we consider cubic prepotentials of 5d supersymmetric gauge theories associated with whose Type IIB 5-brane configurations can be constructed with an O7$^+$-plane or with an O7$^-$-plane. For instance, SU($N$) gauge theories with one symmetric hypermultiplet (SU($N$)+1$\mathbf{Sym}$) can be described by a 5-brane configuration with an O7$^+$-plane where a half NS5-brane is stuck, while SU($N$) gauge theories with one antisymmetric hypermultiplet (SU($N$)+1$\mathbf{AS}$) can be described by a 5-brane configuration with an O7$^-$-plane where a half NS5-brane is stuck as well. Other than the SU-type gauge group,  SO($2N$) or Sp($N$)  gauge theories can be described by a 5-brane web respectively with an O7$^+$-plane or an O7$^-$-plane where none of the 5-branes is stuck as shown in figure \ref{fig:webO7+-}. We note that one can introduce D7-branes into the brane configurations with an O7$^+$ or O7$^-$-plane to describe fundamental hypermultiplets. We also note that an O7$^-$-plane can be resolved into a pair of two 7-branes having the same monodromy, and two distinct sets account for
discrete theta parameters $\theta=0, \pi$ of 5d Sp($N$) gauge theories without fundamental hypermultiplets, which we denote as Sp($N$)$_{\theta}$. 

We now write down the cubic prepotential explicitly for theories involving an O7-plane and compare the one associated with O7$^+$-plane to the other with O7$^+$-plane and 8 D7-branes. For SU($N$) theory of the Chern-Simons level $\kappa$ with $N_f$ flavors ($\mathbf{F}$), $N_{a}$ antisymmetric hypers ($\mathbf{AS}$), and $N_{s}$ symmetric hypers ($\mathbf{Sym}$), the cubic prepotential is expressed in terms of the Coulomb branch parameters $a_i$ and mass parameters. In the Weyl chamber   $a_1>a_2>\cdots>a_{N-1}\ge0$ and $a_N=-\sum_{i=1}^{N-1}a_i$, it takes the form
\begin{equation}\label{eq:F for SU}
\begin{aligned}
  \mathcal{F}_{\mathrm{SU}(N)_\kappa+N_f\mathbf{F}+N_{a}\mathbf{AS}+N_{s}\mathbf{Sym}}
  =&~ \frac{1}{2g_0^2}\sum_{i=1}^Na_i^2 +\frac{\kappa}{6}\sum_{i=1}^N a_i^3+  \frac{1}{6}\sum_{i<j}^N (a_i-a_j)^3\\
 & -\frac{N_f}{12}\sum_{i=1}^N |a_i|^3-\frac{N_{a}}{12}\sum_{i<j}^N |a_i+a_j|^3\cr
 &- \frac{N_{s}}{12}\Big(\sum_{i=1}^N |2a_i|^3+\sum_{i<j}^N |a_i+a_j|^3\Big) \ , 
\end{aligned}
\end{equation}
where we set all the masses to zero for convenience. It easily follows from the prepotential \eqref{eq:F for SU} that the contribution from $1\mathbf{Sym}$ is equivalent to that from $1\mathbf{AS}+8\mathbf{F}$ \cite{Intriligator:1997pq}. 
In other words,  
\begin{align}
    \mathcal{F}_{\mathrm{SU}(N)_\kappa+1\mathbf{Sym}+N_f\mathbf{F}}=\mathcal{F}_{\mathrm{SU}(N)_\kappa+1\mathbf{AS}+(8+N_f)\mathbf{F}}\ . 
\end{align}
We comment that it is straightforward to check that this relation still holds even for massive cases, if the masses of fundamental, antisymmetric, and symmetric hypermultiplets $m_f, m_a, m_s$ are tuned as follows: 
$m_f=m_s/2$ and $m_{a}=m_s$.

Integrating out massive hypermultiplets, the Chern-Simons level $\kappa$ gets shifted. For example, decoupling a fundamental hypermultiplet induces $\pm1/2$ shift depending on whether it is integrated out along with the positive or negative mass. This means that decoupling hypers from 5d SU(3)$_0$ gauge theories with 10 flavors, one gets 5d pure SU(3)$_\kappa$ of integer Chern-Simons levels with the range $0\le |\kappa| \le 5$. See also 
\cite{Jefferson:2018irk, Hayashi:2018lyv}
for SU(3)$_\kappa$ theories with various matter contents, including antisymmetric and symmetric hypers. It turns out 
The Chern-Simons level shift for 5d SU($N$)$_\kappa$ gauge theory due to decoupling a hypermultiplet is given as follows: 
\begin{align}
    \kappa ~~\rightarrow ~~\kappa\pm \frac12I^{(3)} \ ,  
\end{align}
where $I^{(3)}$ is the cubic Dynkin index for the representation associated with hypermultiplets, which is summarized in table \ref{tab:hyper-Dynkin-index}.
\begin{table}[t]
    \centering
    \begin{tabular}{c|c}
    \hline
         Hypermultiplet & $I^{(3)}$\\  \hline
         $\mathbf{F}$ & $1$\\ 
         $\mathbf{AS}$ & $N-4$\\ 
         $\mathbf{Sym}$ & $N+4$\\\hline 
    \end{tabular}
    \caption{Hypermultiplets of 5d SU($N$) gauge theories and the corresponding cubic Dynkin indices of the representation of the associated Lie algebra $ \mathfrak{su}$($N$).}
    \label{tab:hyper-Dynkin-index}
\end{table}

It follows then that the cubic Dynkin index for the symmetric representation of SU($N$) is equivalent to the sum of the cubic Dynkin index for the antisymmetric representation and eight times the Dynkin index for the fundamental representation, 
\begin{equation}
    I^{(3)}_{\mathbf{Sym}} ~=~ I^{(3)}_{\mathbf{AS}}~+~8\,I^{(3)}_{\mathbf{F}}\ .
\end{equation}
This implies that the Chern-Simons level shift, when a symmetric hyper is decoupled, is the same as the shift from an antisymmetric and 8 fundamental hypers altogether.  

This relation manifests the matter content for KK theories of generic SU($N$)$_\kappa$ as well. If an SU($N)_\kappa$ gauge theory of a symmetric is a KK theory, then so is an SU($N)_\kappa$ gauge theory at the {\it same} Chern-Simons level and with an antisymmetric and eight fundamentals. 
For instance, the following KK theories, listed in Table 1 of \cite{Jefferson:2017ahm}, are pairs of theories obtained by replacing $1\mathbf{Sym}$ with $1\mathbf{AS}+8\mathbf{F}$:
\begin{equation}
\begin{aligned}\label{eq:SU+SandAS}
\mathrm{SU}(N)_0+1\mathbf{Sym}+(N-2)\mathbf{F}
&\quad \longleftrightarrow \quad 
\mathrm{SU}(N)_0+1\mathbf{AS}+(N+6)\mathbf{F}\ , \cr
\mathrm{SU}(N)_0+1\mathbf{Sym}+1\mathbf{AS} 
&\quad\longleftrightarrow \quad
\mathrm{SU}(N)_0+2\mathbf{AS}+8\mathbf{F} \ , \cr
\mathrm{SU}(N)_{\frac{N}{2}}+1\mathbf{Sym} &\quad\longleftrightarrow \quad
\mathrm{SU}(N)_{\frac{N}{2}}+1\mathbf{AS}+8\mathbf{F} \ . 
\end{aligned}    
\end{equation}

We now discuss the cubic prepotentials for 5d $\mathcal{N}=1$ Sp($N$) and SO($2N$) gauge theories and relations between the prepotentials of two theories. 
First, consider Sp($N$) gauge theory with massless $N_f$ flavors. In the Weyl chamber $a_1\ge a_2\ge \cdots\ge a_N\ge0$, the cubic prepotential is expressed as 
\begin{align}\label{eq:FforSpN}
  \mathcal{F}_{\mathrm{Sp}(N)+N_f\mathbf{F}} = \frac{1}{g_0^2}\sum_{i=1}^Na_i^2 +  \frac{1}{6}\bigg(\!\sum_{i<j}^N \big((a_i\!-\!a_j)^3
  +(a_i\!+\!a_j)^3\big)
  +(8\!-\!N_f)\!\sum_{i=1}^N a_i^3\bigg)\ ,  
\end{align}
where the terms cubic in the Coulomb branch parameters $a_i$ account for the vector and hyper contributions. In particular, the coefficient $(8-N_f)$ in front of the term proportional to $\sum a_i^3$ is responsible for the contributions from the roots associated with the long root and from $N_f$ massless flavors. 

The prepotential for SO$(2N$) gauge theory with $N_f$ massless flavors can be written similarly. In the Weyl chamber $a_1\ge a_2\ge \cdots\ge a_N\ge0$, the cubic prepotential takes the form
\begin{align}\label{eq:FforSO2N}
  \mathcal{F}_{\mathrm{SO}(2N)+N_f\mathbf{F}} = \frac{1}{g_0^2}\sum_{i=1}^Na_i^2 +  \frac{1}{6}\bigg(\sum_{i<j}^N\big( (a_i-a_j)^3+(a_i+a_j)^3\big) -N_f\sum_{i=1}^N a_i^3\bigg)\ .   
 \end{align}
By comparing these two prepotentials \eqref{eq:FforSpN} and \eqref{eq:FforSO2N}, one can easily see that the prepotential for Sp($N$) gauge theory with ($8+N_f$) massless flavors is equivalent to that for SO($2N$) gauge theory with $N_f$ flavors, 
\begin{align}
    \mathcal{F}_{\mathrm{Sp}(N)+(8+N_f)\mathbf{F}}=\mathcal{F}_{\mathrm{SO}(2N)+N_f\mathbf{F}}\ .
\end{align}
This relation still holds even with nonzero masses, where 8 extra masses of fundamental hypers from Sp($N$) are set to zero.

As discussed in the case for SU($N$) gauge theory with $1\mathbf{Sym}$ or $1\mathbf{AS}+8\mathbf{F}$, KK theories associated with Sp($N$) and SO($2N$) gauge theories are closely related: 
\begin{align}\label{eq:SpandSO}
 \mathrm{Sp}(N)+(2N+6)\mathbf{F}&~\longleftrightarrow ~\mathrm{SO}(2N)+(2N-2)\mathbf{F}\ .
\end{align}

\begin{figure}
    \centering
    \includegraphics[width=11cm]{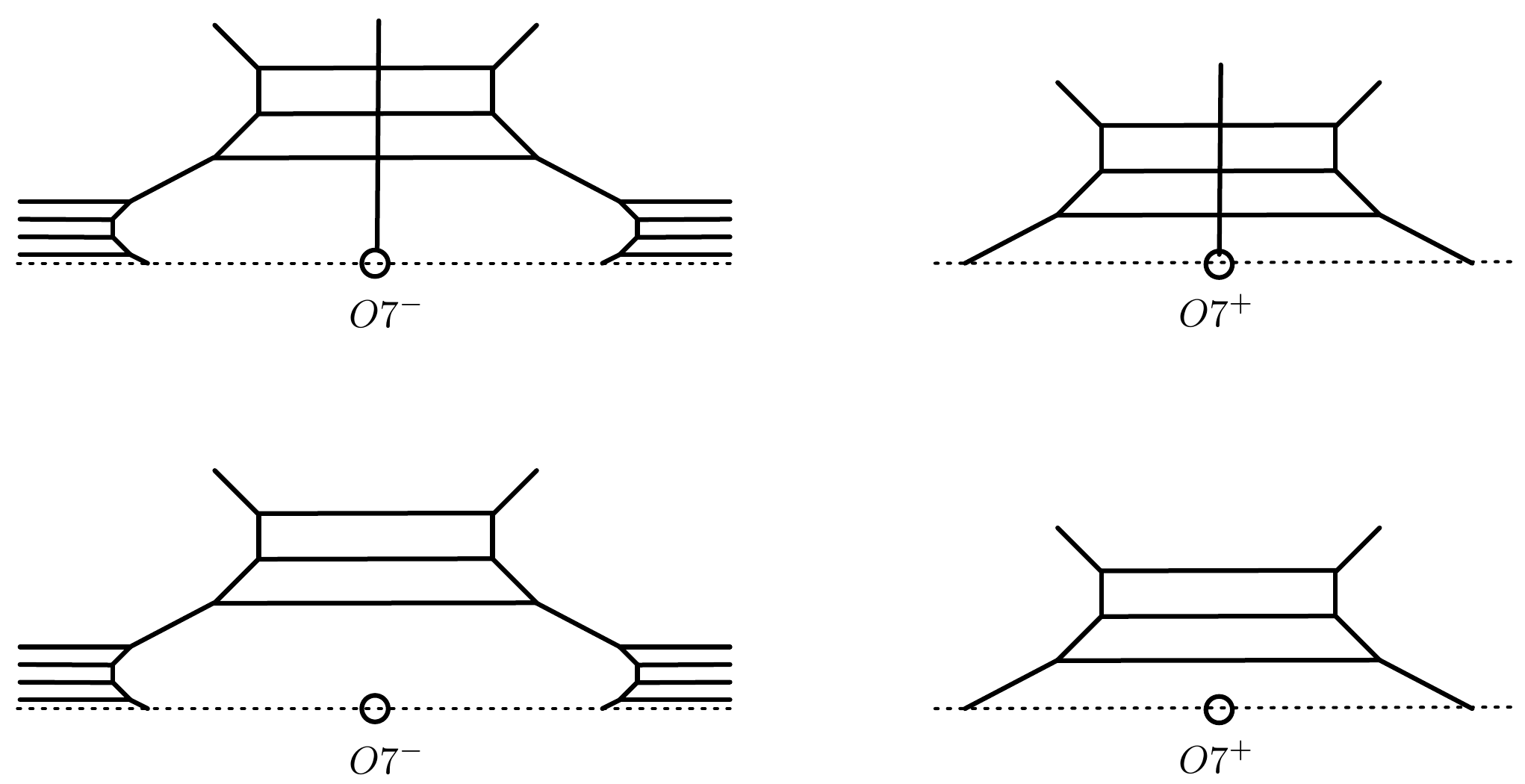}
    \caption{
    5-brane webs with O7$^-+8$D7 or O7$^+$ and Higgs branches.
    From top to bottom: Higgsing from SU($2N)+1\mathbf{AS}$ (O7$^-$)  or $1\mathbf{Sym}$ (O7$^+$) to Sp($N$) or SO($2N$), respectively. From left to right: By tuning mass parameters of 8 fundamental hypers, the shape of the 5-brane webs with O7$^-+8$D7 becomes the same as the shape of the 5-brane webs with an O7$^+$.}
    \label{fig:HiggsSU2N}
\end{figure}

Consider also that two Higgs branches regarding SU($2N$) gauge theories with a symmetric or an antisymmetric:
\begin{align} \label{eq:HiggsBranch}
    \mathrm{SU}(2N)_\kappa+ 1\mathbf{AS}~&\xrightarrow{\text{Higgsing}}~ \mathrm{Sp}(N)\ , \cr
    \mathrm{SU}(2N)_\kappa+ 1\mathbf{Sym}~&\xrightarrow{\text{Higgsing}}~ \mathrm{SO}(2N)\ . 
\end{align}
These Higgsings can be readily seen from the prepotential with the Weyl chamber for SU($2N$) of $a_1\ge\cdots\ge a_N\ge0\ge a_{N+1}\ge\cdots\ge a_{2N}$, by setting 
$a_i=-a_{2N+1-i}$.
It can be also understood from the prepotential or 5-brane webs given in figure \ref{fig:HiggsSU2N}, where the NS5-brane stuck on an O7-plane for 5-brane webs of SU($2N)+1\mathbf{AS}/\mathbf{Sym}$ is Higgsed away together with its reflected mirror brane.\footnote{For a study of Higgs branch for SO($N$) gauge theory via 5-brane webs with an O7$^+$-plane, see \cite{Akhond:2021ffo}.}

These Higgs branches are consistent with the relations \eqref{eq:SU+SandAS} and \eqref{eq:SpandSO} with theories associated with an O7$^+$-plane or with an O7$^-$-plane and 8 D7-branes:
\begin{equation}
\begin{aligned}
&&\mathrm{SU}(2N)_0+1\mathbf{Sym}+(2N-2)\mathbf{F} 
&\quad \longleftrightarrow &&
\mathrm{SU}(2N)_0+1\mathbf{AS}+(2N+6)\mathbf{F}\cr
&&\scalebox{0.8}{Higgsing}~~\big\downarrow \qquad\quad~~~  &  &&\qquad\quad~~~\big\downarrow~~\scalebox{0.8}{Higgsing}\cr
&&\mathrm{SO}(2N)+(2N-2)\mathbf{F} & \quad \longleftrightarrow  &&\mathrm{Sp}(N)+(2N+6)\mathbf{F}\ ,
\end{aligned}    
\end{equation}
and
\begin{equation}
\begin{aligned}
&&\mathrm{SU}(2N)_0+1\mathbf{Sym}+1\mathbf{AS}~~~ 
&\longleftrightarrow &&
\mathrm{SU}(2N)_0+2\mathbf{AS}+8\mathbf{F}\cr
&&\scalebox{0.8}{Higgsing}~~\big\downarrow \qquad\qquad~~~  &  &&\qquad~~~\big\downarrow~~\scalebox{0.8}{Higgsing}\cr
&&\mathrm{SO}(2N)+1\mathbf{Adj}~(=1\mathbf{AS})~~~ &  \longleftrightarrow  &&\mathrm{Sp}(N)+1\mathbf{AS}+8\mathbf{F}\ .
\end{aligned}    
\end{equation}
Here, we note that it follows from \eqref{eq:HiggsBranch} that there is another Higgs branch for $\mathrm{SU}(2N)_0+1\mathbf{Sym}+1\mathbf{AS}$ which is 
\begin{align}
    \mathrm{SU}(2N)_0+1\mathbf{Sym}+1\mathbf{AS}~\xrightarrow{\text{Higgsing}}~ \mathrm{Sp}(N)+1\mathbf{Adj} ~(=1\mathbf{Sym}) \, .
\end{align}

The relation between O7$^+$ and O7$^-+8$D7s discussed in this section can be made more concrete by computing other physical observables. In the next section, we consider the Seiberg-Witten curves which can be obtained from dot (or toric-like) diagrams \cite{Benini:2009gi, Kim:2014nqa}.


\section{Construction of Seiberg-Witten curves with an \texorpdfstring{O7$^+$}{O7+}-plane}\label{sec:O7+construction}

 In this section, we construct Seiberg-Witten curves based on 5-brane webs with an O7$^+$-plane, which includes SO($N$) gauge theory and 5d SU($N$) with a hypermultiplet in symmetric representation. We then check the Higgsing \eqref{eq:HiggsBranch}. 

\subsection{5-brane webs and Seiberg-Witten curves}
We first review briefly how to obtain Seiberg-Witten curves from 5-brane webs. Suppose we have a 5-brane web for a 5d theory of interest, which may or may not allow more than one 5-brane web due to the presence of an orientifold plane, O5-, ON-, or O7-plane. For instance, see \cite{Hayashi:2018lyv}, where many examples of 5d rank-2 theories of different 5-brane configurations are listed. As such an orientifold plane gives rise to different boundary conditions, in this review, we only discuss 5-branes without orientifold planes. More details can be found in  \cite{Kim:2014nqa} or  \cite{Hayashi:2017btw}
for cases with an O5-plane.

Given a 5-brane web diagram without orientifold planes, one can construct its dual toric diagram which is made of dots and edges on a $\mathbb{Z}^2$ lattice. One then associates the position of each dot on the vertices on the lattice with a monomial $t w$ whose power is the coordinates $(m,n)$ of the position on the lattice. By summing over all monomials with coefficients, one can make the characteristic polynomial equation, 
\begin{align}\label{eq:char-poly}
	\sum_{(m,n)\in \text{vertices}} C_{mn}\, t^m\, w^n =0\ .
\end{align}
Here, $C_{mn}$ are coefficients that will be fixed by boundary conditions of the external branes in terms of parameters of the theory, {\it i.e., } the Coulomb branch parameters, instanton factor, mass parameters of hypermultiplets. We remark that three of the coefficients can be chosen by three degrees of freedom of the characteristic equation, that is one from the overall coefficient and two from the shifts of the vertical and horizontal axes. The characteristic equation with the coefficients expressed with the physical parameters is promoted to a complex curve  by associating the coordinates $t$ and $w$ of the $\mathbb{Z}^2$ lattice as 
\begin{align}
	t&= e^{-\frac{1}{R_{M}}(x^6\,+\, i\, x^{11})}, \qquad
	w= e^{-\frac{1}{R_{A}}(x^5\,+\, i\, x^{4})}\ . \label{eq:tw}
\end{align}
where $x^4\sim x^4+2\pi R_A$ with $R_A$ being the radius along the $x^4$-direction and $x^{11}\sim x^{11}+ 2\pi R_{M}$ with $R_{M}$ being the radius of the M-theory circle along the $x^{11}$-direction. This characteristic equation is then the Seiberg-Witten curves describing an M5-brane configuration in $\mathbb{R}^2\times T^2$ for a given 5d theory on a circle. This means that we compactify the theory on a circle along the $x^4$-direction and take a T-duality and then uplift it to the M-theory circle. We comment that the Seiberg-Witten differential $\lambda_{\rm SW}$ is given by \cite{Fayyazuddin:1997by, Henningson:1997hy, Mikhailov:1997jv},
\begin{align}
     \lambda_{\rm SW}=-\frac{i}{(2\pi)^2\ell_p^3}\log t(d\log w)\ ,
\end{align}
where $\ell_p$ is the Planck length and the $\lambda_{\rm SW}$ is proportional to tension of M2-brane.

\begin{figure}
    \centering
    \includegraphics[width=9cm]{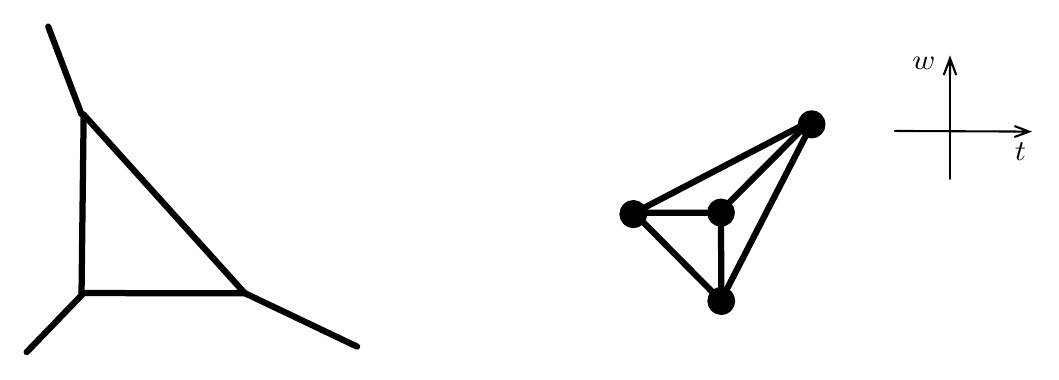}
    \caption{A 5-brane web for a local $\mathbb{P}^2$ and its dual toric diagram.}
    \label{fig:E0-toric}
\end{figure}

The simplest example would be a local $\mathbb{P}^2$, also known as  $E_0$ theory, whose 5-brane web and dual toric diagrams are given in figure \ref{fig:E0-toric}. 
As there are four dots and we have three degrees of freedom to choose, the corresponding characteristic equation can be given with one undetermined coefficient, which will be a function of the Coulomb branch parameter. By associating the Coulomb branch parameter $u$ to the dot corresponding to the compact 4-cycle of the web, one finds the Seiberg-Witten curve for a local $\mathbb{P}^2$ is given by
\begin{align}\label{eq:SWE0}
    w t+ u +w^{-1}+ t^{-1} = 0  \ . 
\end{align}

With various hypermultiplets or higher Chern-Simons levels, 5-brane web diagrams often are 5-brane configurations with external 5-branes bound to a single 7-brane, which lead to 5-brane webs crossing through one another \cite{Benini:2009gi}, whose dual diagram is depicted with white dots to distinguish from usual black dots. Such a diagram is called a dot diagram, toric(-like) diagram, or generalized toric diagram. 
The corresponding Seiberg-Witten curves can be obtained in a similar way by treating white dots as degenerated polynomials \cite{Kim:2014nqa}. We note that for a 5-brane with an O5-plane, one constructs the Seiberg-Witten curve with caution that suitable boundary conditions for an O5-plane need to be imposed to respect a plane project such that the curves should be invariant under the exchange of $w\leftrightarrow w^{-1}$. See \cite{Hayashi:2017btw} for more details. 

\subsection{\texorpdfstring{SO($2N$)}{SO(2N)} gauge theory with 
\texorpdfstring{$N_f$}{Nf} flavors}\label{sec:SOgaugetheory}
We now discuss the Seiberg-Witten curves SO($2N$) theory with a symmetric hypermultiplet with $N_f$ flavors, which can be described with a 5-brane web with an O7-plane. We consider the construction of the Seiberg-Witten curve from pure SO($2N$) theory and extend the construction to the case with hypermultiplets in the fundamental representation.

\paragraph{Pure SO($2N$).}

\begin{figure}
    \centering
    \includegraphics[width=11cm]{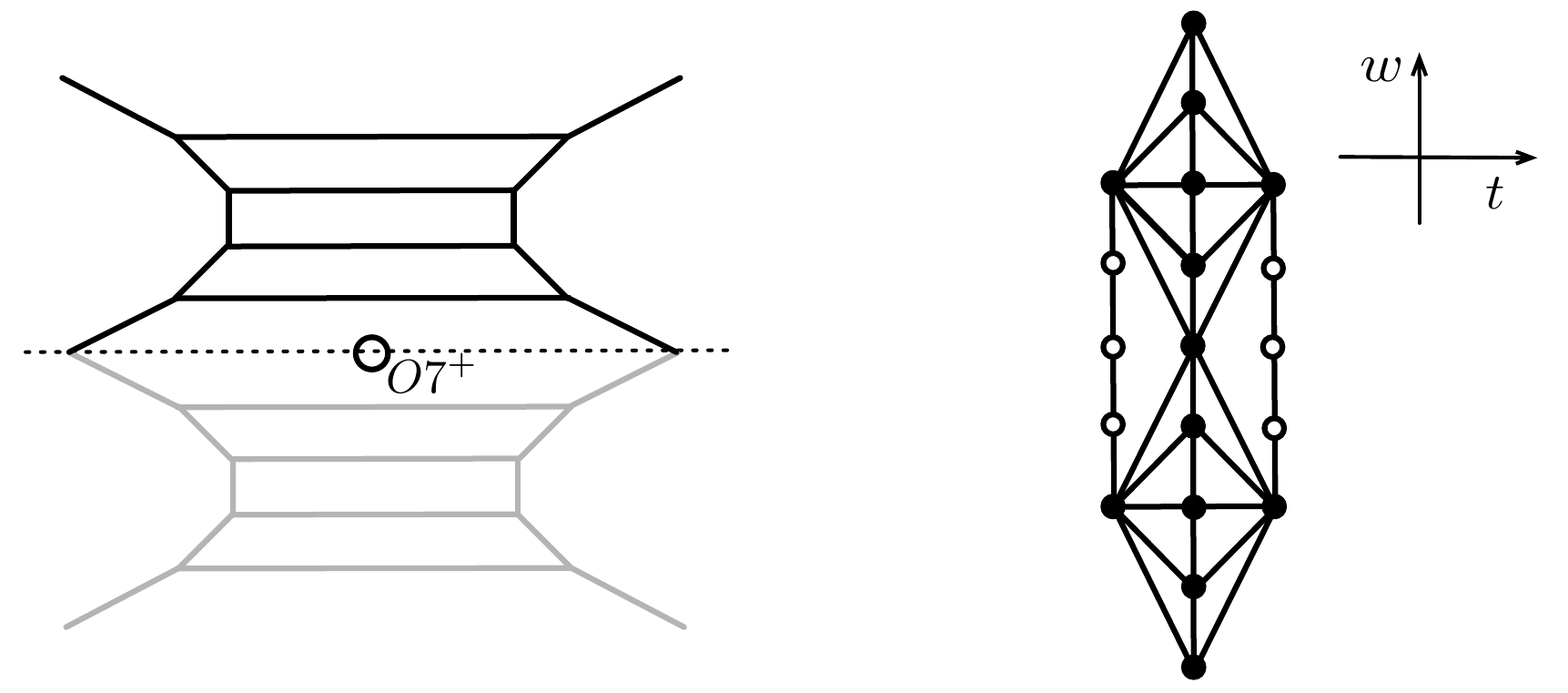}
    \caption{A 5-brane web for pure SO($2N$) and the corresponding dual (non-)toric diagram. Here we chose $N=4$ as a representative example. The gray part below the cut of the O7$^+$-plane represents the reflected mirror image by the O7$^+$-plane.}
    \label{fig:SO}
\end{figure}
Let us first consider SO($2N$) theory without matter, where $N> 2$. The 5-brane web diagram of this theory and its corresponding dual (non-)toric diagram are depicted in figure \ref{fig:SO}. The characteristic equation takes the form 
\begin{align}\label{eq:SWSO2N-char}
p_2(w) t^2+ p_1(w)t + p_0(w)=0\ , 
\end{align}
where $p_0(w)$ and $p_2(w)$ are at most quadratic in $w$ irrespective of $N$. 
It follows from the symmetry due to an O7$^+$-plane that the characteristic equation is invariant under the exchange of $(t, w)\leftrightarrow (t^{-1}, w^{-1})$:
\begin{align}
p_2(w^{-1}) t^{-2} + p_1(w^{-1})t^{-1} + p_0(w^{-1})=0\ . 
\end{align}
which leads to 
\begin{align}
    p_0(w) = p_2(w^{-1}), \qquad p_1(w) = p_1(w^{-1})\ . 
\end{align}
It also follows from the symmetry of the dual toric diagram given in figure \ref{fig:SO} that
\begin{align}
    p_0(w) = p_2(w), 
    \qquad  p_1(w) = \sum_{n=0}^N C_n (w^n+ w^{-n})\ ,\label{eq:p2p0p1forSO}
\end{align}
where $C_n$ ($n=0, \cdots, N-1$) can be set to be the Coulomb moduli parameters.

The boundary conditions that an O7$^+$-plane requires are as follows. Upon T-duality, an O7$^+$-plane becomes two O6$^+$-planes that are located along antipodal points of the T-dual circle. We choose that the positions of two O6$^+$-planes are at $w=\pm 1$. It follows that at $w=\pm 1$, the functions $p_{0,2}(w)$ as $t\to 0$ or $t\to \infty$ satisfy
\begin{equation}\label{boundary_condition}
\begin{aligned}
    p_0(w=\pm 1) = 0 &\quad \Rightarrow \quad p_0 (w) = 
    (w-w^{-1})^2    \ ,\cr
        p_2(w=\pm 1) = 0 &\quad \Rightarrow \quad p_2 (w) = 
    (w-w^{-1})^2    \ .
\end{aligned}    
\end{equation}
This is also consistent with the construction of 4d Seiberg-Witten curves with an O6$^+$-plane in \cite{Landsteiner:1997ei} where the contribution of an O6$^+$-plane to the Seiberg-Witten curve is effectively given by the $\mathbb{Z}_2$ symmetry and two D6-branes at the location of the orientifold. In this case, the boundary condition \eqref{boundary_condition} may be understood effectively as the contribution of two (virtual) D6-branes at each point of the O6$^+$-planes.

For $w$ large, asymptotic behavior of the curves gives rise to the relation to the instanton factor $q$ as $t\sim q^{-1}w^{N-2}$ for $t>\!\!> 1$ and $t\sim q w^{-N+2}$ for $t<\!\!<1$, which leads to
\begin{align}
    C_N = q^{-1} \ .
\end{align}
The Seiberg-Witten curves for pure SO($2N$) gauge theory is then expressed as
\begin{align}\label{pureSOcurve}
    (w-w^{-1})^2 t^2 +  
    \Big( q^{-1} (w^N+w^{-N}) + \sum_{n=0}^{N-1} u_n(w^n+w^{-n})  \Big)   
    t +  (w-w^{-1})^2 = 0,
\end{align}
where $u_n$ corresponds to $N$ Coulomb moduli parameters of the theory. 
We note that in this pure case, the construction of the Seiberg-Witten curves based on an O7$^+$-plane is not much different from that of an O5-plane \cite{Hayashi:2017btw}.

Structurally, Seiberg-Witten curve for $\mathrm{SO}(2N)+N_f\mathbf{F}$ is of the following form
\begin{align}\label{eq:structure of SO(2N)+Nf}
(w-w^{-1})^2\check{p}_2(w)t^2
    +\check{p}_1(w) t 
    + (w-w^{-1})^2\check{p}_2(w^{-1})     
    = 0,
\end{align}
where $\check{p}_{1,2}(w)$ are a Laurent polynomial which is determined by physical parameters of the physics, masses of flavors, Coulomb branch parameters, and the instanton factor. In what follows, we discuss their detailed form according to the number of flavors.

\subsubsection{\texorpdfstring{SO($2N)+N_f(\le 2N-4) \,\mathbf{F}$}{SO(2N)+Nf<2N-4 F}}
\begin{figure}
    \centering
    \includegraphics[width=10cm]{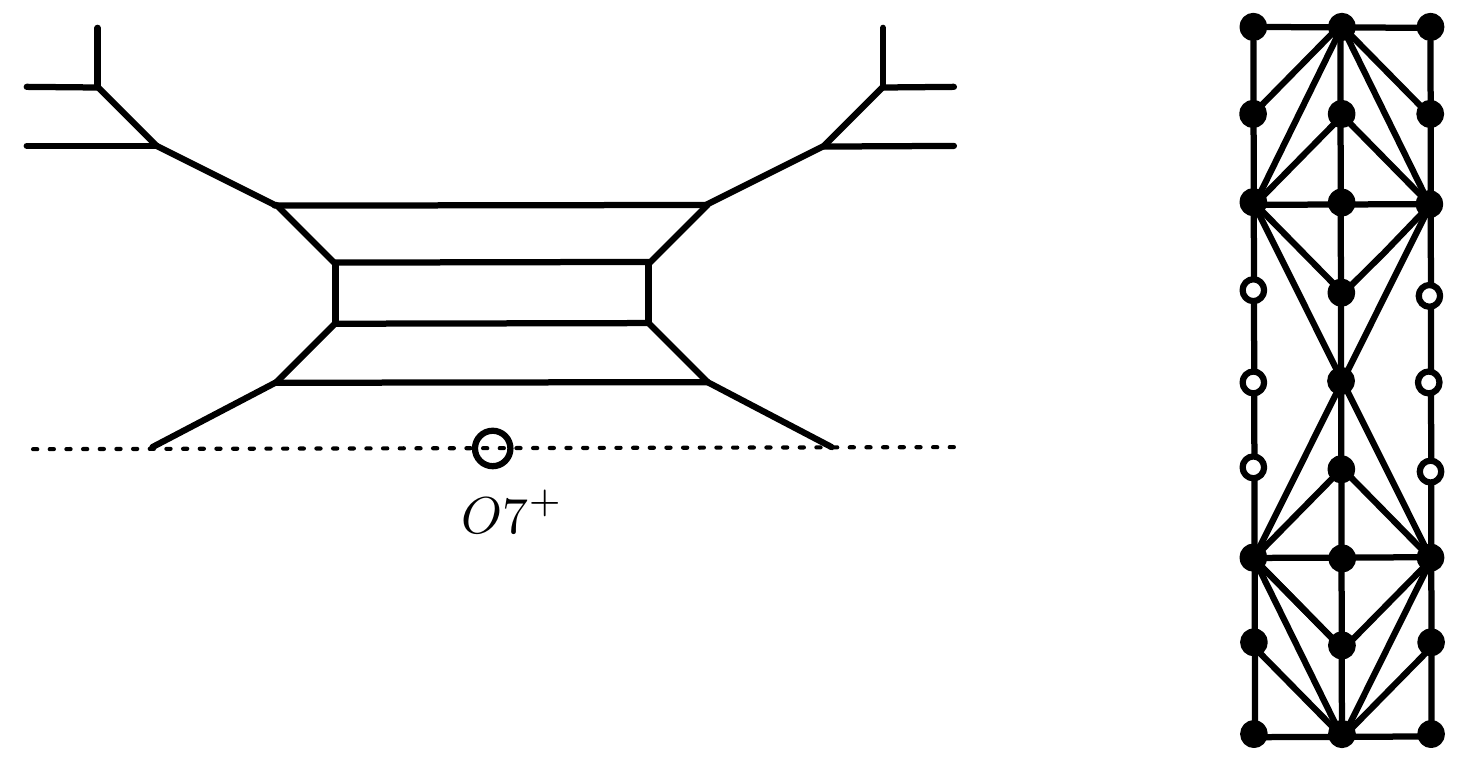}
    \caption{A 5-brane web for pure SO($2N$) with flavors and the corresponding dual (non-)toric diagram. Here $N=4$ and $N_f=4$.}
    \label{fig:SO+F}
\end{figure}
Let us now introduce $N_f\le 2N-4$ hypermultiplets in the fundamental representation (flavors). The 5-brane diagram and the dual (non-)toric diagram are depicted in figure \ref{fig:SO+F}. 
Due to the O7$^+$-plane, we have $N_f$ flavor branes on the left-hand and right-hand sides, respectively, including the mirror images.
We impose the boundary condition at $t \to \infty$ that the flavor branes exist at $w=M_i$ ($i=1,\cdots, N_f$) where $M_i$ are the fugacity of flavor mass parameters. From this condition, we find
\begin{align}\label{eq:SOp0}
    p_0(w) = (w-w^{-1})^2 \prod_{i=1}^{N_f}(w^{-1}-M_i),
\quad  
    p_2(w) = (w-w^{-1})^2 \prod_{i=1}^{N_f}(w-M_i), 
\end{align}
while $p_1(w)$ is the same as the one in \eqref{eq:p2p0p1forSO}. 
The asymptotic behavior of the curves also relates the coefficient $C_N$ to the instanton factor $q$. 
For large $w$, dominant equation is
\begin{align}
w^2 (-1)^{N_f} \prod_{i=1}^{N_f} M_i + C_N w^{N} t + w^{2+N_f} t^2 = 0.
\end{align}
Denoting these two solutions for $t$ as $t_1(w)$ and $t_2(w)$, we denote the ratio of these solutions at $w \to \infty$ as
\begin{align}\label{eq:asymt1t2}
\frac{t_1(w)}{t_2(w)} \sim (-1)^{N_f} q^2 w^{N_f-2N+4} + \mathcal{O}(w^{N_f-2N+3}) \text{ as } w \to \infty \ ,
\end{align}
which leads to 
\begin{align}\label{eq:SONfCN}
C_N = 
\left\{ 
\begin{array}{ll}
q^{-1} \displaystyle \prod_{i=1}^{N_f}M_i^\frac12, & \text{ for } N_f \le 2N-5 ,\\
(1+q^{-1}) \displaystyle \prod_{i=1}^{N_f}M_i^\frac12, & \text{ for } N_f = 2N-4 \ . \\
\end{array}
\right. 
\end{align}
We identify this $q$ as the instanton factor of the gauge theory.\footnote{The square of the instanton factor $q^{2}$ can be interpreted as the distance between the two points where two external edges intersect with the cut of the O7$^+$-plane when they are extrapolated up to a factor $(-1)^{N_f}$. 
Here, the factor $(-1)^{N_f}$ is introduced to make it consistent with the decoupling limit of the flavors: By taking the limit $q \to 0$ while keeping the new instanton factor $q_{N_f-1}^2 = q_{N_f}^2 /M_{N_f}$ finite, we obtain the Seiberg-Witten curve with one less flavor. See also \cite{Li:2021rqr} for a similar discussion with Sp$(N)$ theories where the introduction of the factor $(-1)^{N_f}$ makes the decoupling limit consistent with  \cite{Hayashi:2017btw}. }
The Seiberg-Witten curve for SO($2N$) theory with $N_f\le 2N-5$ flavors is then given by
\begin{align}
     (w-w^{-1})^2 \prod_{i=1}^{N_f}(w^{-1}-M_i)+ p_1(w) t+ \Big((w-w^{-1})^2  \prod_{i=1}^{N_f}(w-M_i)\Big) t^2 = 0\ , 
\end{align}
with
\begin{align}\label{eq:SOFp1}
    p_1(w) = q^{-1}(w^N+w^{-N}) \prod_{i=1}^{N_f}M_i^\frac12 + \sum_{n=0}^{N-1} u_n(w^n+w^{-n})  \ ,
\end{align}
where $u_n$ are the Coulomb moduli parameters. 
For $N_f = 2N-4$, we replace $q^{-1}$ by $1+q^{-1}$ as in \eqref{eq:SONfCN}.

\subsubsection{\texorpdfstring{SO($2N)+ (2N-3)\,\mathbf{F}$}{SO(2N)+2N-3 F}}
\begin{figure}
    \centering
    \includegraphics[width=12cm]{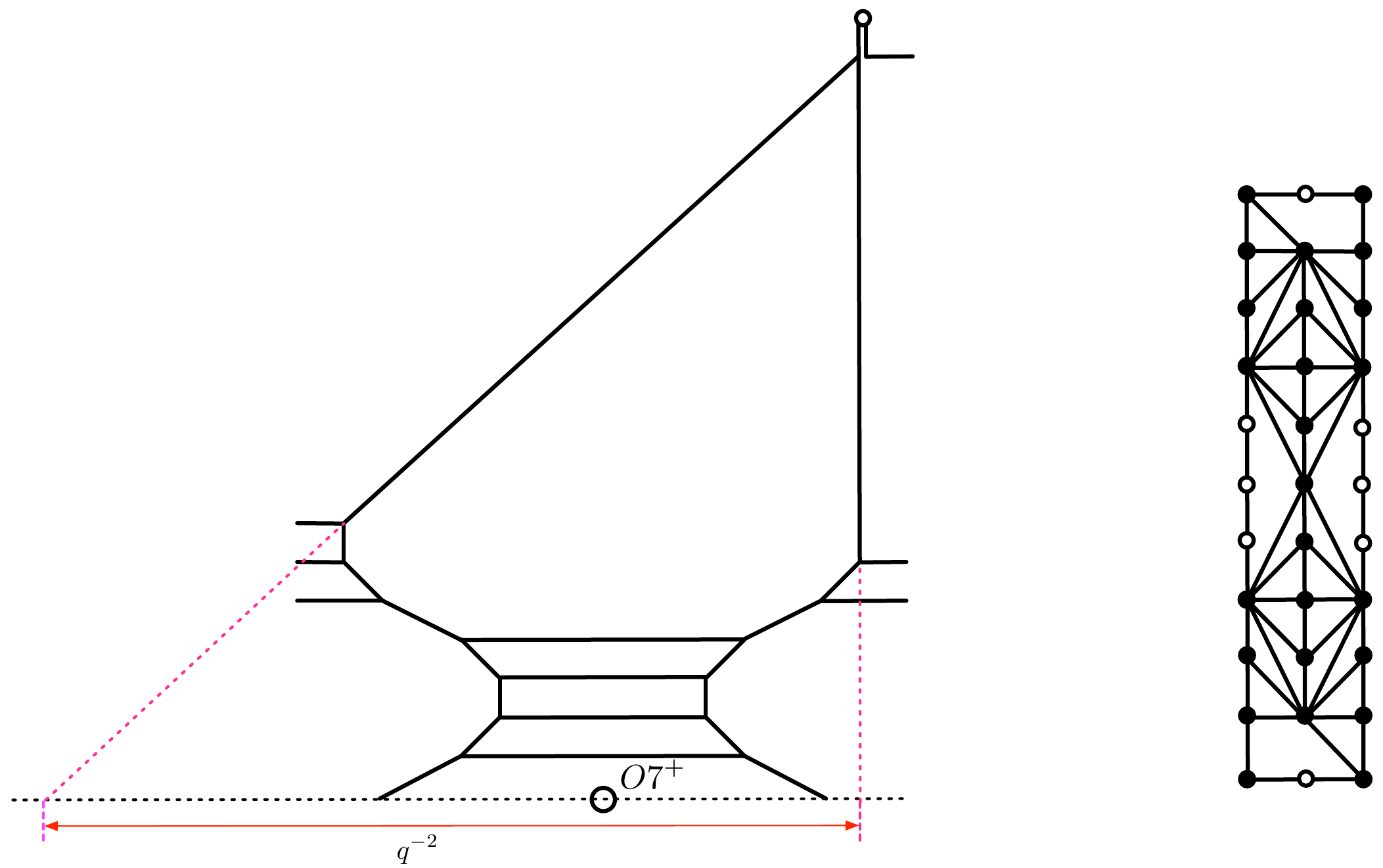}
    \caption{A 5-brane web for pure SO($2N$) with $2N-3$ flavors and the corresponding dual (non-)toric diagram. Here $N=4$.}
    \label{fig:SO+2N-3F}
\end{figure}
As SO($2N$) theory with $2N-2$ flavors is a KK theory \cite{Hayashi:2015vhy}, the 5d SO($2N$) gauge theory with $N_f=2N-3$ flavors is the next to the marginal theory, whose dual toric-like diagram has different feature compared to the cases with less flavors, which is a white dot as shown in figure \ref{fig:SO+2N-3F}. It follows from the 5-brane web given in figure \ref{fig:SO+2N-3F} that there arises an extra flavor associated with the instanton whose mass is given by 
$M_{0} = q^{-2}$. 
It then yields that 
\begin{align}\label{eq:p0p2for SO2N+2N-3}
    p_2(w) &= w^{-N+1} (w-w^{-1})^2 \prod_{i=0}^{2N-3}(w-M_i), \qquad p_0(w) = p_2(w^{-1})\ .     
\end{align}
The $p_1(w)$, in this case, has one more term associated with the white dot in the middle of the figure:
\begin{align}\label{eq:p1for SO2N+2N-3}
p_1(w)=\sum_{n=0}^{N+1} C_n (w^n+ w^{-n})\ , 
\end{align}
where $C_n$ ($n=0, \cdots, N-1$) are the Coulomb branch parameters, while $C_N$ and $C_{N+1}$ are determined in the following.

We note that the terms coming from white dots yield a sequence of a degenerated polynomial, $(t-t_0)^n$, as a 5-brane configuration with white dots indicates that more than one 5-branes are bound to a single 7-brane with the same charge, and,  in this case, we have two NS5-branes are located at $t=t_0$. The order of the degenerated polynomial reduces one by one, as explained in \cite{Kim:2014nqa}. This can be explicitly seen from asymptotic behavior. 
At large $w$, asymptotic behavior gives rise to a quadratic equation in $t$, 
\begin{align}\label{eq:SOwN+1}
    w^{N+1}  \Big(t^2 + C_{N+1} \, t +\prod_{i=0}^{2N-3} M_i \Big) \sim 0 \ .
\end{align}
As the dual diagram has a white dot, this quadratic polynomial in $t$ of \eqref{eq:SOwN+1} should be proportional to the following degenerate polynomial, 
\begin{align}
t^2 + C_{N+1} \, t +\prod_{i=0}^{2N-3} M_i =   (t - t_0 \big)^2 \ ,
\end{align}
and hence 
\begin{align}
 t_0 = - \prod_{i=0}^{2N-3} M_i{}^\frac12  \ , 
\end{align}
from which one finds
\begin{align}\label{eq:CN+1for SO2N+2N-3}
    C_{N+1} = 2 \prod_{i=0}^{2N-3}M_i^\frac12\ . 
\end{align}
The terms of order $w^{N}$ are proportional to one less power of this degenerated polynomial, $(t-t_0)$. A little calculation gives
\begin{align}\label{eq:CNfor SO2N+2N-3}
    C_N = 
    \left( \prod_{k=0}^{2N-3}M_k^{-\frac12} \right) \sum_{i=0}^{2N-3}\Big(
    M_{i} + M_{i}{}^{-1}
    \Big)\ .
\end{align}
Therefore, the Seiberg-Witten curve for 
SO($2N$) theory with $2N-3$ flavors is expressed as
\begin{align}
& \left( w^{-N+1} (w-w^{-1})^2 \prod_{i=0}^{2N-3}(w-M_i) \right) \, t^2 
\cr
& + \left[ 2 \prod_{i=0}^{2N-3}M_i^\frac12 (w^{N+1}+w^{-N-1}) 
\right.
\cr
& \qquad 
\left. 
+ \left( \prod_{k=0}^{2N-3}M_k^{-\frac12} \right) \sum_{i=0}^{2N-3}\Big(
    M_{i} + M_{i}{}^{-1}
    \Big)\ (w^N+w^{-N})
+ \sum_{n=0}^{N-1} u_n (w^n+w^{-n})
\right] t
\cr
& + w^{N-1} (w-w^{-1})^2 \prod_{i=0}^{2N-3}(w^{-1}-M_i)
= 0 \ , 
\end{align}
 with $u_n$ ($n=0, \cdots, N-1$) being the $N$ Coulomb branch parameters of SO(2$N$) gauge theory.

\subsubsection{Higgsing to \texorpdfstring{SO($2N+1$)}{SO(2N+1)} gauge theories}
It is well-known that 5-brane webs for SO($2N+1$) gauge theories can be obtained from a Higgsing from SO($2N+2$)+1$\mathbf{F}$. For instance, for a 5-brane configuration with an O5-plane, a flavor gets zero mass that aligns with one of the color 5-branes which opens this Higgs branch, and as a result,  a half-color brane and half D7-brane cut are stuck at an O5-plane that makes an $\widetilde{\rm O5}$-plane and hence an SO($2N+1$) gauge theory is described by a 5-brane web with an $\widetilde{\rm O5}$-plane \cite{Zafrir:2015ftn}. 

\begin{figure}[t]
    \centering
    \includegraphics[width=15cm]{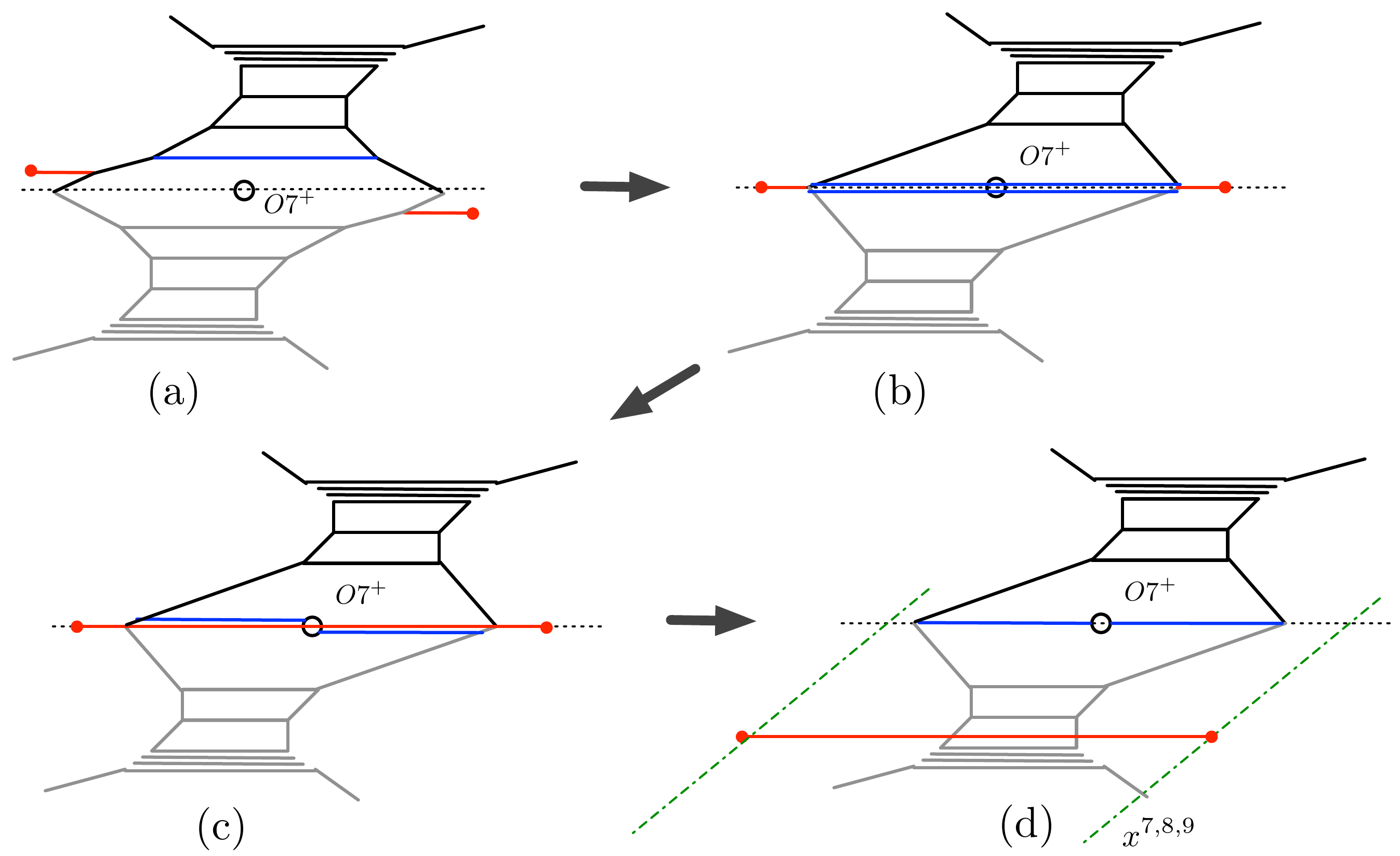}
    \caption{Higgsing of SO($2N+2$)+$1\mathbf{F}$ to SO($2N+1$), where the grey part below an O7$^+$-plane represents the reflected images. (a) A 5-brane web for SO($2N+2$)+$1\mathbf{F}$. (b) A color D5-brane and a flavor D5-brane are brought to the cut of an O7$^
    +$-plane. (c) As the Higgs takes place, the color and flavor 5-brane and its reflected image are combined. (d) After the Higgsing a D5-brane (in red) can be Higgsed away along the $x^{7,8,9}$-direction and the remaining web becomes a 5-brane web for a pure SO($2N+1$) where a half D5-brane is stuck at the point of an O7$^+$-plane.}
    \label{fig:SO2N+1Higgsing}
\end{figure}

For a 5-brane web with an O7$^+$-plane, SO($2N+1$) gauge theories can be understood with a half color brane stuck at the cut of O7$^+$-plane and it is also consistent with the Higgsing from SO($2N+2$)+1$\mathbf{F}$, which can be understood as follows.  When a color D5-brane and a flavor bran are placed along the cut of an O7$^+$-plane, there are two copies of half of their own images due to the point-like projection of an O7$^+$-plane. As a result, along the cut of an O7$^+$-plane there are two half-color D5-branes suspended between two half NS5-branes, and through the Higgsing, half of these two half-color D5-branes are connected to half of these flavor D5-branes which make two copies of half D5-brane. The Higgsing takes place as these two copies depart from the 5-brane web to the $x^{7,8,9}$ directions as depicted in figure \ref{fig:SO2N+1Higgsing}. 
The corresponding Seiberg-Witten curve for SO($2N+1$) can be therefore obtained from that for SO($2N+2$)$+1\mathbf{F}$. 

This Higgsing is realized by tuning one of the mass parameters to be $M_1=1$ and also by imposing $p_1(1)=0$ to the expression in \eqref{eq:SOFp1}, we find the constraint
\begin{align}
q^{-1} + \sum_{n=0}^{N} u_n = 0,
\end{align}
from which we obtain
\begin{align}
p_0(w) 
&= (w-w^{-1})^2 (w^{-1}-1),
\qquad 
p_2(w) 
= (w-w^{-1})^2 (w-1),
\cr
p_1(w) 
&= w^{-1} (w-1)^2 \left( q^{-1} (w^{N}+w^{-N}) + \sum_{n=0}^{N-1} U_n (w^n+w^{-n})) \right), 
\end{align}
where we put 
\begin{align}
U_n = (N-n) q^{-1} + \sum_{k=n+1}^{N-1} (k-n) u_k.
\end{align}
This gives the Seiberg-Witten curve for SO($2N+1$) gauge theory.

Or, at the cost of sacrificing the manifest invariance under $(t,w)\to (t^{-1},w^{-1})$, we can factor out factoring out $(w-1)$, to obtain the following simpler form for the Seiberg-Witten curve:
\begin{align}\label{eq:SW-SOodd}
(w-w^{-1})^2 t^2 + w^{-1}(w-1) \left( q^{-1} (w^{N}+w^{-N}) + \sum_{n=0}^{N-1} U_n (w^n+w^{-n})) \right) t
\cr
- w^{-1}(w-w^{-1})^2  = 0.
\end{align}

\subsection{\texorpdfstring{SU($N$)}{SU(N)} gauge theory with a symmetric and \texorpdfstring{$N_f$}{Nf} flavors}\label{sec:SU+1Sym}

\begin{figure}
    \centering
    \includegraphics[width=14cm]{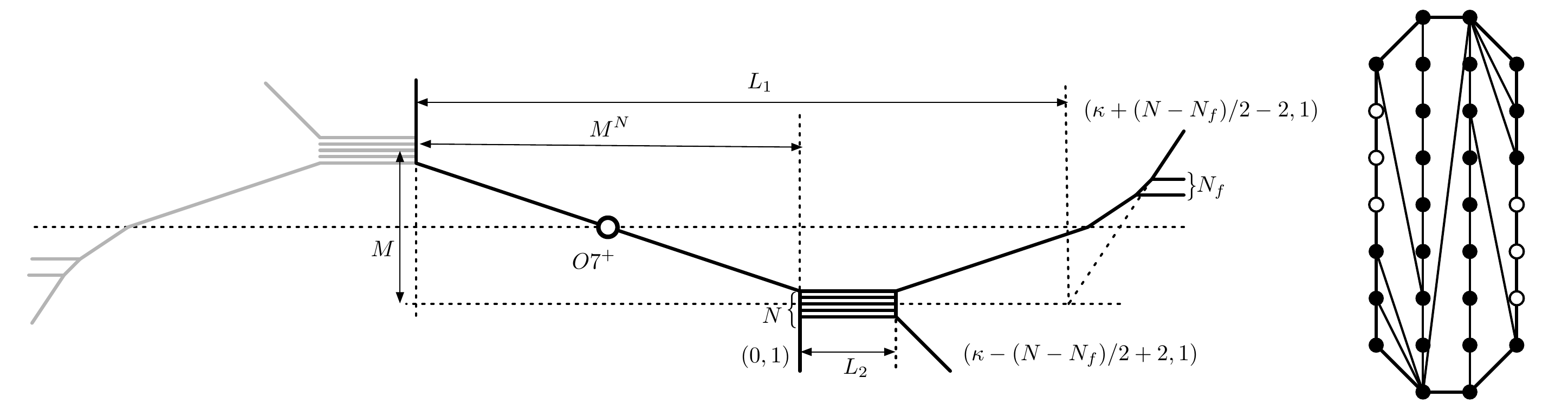}
    \caption{A 5-brane web for SU($N$) gauge theory with a symmetric tensor and with $N_f$ flavors, and the corresponding dual (non-)toric diagram for $N=8$, $N_f=2$, $\kappa=0$.}
    \label{fig:SU+S+F}
\end{figure}

We consider SU($N$)$_\kappa$ gauge theory with a symmetric matter and $N_f$ flavors, whose 5-brane web has an O7$^+$-plane at which an $(p,1)$ 5-brane is stuck as depicted in figure \ref{fig:SU+S+F}.
For $N\ge 3$, $\kappa$ is the Chern-Simons level, while for $N=2$, $\kappa$ plays the role of the $\mathbb{Z}_2$-valued discrete theta parameter $\theta = 0$ or $\pi$, respectively for $\kappa$ even or odd. 

The corresponding characteristic equation leads to a cubic equation in $t$, 
\begin{align}
F_{SW}(t,w) \equiv p_3(w) t^3  + p_2(w)t^2 + p_1(w) t + p_0(w)= 0\ ,
\label{eq:chSW-SUN+1Sym}
\end{align}
where $p_m(w)$ are Laurent polynomial in $w$
\begin{align}\label{eq:pm}
	p_m(w) = \sum_{n=n_m^-}^{n_m^+} C_{m,n} w^n\ . 
\end{align}
Here, $n_m^\pm$ depends on the rank of the gauge group, the number of flavors $N_f$, and the Chern-Simons level $\kappa$.

The $\mathbb{Z}_2$ symmetry due to an O7$^+$-plane demands \eqref{eq:chSW-SUN+1Sym} to be invariant under $(w, t)\leftrightarrow (w^{-1}, t^{-1})$, which yields
\begin{align}
c\, t^3 F_{SW}(t^{-1},w^{-1})
&= 
c \, p_3(w^{-1}) + c\,  p_2(w^{-1})\, t+ c\,p_1(w^{-1})\,t^2+ c\, p_0(w^{-1})\, t^3 \cr
&= F_{SW}(t,w)\ .
\end{align}
Here, $c$ is a multiplicative constant subject to $c^2=1$. We choose $c=-1$ so that the location of the orientifold plane along the $t$-coordinate is given as $t=1$. In other words, upon T-duality, an O7$^+$-plane is split into two O6$^+$-planes which are located at $(w, t) = (\pm1, 1)$.  
With $c=-1$, one readily finds that 
\begin{align}\label{eq:O7p-po123}
	p_0(w) = - p_3(w^{-1}), \qquad 	p_1(w) = - p_2(w^{-1})\ . 
\end{align}

As discussed in section \ref{sec:SOgaugetheory}, asymptotic behaviors as $t\to\infty$ and $t\to 0$ give rise to the following boundary conditions that the curves should satisfy $p_{0} (w=\pm1)=0$ due to the O7$^+$ plane. 
From this constraint, one finds that $p_{3}(w)$ are of the following form, 
\begin{align}\label{eq:p0 and p3 behaviors} 
	p_3(w) & = (w-w^{-1})^2 \,\check{p}_3(w)\ , 
\end{align}
where $\check{p}_{3}(w)$ is a Laurent polynomial. From the constraints \eqref{eq:O7p-po123} and \eqref{eq:p0 and p3 behaviors}, we find that the  Seiberg-Witten curve \eqref{eq:chSW-SUN+1Sym} for $\mathrm{SU}(N)_{\kappa}+N_f\mathbf{F}+1\mathbf{Sym}$ has following structure: 
\begin{align}
(w-w^{-1})^2 \,\check{p}_3(w) t^3 + p_2(w)\, t^2 - p_2(w^{-1})\,t - (w-w^{-1})^2 \,\check{p}_3(w^{-1})\,  = 0\ .
\label{eq:SW-SUN+1Sym-structure}
\end{align}

In the following, we relate the coefficients in $p_i(w)$ with mass parameters. For simplicity, we assume $N_f < N-4$ and $|\kappa| < \frac{1}{2}(N-4-N_f)$. In this parameter region, no external branes intersect with each other. 

First, analogous to \eqref{eq:SOp0}, $\check{p}_3(w)$ is written as 
\begin{align}\label{eq:SU+S-p1}
    \check{p}_3(w) = w^{\alpha} \prod_{i=1}^{N_f}(w-M_i), \quad  
\end{align}
where $M_i$ is the mass parameters of the $N_f$ flavors while $\alpha = n_3^+ - N_f - 2 = n_3^- + 2$. 

Next, we denote the three solutions of \eqref{eq:SW-SUN+1Sym-structure} for $t$ as $t_i(w)$ ($i=1,2,3$) and consider their asymptotic behavior. 
Without the loss of generality, we assume $|t_1(w)|< |t_2(w)|<|t_3(w)|$ at large $w$.
Since they satisfy 
\begin{align}\label{eq:symti}
t_i(w^{-1})^{-1} = t_i (w) \qquad (i=1,2,3)
\end{align}
due to the $\mathbb{Z}_2$ symmetry $(w, t)\leftrightarrow (w^{-1}, t^{-1})$, 
they are ordered 
$|t_3(w)|< |t_2(w)|<|t_1(w)|$ at small $w$.
Due to the parameterization in figure \ref{fig:SU+S+F}, the mass parameter $M$ of the symmetric tensor, which is given by the distance between the center of mass of the color branes and their mirror image, is related to the asymptotic behavior of $t_2(w)$ as 
\begin{align}\label{eq:SUsym-t2sim}
t_2(w) 
\sim 
\frac{p_2(w^{-1})}{p_2(w)}
\sim
(-1)^N M^{\frac{N}{2}} w^{-n_2^{+}-n_2^{-}}\quad 
\text{ as } w \to \infty\ .  
\end{align}
The factor $(-1)^N$ is introduced to make it consistent with Higgsing from SU($N$) to SU($N-1$) as we will see later.
From \eqref{eq:SUsym-t2sim}, we find that the relation among the coefficients in $p_2(w)$ as
\begin{align}\label{eq:symMC}
\frac{C_{2,n_2^-}}{C_{2,n_2^+}} = (-1)^N M^{\frac{N}{2}}.
\end{align}

We also claim that the instanton factor is given by the geometric average of the two distances $L_1$, $L_2$ of the two external 5-branes above and below, respectively, 
by interpolating the asymptotic behavior of them at the center of mass of the color branes $w=M^{\frac{1}{2}}$.
This is readily proposed in various past literature including \cite{Bao:2011rc}.
Expressing this condition in terms of $t_i(w)$, we write 
\begin{align}
q^2 = (-1)^{N_f+N} L_1 L_2\ , 
\end{align}
with
\begin{align}\label{eq:L1L2}
\frac{t_2(w)}{t_3(w)} 
&\sim
- \frac{p_3(w) p_2(w^{-1})}{p_2(w)^2}
\sim L_1 \left( \frac{w}{M^{\frac{1}{2}}} \right)^{ n_3^+ - n_2^- - 2n_2^+} 
~\text{ as }~~ w \to \infty,
\cr
\frac{t_2(w)}{t_1(w)} 
&\sim 
- \frac{p_3(w) p_2(w^{-1})}{p_2(w)^2}
\sim L_2 \left( \frac{w}{M^{\frac{1}{2}}} \right)^{n_3^- - n_2^+ - 2n_2^-}
~\text{ as }~~ w \to 0 .
\end{align}
Here, we introduce the sign factor $(-1)^{N_f+N}$ analogous to \eqref{eq:asymt1t2}.
Rewriting the second expression in \eqref{eq:L1L2} by rewriting $w \to w^{-1}$, this condition leads to
\begin{align}
\frac{t_1(w)}{t_3(w)} 
&\sim 
\frac{p_3(w) p_3(w^{-1})}{p_2(w) p_2(w^{-1})}\cr
&\sim 
(-1)^{N_f+N} q^2 M^{\frac{1}{2} (-n_3^+ - n_3^- + 3 n_2^+ + 3 n_2^- )}
w^{n_3^+ - n_3^- + n_2^- - n_2^+}
~~
\text{ as }~~ w \to \infty.
\end{align}
From this, we find another relation among the coefficients in $p_2(w)$
\begin{align}
C_{2,n_2^+} C_{2,n_2^-} =
(-1)^{N} q^{-2} M^{\frac{1}{2} (n_3^+ + n_3^- - 3 n_2^+ - 3 n_2^- )}
\prod_{i=1}^{N_f} M_i.
\end{align}
Combined with \eqref{eq:symMC}, we find
\begin{align}\label{eq:C1n1pm}
C_{2,n_2^+} &= q^{-1} M^{\frac{1}{4} (n_3^+ + n_3^- - 3 n_2^+ - 3 n_2^-  - N)} \prod_{i=1}^{N_f} M_i^{\frac{1}{2}}\ ,
\cr
C_{2,n_2^-} &= (-1)^{N} q^{-1} M^{\frac{1}{4} (n_3^+ + n_3^- - 3 n_2^+ - 3 n_2^-  + N) } \prod_{i=1}^{N_f} M_i^{\frac{1}{2}} \ .
\end{align}

With this result, we find that the Seiberg-Witten curve can be written more explicitly.
As we see below, the explicit form depends on whether $N$ is even or odd. 

\subsubsection{\texorpdfstring{$\mathrm{SU}(2n)$}{SU(2n)} cases}

We consider the Seiberg-Witten curve for $\mathrm{SU}(2n)_{\kappa}$ with $N_f$ flavors,
where we assume $N_f < 2n-4$ and $|\kappa| < n-2-\frac{1}{2}N_f$ for simplicity.  
In this case, the summation region $n_m^\pm$ in \eqref{eq:pm} can be read off from figure \ref{fig:SU+S+F} and is given by
\begin{align}\label{eq:sumrange-even}
n_3^{\pm} = \pm \frac{N_f+4}{2} - \kappa\ , 
\quad
n_1^{\pm} = n_2^{\pm} = \pm n\ , 
\quad
n_0^{\pm} = \pm \frac{N_f+4}{2} + \kappa\ .
\end{align}
Note that the Chern-Simons level $\kappa$ is an integer if $N_f$ is even while it is half-integer if $N_f$ is odd so that $n_0^{\pm}$ and $n_3^{\pm}$ above are always integers.

With these, we find the Seiberg-Witten curve is given explicitly as
\begin{align}\label{eq:SW-SUeven-1Sym}
& \left( w^{-\frac{N_f}{2}-\kappa} (w-w^{-1})^2 \,\prod_{i=1}^{N_f} (w - M_i) \right) \, t^3 
\cr
& + 
q^{-1} M^{-\frac{\kappa+n}{2}} \left( \prod_{i=1}^{N_f} M_i{}^{\frac{1}{2}} \right) 
w^{-n} \left( w^{2n} 
+ \sum_{k=1}^{2n-1} U_k w^k + M^{n}
\right) \, t^2
\cr
& - q^{-1} M^{-\frac{\kappa+n}{2}} \left( \prod_{i=1}^{N_f} M_i{}^{\frac{1}{2}} \right) 
w^{n} \left( w^{-2n} 
+ \sum_{k=1}^{2n-1} U_k w^{-k} + M^{n}
\right)  \, t
\cr
& - w^{\frac{N_f}{2}+\kappa} \, (w-w^{-1})^2 \, \prod_{i=1}^{N_f} (w^{-1}-M_i) = 0\ .
\end{align}

\subsubsection*{Higgsing from \texorpdfstring{$\mathrm{SU}(2n)+1\mathbf{Sym}$}{SU(2n)+1Sym} to \texorpdfstring{$\mathrm{SO}(2n)$}{SO(2n)}}

In the limit where the symmetric hypermultiplet is massless and the Coulomb branch parameters are specially tuned, there arises the Higgs branch that $\mathrm{SU}(2n)_\kappa+1\mathbf{Sym}+N_f\mathbf{F}\to \mathrm{SO}(2n)+N_f\mathbf{F}$. 

It follows from \eqref{eq:SW-SUN+1Sym-structure} that
the curve factorizes at $w=\pm1$ as
\begin{align}\label{eq:facorized at w=pm1}
(t-1) t \ p_{2}(w=\pm 1) =  0\ . 
\end{align}
This implies that three solutions for $t$ at $w=\pm 1$ is $t = 0, 1, \infty$, respectively, assuming that $p_{2}(w=\pm 1) \neq 0$.
The 5-brane web diagrams in figure \ref{fig:HiggsSU2N} imply that we can go to the Higgs branch if one of the solutions is given by $t_2(w)=1$ for an arbitrary value of $w$. 
This is possible only if the condition 
\begin{align}\label{eq:HiggsingConditionto SO}
\check{p}_2(w) \equiv p_3(w)+p_2(w) = p_3(w^{-1})+p_2(w^{-1})
= \check{p}_2(w^{-1})
\end{align}
is satisfied. 
We find from \eqref{eq:symMC} that 
this condition is consistent with the fact that the mass parameter of the symmetric matter should be massless, $M=1$, to go to the Higgs branch since $n_2^+ > n_3^+$ from the assumption.
Under the condition \eqref{eq:HiggsingConditionto SO}, 
the curve is factorized as 
\begin{align}
\label{eq:facorizedtobeSO2N}
(t-1) \Big(\,p_3(w) t^2 +  \check{p}_2(w) t 
+p_3(w^{-1}) \, \Big) =  0 \ .
\end{align}

Taking into account \eqref{eq:p0 and p3 behaviors} and \eqref{eq:HiggsingConditionto SO}, we find that the terms in the bigger parenthesis in \eqref{eq:facorizedtobeSO2N} 
reproduce the structure \eqref{eq:structure of SO(2N)+Nf} of SO($2n$) Seiberg-Witten curve. 
By comparing 
\eqref{eq:SOp0} with \eqref{eq:SU+S-p1} and 
\eqref{eq:SONfCN} with \eqref{eq:C1n1pm} under $M=1$,
we find the dependence of the coefficients in $p_i(w)$ on the mass parameters and on the instanton factor also agrees after the coordinate change $t \to w^{\frac{N_f}{2}+\kappa} \, t$.

Thus, we have found the Higgs mechanism
$\mathrm{SU}(2n)_\kappa+1\mathbf{Sym}+N_f\mathbf{F}\to \mathrm{SO}(2n) + N_f\mathbf{F}$
at the level of the Seiberg-Witten curve.

\subsubsection{\texorpdfstring{$\mathrm{SU}(2n+1)$}{SU(2n+1)} case}

\begin{figure}
    \centering
    \includegraphics[width=15cm]{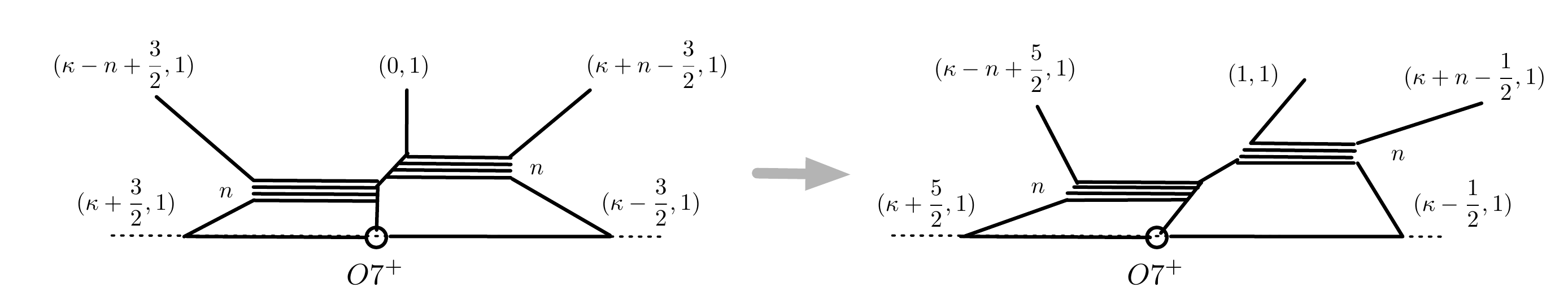}
    \caption{SU($2n+1)_\kappa$+1$\mathbf{Sym}$ and an SL$(2,\mathbb{Z})$~ T-transformed web.}
    \label{fig:SUoddTransf}
\end{figure}

For SU($2n+1$) case, the number of terms in the Laurent polynomials $p_{1,2}$ in \eqref{eq:pm} is, in general, different from those cases for SU($2n$).  
If we try to generalize the result for SU($2n$) naively, we either find the half-integer power of $w$ appears, or otherwise, we need to sacrifice the manifest $\mathbb{Z}_2$ symmetry of the form $F_{SW}(w, t) = t^3 F_{SW} (w^{-1}, t^{-1})$.

To make the power of the monomial to be an integer while keeping the manifest $\mathbb{Z}_2$ symmetry $(w, t) \leftrightarrow (w^{-1}, t^{-1})$, one can take a transformation 
\begin{align}
t \to w^{-1} t,
\end{align}
which is the SL($2,\mathbb{Z}$) T-transformation on 5-brane webs that shifts dual toric-like diagrams. As a result, the middle external NS5-brane is transformed into a ($1, 1$) 5-brane, as depicted in figure \ref{fig:SUoddTransf}.

After this T-transformation, the structure 
\eqref{eq:SW-SUN+1Sym-structure} of the Seiberg-Witten curve is still preserved. However, the summation region $n_m^\pm$ in \eqref{eq:pm} is now given by
\begin{align}\label{eq:sumrange-odd}
&n_0^{\pm} = \pm \left( \frac{N_f}{2}+2 \right) + \kappa + \frac{3}{2}, 
\qquad
n_1^{+} = n+1, 
\qquad
n_1^{-} = -n, 
\cr
&
n_2^{+} = n,
\qquad 
n_2^{-} = -n-1,
\qquad 
n_3^{\pm} = \pm \left( \frac{N_f}{2}+2 \right) - \kappa - \frac{3}{2},
\end{align}
which are slightly different from \eqref{eq:sumrange-even}. 
Note that the Chern-Simons level $\kappa$ is an integer if $N_f$ is odd while it is half-integer if $N_f$ is even so that $n_0^{\pm}$ and $n_3^{\pm}$ above are always integers.

With these, we find the Seiberg-Witten curve for SU($2n+1$) is given explicitly as
\begin{align}\label{eq:SW-SUodd-1Sym}
& \left( w^{-\frac{N_f}{2}-\kappa-\frac{3}{2}} (w-w^{-1})^2 \,\prod_{i=1}^{N_f} (w - M_i) \right) \, t^3
\cr
& +  q^{-1} M^{-\frac{2\kappa+2n+1}{4}} \left( \prod_{i=1}^{N_f} M_i{}^{\frac{1}{2}} \right) w^{-n-1} 
 \left(  w^{2n+1} 
+ \sum_{k=1}^{2n} U_{k} w^k - M^{\frac{2n+1}{2}} 
\right)\,t^2
\cr
& - q^{-1} M^{-\frac{2\kappa+2n+1}{4}} \left( \prod_{i=1}^{N_f} M_i{}^{\frac{1}{2}} \right) w^{n+1} 
 \left(  w^{-2n-1} 
+ \sum_{k=1}^{2n} U_{k} w^{-k} - M^{\frac{2n+1}{2}} 
\right)\,t
\cr
& - w^{\frac{N_f}{2}+\kappa+\frac{3}{2}} \, (w-w^{-1})^2 \, \prod_{i=1}^{N_f} (w^{-1}-M_i) 
= 0\ .
\end{align}
 
 \subsubsection*{Higgsing from \texorpdfstring{$\mathrm{SU}(2n+1)+1\mathbf{Sym}$}{SU(2n+1)+1S} to \texorpdfstring{$\mathrm{SO}(2n+1)$}{SO(2n+1)}}
 
 It follows then that when both $w$ and $t$ are large, an asymptotic boundary condition that the curve satisfies is proportional to $t=w$. The Higgs branch that $\mathrm{SU}(2n+1)_\kappa+1\mathbf{Sym}+N_f\mathbf{F}\to \mathrm{SO}(2n+1)$, leads to the following factorized form
\begin{align}\label{eq:facorizedtobeSO2n+1}
	(t-w) \Big(\, p_3(w)t^2 + \check{p}_2(w) t + w^{-1} p_3(w^{-1})  \, \Big) =  0\ , 
\end{align}
which is subject to the condition, 
\begin{align}\label{eq:HiggsingConditionto SO2n+1}
\check{p}_2(w)\ \equiv w\, p_3(w) + p_2(w) = w^{-2} p_3(w^{-1})+ w^{-1} \, p_2(w^{-1}) = w^{-1} \check{p}_2(w^{-1}) \ ,
\end{align}
where we have assumed $\check{p}_1( \pm 1 ) \neq 0$. 

We find that the bigger parenthesis in \eqref{eq:facorizedtobeSO2n+1} reproduces 
the Seiberg-Witten curve
\eqref{eq:SW-SOodd} for SO($2n+1$) up to the convention change $w \to -w$.
Thus, we have found the Higgs mechanism
$\mathrm{SU}(2n+1)_\kappa+1\mathbf{Sym}+N_f\mathbf{F}\to \mathrm{SO}(2n+1) + N_f\mathbf{F}$
at the level of the Seiberg-Witten curve.

\subsubsection*{Higgsing from \texorpdfstring{$\mathrm{SU}(N)$}{SU(N)} to \texorpdfstring{$\mathrm{SU}(N-1)$}{SU(N-1)} }
We also briefly comment on 
Higgsing from $\mathrm{SU}(N)_{\kappa}+N_f\mathbf{F}+1\mathbf{Sym}$ to $\mathrm{SU}(N-1)_{\kappa}+(N_f-1)\mathbf{F}+1\mathbf{Sym}$ to check the consistency. 
In order to realize this Higgsing, we tune the parameters such that $p_3(1)=p_2(1)=0$ is satisfied. 
Under this tuning, we can check explicitly from 
\eqref{eq:SW-SUeven-1Sym} and \eqref{eq:SW-SUodd-1Sym} that 
\begin{align}
\left. F_{SW}^{SU(2n+2)+N_f \mathbf{F}}(w,t) \right|_{p_3(1)=p_2(1)=0} & = w^{-2}(1-w) F_{SW}^{SU(2n+1)+(N_f-1) \mathbf{F}}(w,-wt) \ ,
\cr
\left. F_{SW}^{SU(2n+1)+N_f \mathbf{F}}(w,t)\right|_{p_3(1)=p_2(1)=0} &= w(1-w) F_{SW}^{SU(2n)+(N_f-1) \mathbf{F}}(w,-w^{-1}t) \ ,
\end{align}
are satisfied after properly redefining the mass of the symmetric tensor and the instanton factor
as
\begin{align}
M_{{\rm SU}(N)} = (M_{{\rm SU}(N-1)})^{\frac{N-1}{N}},
\qquad
q_{{\rm SU}(N)} = q_{{\rm SU}(N-1)} (M_{{\rm SU}(N-1)})^{\frac{\kappa}{2N}}.
\end{align}
They are the Higgsing from $\mathrm{SU}(N)$ to $\mathrm{SU}(N-1)$ for the case of even $N$ and odd $N$, respectively,.

\subsubsection{\texorpdfstring{$\mathrm{SU}(N)_\kappa+1\mathbf{Sym}$ and 4d limit}{SU(N)k+1Sym and 4d limit}}
An SU($N$)$_\kappa$ gauge theory with a symmetric but without flavors has the marginal Chern-Simons level $\kappa=N/2$, which corresponds to a 6d theory on a circle with or without a twist. Here, we consider generic 5d cases. For $\kappa < \frac{N-4}{2}$, $p_{0,1}(w)$ takes the form
\begin{align}
	p_{3}(w) = w^{\alpha} (w-w^{-1})^2 ,\qquad 
	p_{2}(w)= q^{-1} M^{-\frac{\kappa}{2}-\frac{N}{4}} w^{\alpha'}\prod_{i=1}^{N} (w-A_i)\ .
\end{align}
Here, $\alpha = - \kappa$, $\alpha'=-\frac{N}{2}$ for even $N$ while $\alpha = -\kappa-\frac{3}{2}$, $\alpha'=-\frac{N+1}{2}$ for odd $N$. The K\"ahler parameters $A_i$ denote the vertical positions of the color D5-branes from the cut of the O7$^+$-plane in the classical limit. They are related to the mass parameter of the symmetric hypermultiplet by 
\begin{align}
\prod_{i=1}^{N} A_i = M^{\frac{N}{2}} \ ,
\end{align}
which is geometrically the square of the distance between the O7$^+$-plane cut and the center of the Coulomb branch \cite{Hayashi:2018lyv}. The curve is then given by
\begin{equation}
\begin{aligned}
&w^{\alpha} (w-w^{-1})^2 \, t^3 
+ \Big(q^{-1} M^{-\frac{\kappa}{2}-\frac{N}{4}} w^{\alpha'}\prod_{i=1}^{N} (w-A_i)
	\Big)\, t^2
\\
& - \Big(q^{-1} M^{-\frac{\kappa}{2}-\frac{N}{4}} w^{-\alpha'}\prod_{i=1}^{N} (w^{-1}-A_i)
	\Big)\, t
- w^{-\alpha} (w-w^{-1})^2 
 = 0\ .
 \end{aligned}
\end{equation}

To take the 4d limit, we set 
\begin{align}
w=e^{-\beta v}, \qquad A_i=e^{-\beta a_i}, \quad \mathrm{and}\quad q=\frac{(-1)^{N-1}}{4} (\beta\Lambda)^{N-2} \ .	
\end{align}
Here, $\beta$ is the radius of the circle that goes to zero in the 4d limit and $\Lambda$ is the dynamical scale in 4d. It is easy to see that the contributions from the Chern-Simons term are subleading and the curve at order $\beta^2$ leads to the Seiberg-Witten curve for 4d SU($N$) gauge theory with a symmetric hypermultiplet, 
\begin{align}\label{eq:4dSWfor SU(N)+1Sym}
v^2t^3 - \left( \Lambda^{-N+2}\prod_{i=1}^{N}(v-a_i) \right) \, t^2 
 + \left( (-1)^N \Lambda^{-N+2} \prod_{i=1}^{N}(v+a_i) \right) \, t - v^2 = 0 \ . 
\end{align}
By $t = -\Lambda^{-N+2}\, y$, one finds that this curve agrees with \cite{Landsteiner:1997ei}, where the mass of the symmetric hypermultiplet is given by $m =\frac{2}{N}\sum_{i=1}^{N}a_i$, which is also consistent with $M=e^{-\beta m}$.

\bigskip

\section{Construction of Seiberg-Witten curves with an \texorpdfstring{O7$^-$}{O7-}-plane}\label{sec:O7-construction}

In this section, we propose how to compute the Seiberg-Witten curves for 5d $\mathcal{N}=1$ theories which have a construction of a 5-brane web with an O7$^-$-plane. The theories include Sp($N$) gauge theories with fundamental hypers and SU($N$) theories with an antisymmetric and fundamental hypers. Sp($N$) gauge theories can also be engineered from a 5-brane web with an O5-plane, and we show that our construction of the Seiberg-Witten curve from a 5-brane with an O7$^-$-plane agrees with that from an O5-plane. We also discuss that Seiberg-Witten curves for Sp($N$) gauge theories are consistent with the curves from the Higgsing of SU($2N$) gauge theories with an antisymmetric hyper.

\subsection{\texorpdfstring{Sp($N$)}{Sp(N)} gauge theory with \texorpdfstring{$N_f$}{Nf} flavors}\label{sec:Sp+Nf}

\begin{figure}
    \centering
    \includegraphics[width=11cm]{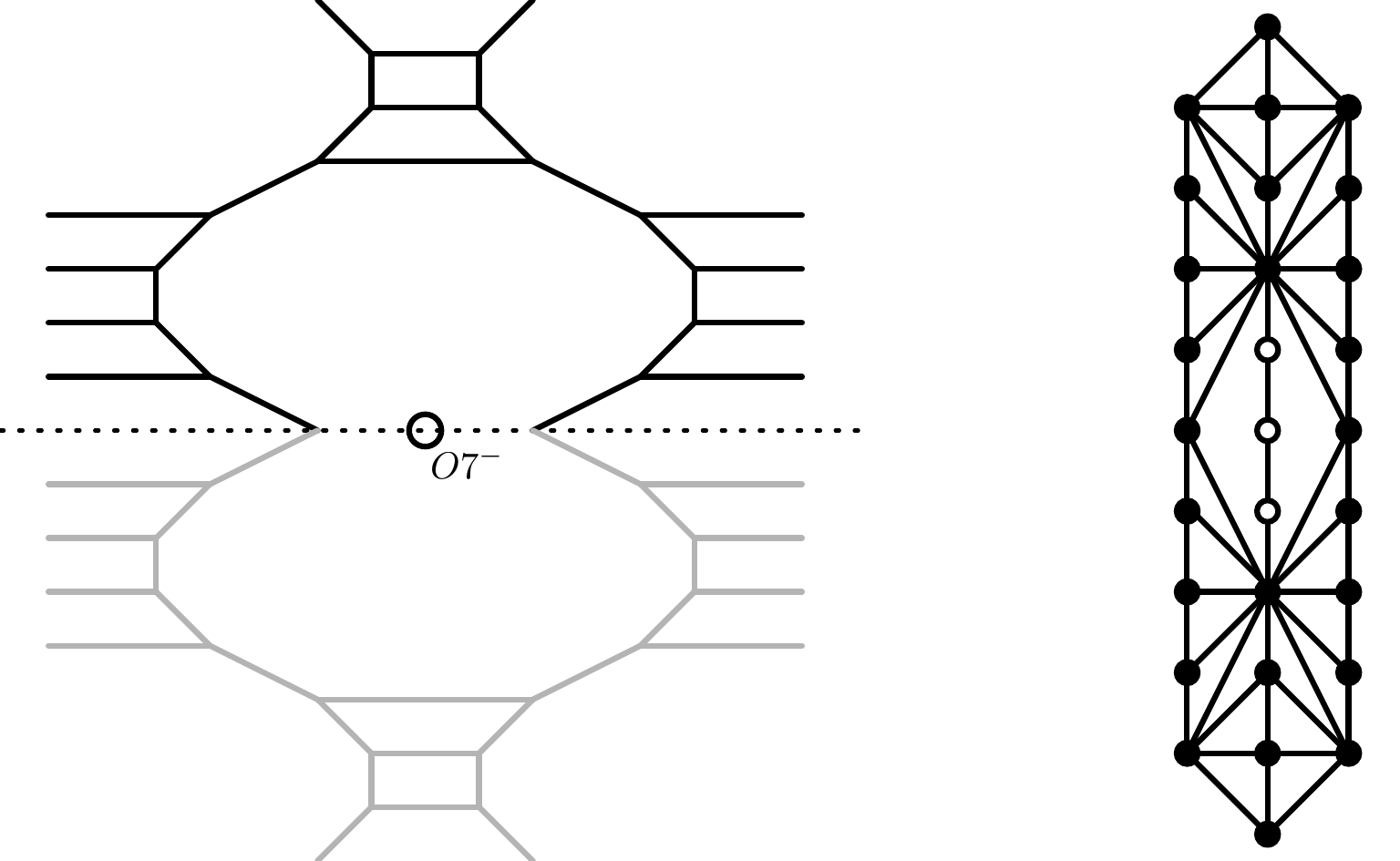}
    \caption{A 5-brane web Sp($N$) gauge theory with $N_f$ flavor and its corresponding dual toric (-like) diagram. Here we chose $N=3$, $N_f=8$ as a representative example. The gray part below the cut of the O7$^-$-plane represents the reflected mirror image by the O7$^-$-plane.}
    \label{fig:Sp+Nf}
\end{figure}
We consider the Seiberg-Witten curve for 5d Sp($N$) gauge theory with $N_f$ flavors. 
From the dual graph of the 5-brane web given in figure \ref{fig:Sp+Nf}, the Seiberg-Witten curve takes the form
\begin{equation}
	F_{\rm Sp}(t,w) \equiv \sum_{m=0}^2 p_m(w) t^m = 0\ , \label{eq:SW-Sp}
\end{equation}
where $p_m(w)$ are Laurent polynomials of the form
\begin{align}
p_m(w) = \sum_{n=n_m^-}^{n_m^+} C_{m,n} w^n \ , 
\end{align}
where $C_{m,n}$ will be fixed by boundary conditions which we will discuss below. 
By multiplying certain monomial to \eqref{eq:SW-Sp}, we can assume that 
\begin{align}
n_1^- = - n_1^+ ( \le -3 )
\end{align}
is satisfied.
For $N_f \le 2N+4$,
the summation ranges $n_m^{\pm}$ are given by
\begin{align}\label{eq:sumrangeSp}
n_0^{\pm} = \pm \frac{N_f}{2} + \kappa, 
\qquad
n_1^{\pm} = \pm (N+2), 
\qquad
n_2^{\pm} = \pm \frac{N_f}{2} - \kappa.
\end{align}
For $N_f=2N+5$, they are given by
\begin{align}
n_0^{\pm} = \pm (N+3) + \kappa, 
\qquad
n_1^{\pm} = \pm (N+3), 
\qquad
n_2^{\pm} = \pm (N+3) - \kappa,
\end{align}
as will be seen later.
Here $\kappa$ is the parameter related to the discrete theta angle.

Due to an O7$^-$-plane, we impose that the Seiberg-Witten curve is invariant under the symmetry $(t,w) \to (t^{-1},w^{-1})$. Imposing that the right-hand side of \eqref{eq:SW-Sp} 
is invariant under this transformation up to an overall multiplication of the monomial $c t^2$, this gives the constraints:
\begin{align}
p_0(w) = c\, p_2(w^{-1}), \qquad
p_1(w) = c\, p_1(w^{-1}), \qquad
c^2= 1.
\end{align}
Out of two choices $c=\pm 1$, we would like to choose $c=1$.
At this stage, the Seiberg-Witten curve \eqref{eq:SW-Sp}  
reduces to the form
\begin{align}\label{eq:p010}
p_2(w) t^2 + p_1(w) t + p_2(w^{-1}) =0\ ,
\end{align}
with $p_1(w)$ satisfying 
\begin{align}\label{eq:constrSpp1}
p_1(w^{-1}) = p_1(w).
\end{align}

In the following, we impose the boundary condition at $(w,t) = (1,1), (-1,1)$, where the two O6$^-$-planes are supposed to exist. 
Our claim is that $F_{SW}(t, w)$ has a double root both at $t=1$ when $w = \pm 1$, which indicates
\begin{align}\label{eq:constrSpO7}
p_1(1) = - 2 p_2(1), 
\qquad
p_1(-1) = - 2 p_2(-1).
\end{align}
The solution to the constraints \eqref{eq:constrSpp1} and \eqref{eq:constrSpO7} is given by
\begin{align}\label{eq:solp1-Sp}
p_1(w) = - \frac{p_2(1)(w+1)^2}{2w} + \frac{p_2(-1) (w-1)^2}{2 w} 
+ (w-w^{-1})^2 \hat{p}_1(w) \ , 
\end{align}
with $\hat{p}_1(w)$ being a Laurent polynomial of the form
\begin{align}\label{eq:p1hat-Sp}
\hat{p}_1(w) = \sum_{n=0}^{n_1^+ - 2} \hat{C}_{n} (w^n+w^{-n}).
\end{align}

In the following, we determine the Seiberg-Witten curve more in detail, which depends on the region of $N_f$ and $\kappa$.

\subsubsection{\texorpdfstring{$\mathrm{Sp}(N)_{\kappa \pi}$}{Sp(N)kpi}}
We first consider the special case, $N_f=0$. 
From the solution \eqref{eq:solp1-Sp} for $p_1(w)$ as well as the summation range given in \eqref{eq:sumrangeSp}, leads to
\begin{align}\label{eq:SpNkp01}
p_2(w) 
&= w^{-\kappa}
\cr
p_1(w) 
&=
\left\{ 
\begin{array}{ll}
 - 2 + (w-w^{-1})^2 \hat{p}_1(w) & \quad \text{ for even } \kappa \\
- (w+w^{-1}) + (w-w^{-1})^2 \hat{p}_1(w) & \quad \text{ for odd } \kappa\ ,
\end{array}
\right.
\end{align}
where $\hat{p}_1(w)$ is given in the form 
\begin{align}\label{eq:SpNkp1h}
\hat{p}_1(w) = \sum_{n=0}^N \hat{C}_n (w^n+w^{-n}).
\end{align}
By imposing the asymptotic behavior as $t \sim q^{-1} w^{N+2}$ and $t \sim q w^{-N-2}$ at large $w$, we find 
\begin{align}
\hat{C}_N = q^{-1},
\end{align}
where $q$ is identified as the instanton factor. 
The remaining parameters $\hat{C}_n$ $(n=0,1,\cdots, N-1)$ correspond to Coulomb moduli. 

In order to compare this result with the past literature, we expand $p_1(w)$ as
\begin{align}
p_1(w) = &\sum_{n=3}^{N+2} (\hat{C}_{n+2} - 2\hat{C}_n + \hat{C}_{n-2} ) (w^n+w^{-n})
\!+\! (\hat{C}_{4} - 2\hat{C}_2 + 2\hat{C}_{0} ) (w^2+w^{-2})
\cr
& \quad + \left\{
\begin{array}{ll}
(\hat{C}_3 - \hat{C}_1) (w+w^{-1}) + 2 (\hat{C}_2 - 2\hat{C}_0 - 1 ) & \quad \text{ for even } \kappa \\
(\hat{C}_3 - \hat{C}_1 + 1) (w+w^{-1}) + 2 (\hat{C}_2 - 2\hat{C}_0 ) & \quad \text{ for odd } \kappa\ , 
\end{array}
\right.
\end{align}
with the convention
\begin{align}
\hat{C}_{N+4}= \hat{C}_{N+3}= \hat{C}_{N+2} = \hat{C}_{N+1} = 0.
\end{align}
Motivated by this expansion, we introduce a new parameterization:
\begin{align}\label{eq:UC-rel}
U_{n} &\equiv q (\hat{C}_{n+2} - 2 \hat{C}_{n} + \hat{C}_{n-2})
\qquad n=3,\cdots N+1,
\cr
U_{2} &\equiv q (\hat{C}_{4} - 2 \hat{C}_{2} + 2\hat{C}_{0}).
\end{align}
With these new parameters, we can rewrite $p_1(w)$ in the form
\begin{align}
p_1(w) &= q^{-1}
\left( 
(w^{N+2}+w^{-(N+2)})
+ \sum_{n=2}^{N+1} U_n (w^{n}+w^{-n})
+ q (w)
\right),
\\
q(w)
&\equiv \left\{
\begin{array}{ll}
\left( - \displaystyle \sum_{k=1}^{\frac{N}{2}} U_{2k+1} \right) (w+w^{-1}) 
- 2 \left( q + 1 + \displaystyle \sum_{k=1}^{\frac{N}{2}} U_{2k} \right)
& \quad  N \text{ even, } \kappa \text{ even } 
\\
\left( q - \displaystyle \sum_{k=1}^{\frac{N}{2}} U_{2k+1}  \right)(w+w^{-1}) 
- 2 \left( 1 + \displaystyle \sum_{k=1}^{\frac{N}{2}} U_{2k} \right)
& \quad  N \text{ even, } \kappa \text{ odd } 
\\
\left( - 1 - \displaystyle \sum_{k=1}^{\frac{N-1}{2}} U_{2k+1} \right) (w+w^{-1}) 
- 2 \left( q + \displaystyle \sum_{k=1}^{\frac{N+1}{2}} U_{2k} \right)
& \quad N \text{ odd, } \kappa \text{ even }  
\\
\left(q - 1 - \displaystyle \sum_{k=1}^{\frac{N-1}{2}} U_{2k+1} \right)(w+w^{-1}) 
- 2 \displaystyle \sum_{k=1}^{\frac{N+1}{2}} U_{2k}
& \quad N \text{ odd, } \kappa \text{ odd }. \nonumber  
\end{array}
\right.
\end{align}

After the coordinate transformation
\begin{align}\label{eq:coordtwt}
t \to w^{-\kappa} t\ ,
\end{align}
we find that this agrees with the result given in \cite{Li:2021rqr}.
Here, we have identified that even $\kappa$ corresponds to the discrete theta angle $0$ while odd $\kappa$ corresponds to the discrete theta angle $\pi$ if $N$ is even. 
If $N$ is odd, the identification is the opposite:
even $\kappa$ corresponds to the discrete theta angle $\pi$ while odd $\kappa$ corresponds to the discrete theta angle $0$.

In summary, we find that the Seiberg-Witten curve for $\mathrm{Sp}(N)$ gauge theory obtained from the 5-brane web diagram with O7$^-$-plane is given by 
\begin{align}
(t-1)^2 + (w-w^{-1})^2 \left( q^{-1} (w^N + w^{-N}) + \sum_{n=0}^{N-1} \hat{C}_n (w^n + w^{-n}) \right)\! t = 0 \ , 
\end{align}
for even $N$ and with discrete theta angle 0, or for odd $N$ and with discrete theta angle $\pi$.
For even $N$ and with discrete theta angle $\pi$, or for odd $N$ and with discrete theta angle 0, the Seiberg-Witten curve is 
\begin{align}
(t-w)(t - w^{-1}) + (w-w^{-1})^2 \left( q^{-1} (w^N + w^{-N}) + \sum_{n=0}^{N-1} \hat{C}_n (w^n + w^{-n}) \right)\! t = 0. 
\end{align}
Here, we have performed the coordinate transformation \eqref{eq:coordtwt} from the expression \eqref{eq:SpNkp01}.

\subsubsection{\texorpdfstring{$\mathrm{Sp}(N)+N_f( \le 2N+4)\, \mathbf{F}$}{Sp(N)+0< Nf < 2N+4 F}}
\label{subsec:Sp+2N+4F}

When the theory has $1 \le N_f \le 2N+4$ flavors, 
we impose the boundary condition at $t \to \infty$ that the flavor branes exist at $w=M_i$ ($i=1,\cdots, N_f$). From this condition, we find
\begin{align}\label{eq:p0Sp2n+3}
p_2(w) = w^{-\frac{N_f}{2}-\kappa} \prod_{i=1}^{N_f} (w - M_i).
\end{align}
where we imposed $C_{2,{N_f}/{2}-\kappa}=1$ by using the overall multiplication of the Seiberg-Witten curve.

Denoting the two solutions of \eqref{eq:p010} as an equation for $t$ as $t_1(w)$ and $t_2(w)$, where we assume $|t_1(w)| < |t_2(w)|$ in the region $w\to \infty$ without the loss of generality, we denote the ratio of these solutions at $w \to \infty$ as

\begin{align}
\frac{t_1(w)}{t_2(w)} \sim (-1)^{N_F} q^2 w^{N_f - 2N - 4} + \mathcal{O}(w^{N_f - 2N - 5}) \quad \text{ as } w \to \infty\ . 
\end{align}
We identify this $q$ as the instanton factor of the gauge theory.
With this notation, we find from \eqref{eq:solp1-Sp} that
\begin{align}\label{eq:SpNCN}
\hat{C}_N = 
\left\{
\begin{array}{ll}
q^{-1} \displaystyle \prod_{i=1}^{N_f} M_i{}^{\frac{1}{2}} & \text{ for } N_f \le 2N+3 \\
(1+q^{-1}) \displaystyle \prod_{i=1}^{2N+4} M_i{}^{\frac{1}{2}} &  \text{ for } N_f = 2N+4 \, .
\end{array}
\right. 
\end{align}
Thus, the Seiberg-Witten curve for $\mathrm{Sp}(N)+N_f( \le 2N+3)\, \mathbf{F}$ is given in the form \eqref{eq:p010} with 
\begin{align}\label{eq:SpNfp1Nf}
p_1(w) 
=& - \frac{(w+1)^2 \prod_{i=1}^{N_f}(1-M_i)}{2w}
+ \frac{(-1)^{N_f}(w-1)^2 \prod_{i=1}^{N_f}(1+M_i)}{2w}
\\
& + (w - w^{-1})^2 \left( q^{-1} \left( \prod_{i=1}^{N_f} M_i{}^{\frac{1}{2}} \right) (w^{N} + w^{-N}) + \sum_{n=0}^{N-1} \hat{C}_n (w^n+w^{-n}) \right),\nonumber
\end{align}
while $p_0(w)$ being given in \eqref{eq:p0Sp2n+3}.
For $N_f = 2N+4$,  $q^{-1}$ should be replaced by $1+q^{-1}$ according to \eqref{eq:SpNCN}.

In order to compare this curve with the known result in \cite{Li:2021rqr}, 
where the cases $N_f \le 2N+3$ are studied,
we consider the coordinate change
\begin{align}
t \quad \to \quad p_2(w)^{-1} t.
\end{align}
Under the new coordinate, we obtain the Seiberg-Witten curve of the form
\begin{align}
t^2 + p_1(w) t + Q(w) =0,
\end{align}
where we put
\begin{align} 
Q(w) &\equiv p_2(w) p_2(w^{-1})
= \prod_{i=1}^{N_f} (w - M_i)(w^{-1} - M_i) , 
\end{align}
while $p_1(w)$ is the one given in \eqref{eq:SpNfp1Nf}. 

Here note that the boundary condition \eqref{eq:constrSpO7} is rewritten in this terminology as
\begin{align}
p_1( \pm 1 ){}^2 -4 Q( \pm 1) = 0.
\end{align}
This is exactly the boundary condition imposed when we compute the Seiberg-Witten curve based on the 5-brane web with O5-plane.

Indeed, introducing the parameterization 
\begin{align}
U_{n} &\equiv q \left( \prod_{i=1}^{N_f} M_i{}^{-\frac{1}{2}} \right)
(\hat{C}_{n+2} - 2 \hat{C}_{n} + \hat{C}_{n-2})
\qquad n=3,\cdots N+1,
\cr
U_{2} &\equiv q \left( \prod_{i=1}^{N_f} M_i{}^{-\frac{1}{2}} \right)
(\hat{C}_{4} - 2 \hat{C}_{2} + 2\hat{C}_{0})\ ,
\end{align}
and assuming that $\frac{1}{2} N_f + \kappa$ is even, we obtain
\begin{align}
p_1(w) &= q^{-1} \left( \prod_{i=1}^{N_f} M_i{}^{\frac{1}{2}} \right)
\left( 
(w^{N+2}+w^{-(N+2)})
+ \sum_{n=2}^{N+1} U_n (w^{n}+w^{-n})
+ q (w)
\right),
\cr
q(w)
&\equiv \left\{
\begin{array}{ll}
\left( q \chi_s - \displaystyle \sum_{k=1}^{\frac{N}{2}} U_{2k+1} \right) (w+w^{-1}) 
-2 \left( q \chi_c + 1 + \displaystyle \sum_{k=1}^{\frac{N}{2}} U_{2k} \right)
& \quad  N \text{ even } 
\\
\left( q \chi_s - 1 - \displaystyle \sum_{k=1}^{\frac{N-1}{2}} U_{2k+1} \right) (w+w^{-1}) 
- 2 \left( q \chi_c + \displaystyle \sum_{k=1}^{\frac{N+1}{2}} U_{2k} \right)
& \quad N \text{ odd } 
\\
\end{array}
\right. ,
\cr
\end{align}
where we have defined 
\begin{align}
\chi_s &\equiv \frac{1}{2} \left( \prod_{i=1}^{N_f} (M_i^{-\frac{1}{2}} + M_i^{\frac{1}{2}}) + \prod_{i=1}^{N_f} (M_i^{-\frac{1}{2}}-M_i^{\frac{1}{2}})\right),
\cr
\chi_c &\equiv \frac{1}{2} \left( \prod_{i=1}^{N_f} (M_i^{-\frac{1}{2}}+M_i^{\frac{1}{2}})- \prod_{i=1}^{N_f} (M_i^{-\frac{1}{2}}-M_i^{\frac{1}{2}}) \right).
\end{align}
Also, in the case of $\frac{1}{2} N_f + \kappa$ odd, the results above can be reproduced by the redefinition $M_1 \to M_1{}^{-1}$, $t \to M_1^{-1} t$. Thus, the value of $\kappa$ does not play any significant role, unlike the case for $N_f=0$. 
The results above indicate that the Seiberg-Witten curve that we have obtained is identical to the result in \cite{Hayashi:2017btw, Li:2021rqr}.

\subsubsection{\texorpdfstring{$\mathrm{Sp}(N)+(2N+5)\, \mathbf{F}$}{Sp(N)+ Nf = 2N+5 F}}
\label{subsec:Sp+2N+5F}

\begin{figure}
    \centering
    \includegraphics[width=12cm]{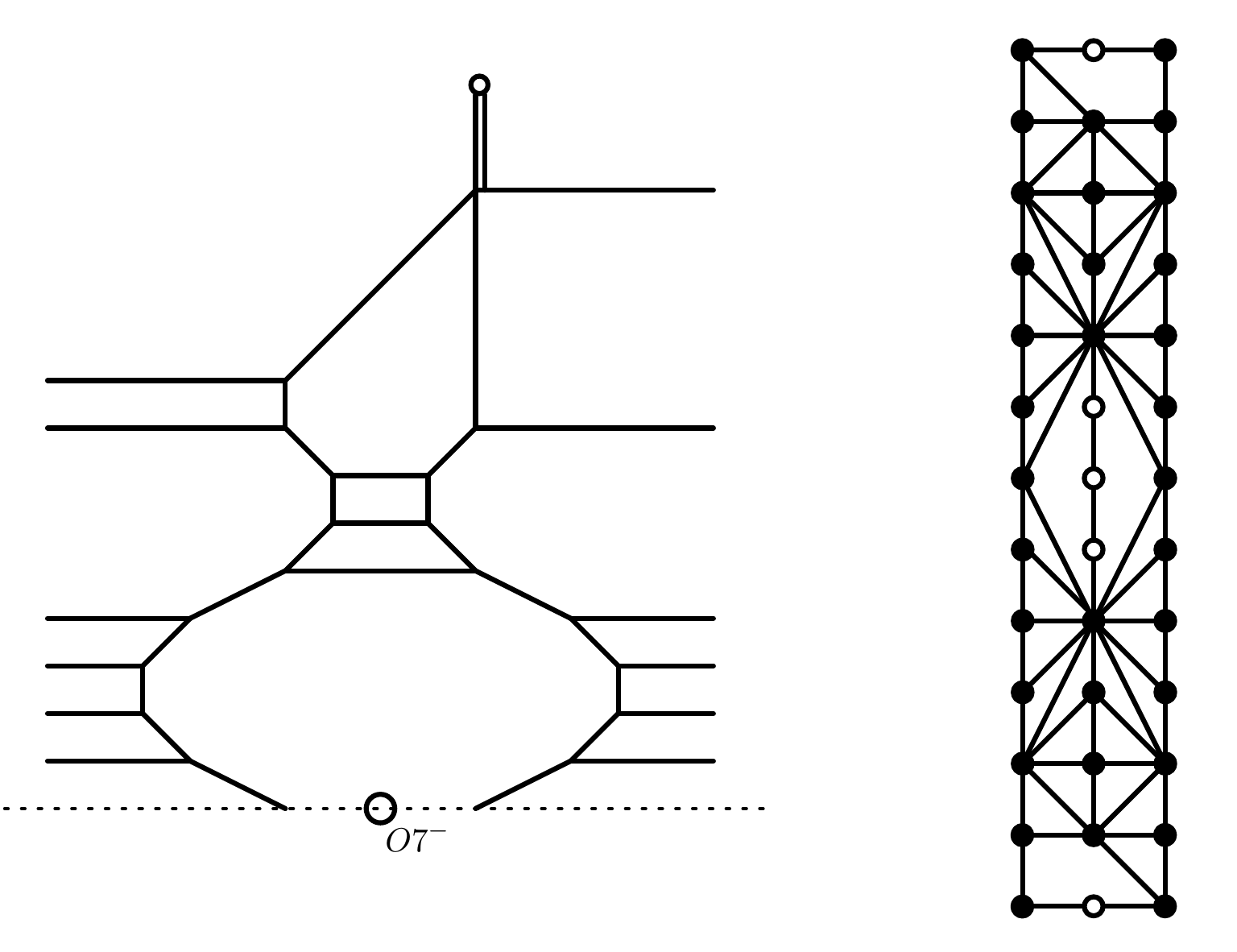}
    \caption{A 5-brane web for pure Sp($N$) with $2N+5$ flavors and the corresponding dual toric(-like) diagram. Here $N=4$.}
    \label{fig:Sp+2N+5F}
\end{figure}

Analogous to the case with SO($2N$) theory with $N_f=2N-3$ flavors, Sp($N$) theory with $N_f=2N+5$ flavors, which is the next to the marginal theory, has different features compared to the cases with less flavors. The 5-brane web diagram for this case is given in figure \ref{fig:Sp+2N+5F}.
Denoting $M_0 = q^{-2}$ where $q$ is the instanton factor, from the boundary condition at $t \to 0$, we find
\begin{align}\label{eq:p0Sp2n+5}
p_2(w) 
=& w^{-N-3} \prod_{i=0}^{2N+5} (w - M_i).
\end{align}
analogous to \eqref{eq:p0Sp2n+3}. However, in this case, the index $i$ runs from $i=0$ instead of from $i=1$, unlike the case of \eqref{eq:p0Sp2n+3}.
Also, from \eqref{eq:SpNkp01}, we have
\begin{align}
p_1(w) 
=&
- \frac{(w+1)^2}{2w}\prod_{i=0}^{2N+5} (1 - M_i) 
+ \frac{(w-1)^2}{2 w} (-1)^{N+1}\prod_{i=0}^{2N+5} (1 + M_i)
\cr
& + (w-w^{-1})^2 \sum_{n=0}^{N+1}  \hat{C}_{n} (w^n+w^{-n}) .
\end{align}

To this expression, we impose the boundary condition at $w \to \infty$ that the solution for $t$ has a double root, which leads to 
\begin{align}
\hat{C}_{N+1} = - 2\prod_{i=0}^{2N+5} M_i{}^{\frac{1}{2}}.
\end{align}
With this coefficient, we have the double root at
\begin{align}\label{eq:O7mroot}
t = \prod_{i=1}^{2N+5} M_i{}^{\frac{1}{2}},
\end{align}
which corresponds to the two coincident external NS5-branes attached to an identical $(0,1)$ 7-brane. Furthermore, impose that the subleading contribution in $w$ expansion also has an identical root \eqref{eq:O7mroot}, which leads to
\begin{align}
\hat{C}_{N} = \sum_{i=0}^{2N+5} (M_i + M_i^{-1}) \prod_{j=0}^{2N+5} M_j{}^{\frac{1}{2}}.
\end{align}

From the discussion above, the Seiberg-Witten curve for 
$\mathrm{Sp}(N)+(2N+5)\, \mathbf{F}$ is given by
\begin{align}\label{eq:SW-SpN-2N+5F}
& w^{-N-3} \prod_{i=0}^{2N+5} (w - M_i) \cdot t
+ w^{N+3} \prod_{i=0}^{2N+5} (w^{-1} - M_i) \cdot t^{-1}
\cr
& - \frac{(w+1)^2}{2w}\prod_{i=0}^{2N+5} (1 - M_i) 
+ \frac{(w-1)^2}{2 w} (-1)^{N+1}\prod_{i=0}^{2N+5} (1 + M_i)
\cr
& + (w-w^{-1})^2 \biggl( - 2 \prod_{i=0}^{2N+5}  M_i{}^{\frac{1}{2}} \cdot  (w^{N+1} + w^{-N-1}) 
\cr
& \quad
+ \!\sum_{i=0}^{2N+5} (M_i + M_i^{-1}) \prod_{j=0}^{2N+5} M_j{}^{\frac{1}{2}} \cdot (w^{N} + w^{-N}) 
+ \!\sum_{n=0}^{N-1} \hat{C}_{n} (w^n+w^{-n})\! \biggr) = 0.\quad 
\end{align}

\subsection{\texorpdfstring{$\mathrm{SU}(N)_{\kappa}$}{SU(N)k} gauge theory with an antisymmetric hypermultiplet}\label{sec:SU+1AS}

\begin{figure}
    \centering
    \includegraphics[width=14cm]{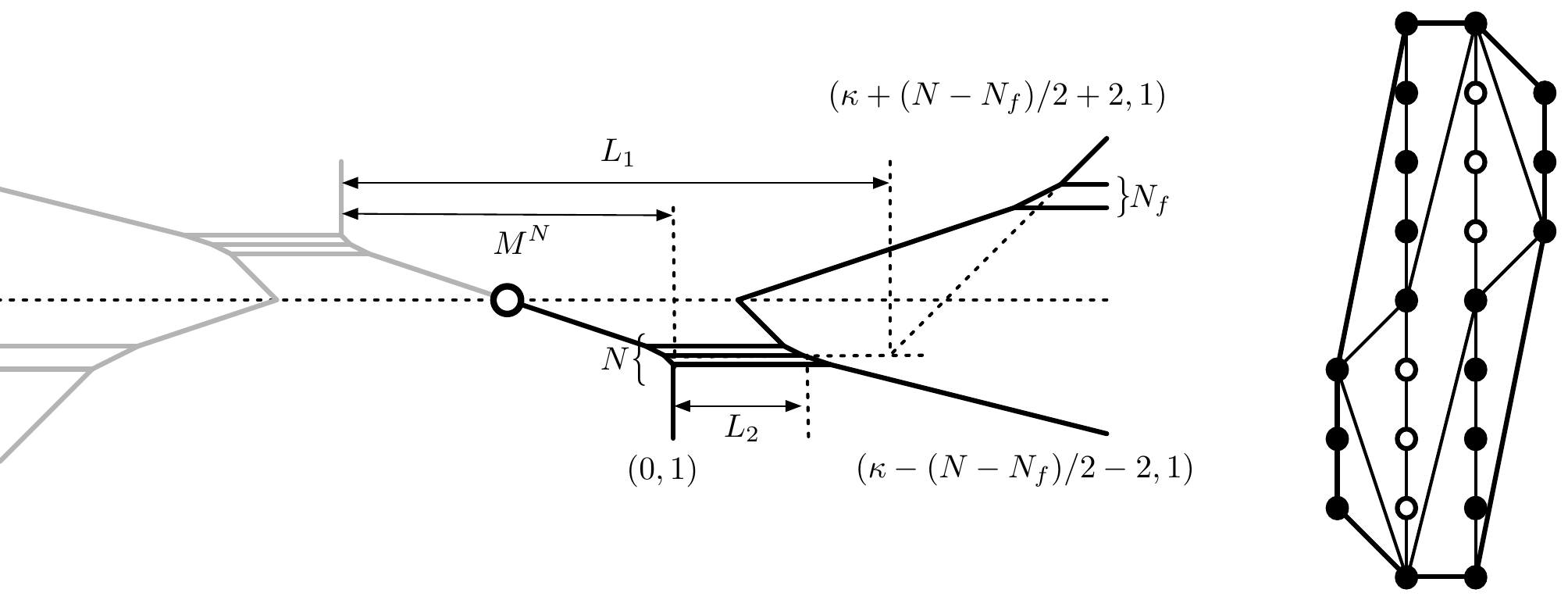}
    \caption{Left: A 5-brane web for SU($N$) gauge theory with Chern-Simons level $\kappa$ and with antisymmetric tensor and $N_f$ flavors. Right: the corresponding dual toric(-like) diagram for $N=4$, $N_f=2$, $\kappa=-2$.}
    \label{fig:SU+AS+Nf}
\end{figure}

We consider the Seiberg-Witten curve for 5d SU($N$)$_\kappa$ gauge theory at the Chern-Simons level $\kappa$ with a hypermultiplet in the antisymmetric representation and $N_f$ flavors. 
We note that the Chern-Simons level exists only when $N\ge3$. For the SU($2$) case, $\kappa$ means the discrete theta angle parameter $\theta= \kappa \pi$ (mod $2 \pi$) instead of the Chern-Simons level. 

From the dual graph of the 5-brane web in figure \ref{fig:SU+AS+Nf}, the Seiberg-Witten curve takes the form
\begin{align}
F_{{\rm SU}+1\mathbf{AS}}(t,w) \equiv 
\sum_{m=0}^3 p_m(w) t^m = 0 \ ,\label{eq:SW-SU-AS}
\end{align}
where $p_m(w)$ are Laurent polynomials of the form
\begin{align}
p_m(w) = \sum_{n=n_m^-}^{n_m^+} C_{m,n} w^n.
\end{align}

Due to an O7$^-$-plane, we impose that the Seiberg-Witten curve is invariant under the symmetry $(t,w) \to (t^{-1},w^{-1})$. Imposing that the right-hand side of \eqref{eq:SW-SU-AS} is invariant under this transformation up to an overall multiplication of the monomial $c t^3$, this gives the constraints:
\begin{align}
p_0(w) = c\, p_3(w^{-1}), \qquad
p_1(w) = c\, p_2(w^{-1}), \qquad
c^2= 1.
\end{align}
Out of two choices $c=\pm 1$, we would like to choose $c=-1$ so that we have $t=1$ instead of $t=-1$ at $w \to 0, \infty$ in the massless limit of the antisymmetric tensor.

At this stage, Seiberg-Witten curve \eqref{eq:SW-SU-AS} reduces to the form
\begin{align}\label{eq:p0110}
p_3(w) t^3  + p_2(w) t^2 - p_2(w^{-1}) t - p_3(w^{-1}) =0.
\end{align}

\paragraph{Boundary conditions.} 
In the following, we impose the boundary condition at $(w,t) = (1,1), (-1,1)$, where the two O6$^-$-planes are supposed to exist.
Substituting $w=\pm 1$, we see that the Seiberg-Witten curve factorizes as 
\begin{align}
&  p_3(\pm 1) t^3 
+ p_2(\pm 1) t^2
- p_2(\pm 1) t
- p_3(\pm 1)
\cr
&=
(t - 1) \Big( p_3(\pm 1) t^2 + (p_2(\pm 1)+ p_3(\pm 1)) t + p_3(\pm 1) \Big).
\end{align}
Based on this observation, we further impose that this curve has a triple root at $t=1$ when $w = \pm 1$, which gives the constraints
\begin{align}\label{eq:constr1-AS}
p_2(\pm 1) = - 3 p_3(\pm 1).
\end{align}

In order to obtain further constraints, we differentiate \eqref{eq:p0110} in $w$ regarding $t$ as a function of $w$ that satisfies \eqref{eq:p0110}, which gives
\begin{align}\label{eq:diffSW}
&\left[ 3 p_3(w) t(w)^2 + 2 p_2(w) t(w) - p_2(w^{-1})   \right] \frac{\partial t(w)}{\partial w}
\cr
&+ \left[ 
p'_3(w) t(w)^3 + p'_2(w) t(w)^2
+ w^{-2}p'_2(w^{-1}) t(w) 
+ w^{-2} p'_3(w^{-1}) \right] = 0\ , 
\end{align}
at generic value of $w$, where we denote $p_{2,3}'(w)\equiv dp_{2,3}(w)/dw$.
Taking into account the constraints \eqref{eq:constr1-AS} as well as the fact that they were obtained by imposing $t(w=\pm 1)=1$, we find that the factor in the square bracket of the first term in \eqref{eq:diffSW} vanishes at $w = \pm 1$.
Thus, under the mild assumption that $\partial t(w) / \partial w$ is finite at $w = \pm 1$ for at least one out of three solutions $t(w)$ of \eqref{eq:p0110}, we see that the second term in \eqref{eq:diffSW} should also vanishes at $w=\pm 1$, which leads to
\begin{align}\label{eq:constr2-AS}
p'_3( \pm 1 ) + p'_2( \pm 1) = 0 \ .
\end{align}
The constraints \eqref{eq:constr1-AS} and \eqref{eq:constr2-AS} are satisfied if $p_1(w)$ is of the form
\begin{align}\label{eq:solforp1}
p_2(w) = - p_3 (w) - \frac{1}{2} p_3(1) \frac{(w+1)^2}{w} + \frac{1}{2} p_3(-1) \frac{(w-1)^2}{w}
+ (w-w^{-1})^2 \hat{p}_2(w) , 
\end{align}
where $\hat{p}_2(w)$ is a Laurent polynomial of the form
\begin{align}
\hat{p}_2(w) = 
\sum_{n=n_2^- + 2}^{n_2^+ - 2} \hat{C}_n w^n.
\end{align}

In the following, we relate the coefficients in $p_i(w)$ with mass parameters in the gauge theory. Analogous to \eqref{eq:p0Sp2n+3}, the masses of the flavors appear in $p_0(w)$ as
\begin{align}
p_3(w) = w^{\alpha} \prod_{i=1}^{N_f} (w-M_i),
\end{align}
where $\alpha = n_3^+ - N_f  = n_3^-$.
Also, discussion analogous to section \ref{sec:SU+1Sym} leads to the 
\begin{align}
\hat{C}_{n_2^+-2} 
&= q^{-1} M^{\frac{1}{4} (n_3^+ + n_3^- - 3 n_2^+ - 3 n_2^-  - N)} \prod_{i=1}^{N_f} M_i^{\frac{1}{2}}, 
\cr
\hat{C}_{n_2^-+2} 
&= (-1)^N q^{-1} M^{\frac{1}{4} (n_3^+ + n_3^- - 3 n_2^+ - 3 n_2^-  + N)  } \prod_{i=1}^{N_f} M_i^{\frac{1}{2}} \ , 
\end{align}
where we used \eqref{eq:solforp1}.
With this result, we write the Seiberg-Witten curve more explicitly depending on whether $N$ is even or even.

\subsubsection{\texorpdfstring{$\mathrm{SU}(2n)$}{SU(2n)} cases}

First, we consider the case $N=2n$.
Assuming $N_f<2n+4$ and $|\kappa| < n+2-\frac{1}{2}N_f$,
the summation ranges $n_m^{\pm}$ can be read off from figure \ref{fig:SU+AS+Nf} to be given by
\begin{align}\label{eq:sumrange-even-AS}
&n_0^{\pm} = \pm \frac{N_f}{2} + \kappa, 
\qquad
n_1^{\pm} = n_2^{\pm} = \pm ( n+2 ), 
\qquad
n_3^{\pm} = \pm \frac{N_f}{2} - \kappa.
\end{align}
Note that the Chern-Simons level $\kappa$ is an integer if $N_f$ is even while it is half-integer if $N_f$ is odd so that $n_0^{\pm}$ and $n_3^{\pm}$ above are always integers.

With these, we find the Seiberg-Witten curve is given explicitly as
\begin{align}
& p_3(w)t^3 + p_2(w) t^2 - p_2(w^{-1}) t - p_3(w^{-1}) = 0, 
\cr
& p_3 (w) 
= w^{-\frac{N_f}{2}-\kappa} \, \prod_{i=1}^{N_f} (w-M_i),
\cr
& p_2(w) 
= 
- w^{-\frac{N_f}{2}-\kappa} \, \prod_{i=1}^{N_f} (w-M_i)  - \frac{(w+1)^2}{2w} \prod_{i=1}^{N_f} (1-M_i) 
\cr
& \qquad \qquad
+ \frac{(-1)^{\frac{N_f}{2}-\kappa} (w-1)^2}{2w} \prod_{i=1}^{N_f} (1+M_i)
+ (w-w^{-1})^2 \hat{p}_2(w), 
\cr
& \hat{p}_2(w) = 
q^{-1} M^{-\frac{\kappa+n}{2}}
\left( \prod_{i=1}^{N_f} M_i{}^{\frac{1}{2}} \right)
w^{-n} \left( w^{2n} 
+ \sum_{k=1}^{2n-1} \hat{C}_k w^k + M^{n} \right).
\end{align}

\subsubsection*{Higgsing from \texorpdfstring{$\mathrm{SU}(2n)+1\mathbf{AS}$}{SU(2n)+1AS} to \texorpdfstring{$\mathrm{Sp}(n)$}{Sp(n)}}

We consider the special case 
\begin{align}
M=1, \qquad \hat{C}_k = \hat{C}_{2n-k}\ ,
\end{align}
which indicates 
\begin{align}\label{eq:cond-p1-p1inv}
p_3(w^{-1}) + p_2(w^{-1}) = p_3(w) + p_2(w).
\end{align}
In this case, we find that the Seiberg-Witten curve \eqref{eq:p0110} factorizes as
\begin{align}
F_{{\rm SU}(2n)+1\mathbf{AS}} = (t-1) F_{{\rm Sp}(n)} = 0,
\end{align}
where we denote
\begin{align}
F_{{\rm Sp}(n)} = 
&\left[ p_3(w) t^2 + \left(- \frac{1}{2} p_3(1) w^{-1}(w+1)^2 + \frac{1}{2} p_3(-1) w^{-1}(w-1)^2
\right. \right.
\cr
&
\qquad\qquad
+ (w-w^{-1})^2 \hat{p}_2(w) \biggl) t + p_3(w^{-1})  \biggl].
\end{align}
This is identical to the one obtained previously in 
\eqref{eq:p010} with $p_1(w)$ given in \eqref{eq:solp1-Sp}.

\subsubsection{\texorpdfstring{$\mathrm{SU}(2n+1)$}{SU(2n+1)} cases}

Suppose $N=2n+1$.
By using SL(2,$\mathbb{Z}$) transformation to the coordinate system $(t,w)$ and by multiplying certain monomial to \eqref{eq:SW-SU-AS}, we can assume that 
\begin{align}
n_1^- = - n_1^+ + 1, 
\qquad 
n_2^- = - n_2^+ - 1
\end{align}
are satisfied.
Assuming $N_f<2n+5$ and $|\kappa| < n+\frac52-\frac{1}{2}N_f$,
the summation ranges $n_m^{\pm}$ are given by
\begin{align}\label{eq:sumrange-odd-AS}
&n_0^{\pm} = \pm \frac{N_f}{2} + \kappa + \frac{3}{2}, 
&
n_1^+ &= n+3, 
& 
n_1^- &= -n-2,
\cr
&
n_2^+ = n+2, 
&
n_2^- &= -n - 3,
&
n_3^{\pm} &= \pm \frac{N_f}{2} - \kappa - \frac{3}{2}\ .
\end{align}
With these, we find the Seiberg-Witten curve is given explicitly as
\begin{align}
& p_3(w)t^3 + p_2(w) t^2 - p_2(w^{-1}) t - p_3(w^{-1})  = 0, 
\cr
& p_3 (w) 
= w^{-\frac{N_f}{2}-\kappa-\frac{3}{2}} \, \prod_{i=1}^{N_f} (w-M_i),
\cr
& p_2(w) 
= 
- w^{-\frac{N_f}{2}-\kappa-\frac{3}{2}} \, \prod_{i=1}^{N_f} (w-M_i)  - \frac{(w+1)^2}{2w} \prod_{i=1}^{N_f} (1-M_i) 
\cr
& \qquad \qquad
+ \frac{(-1)^{\frac{N_f}{2}-\kappa-\frac{3}{2}} (w-1)^2}{2w} \prod_{i=1}^{N_f} (1+M_i)
+ (w-w^{-1})^2 \hat{p}_2(w), 
\cr
& \hat{p}_2(w) = 
q^{-1} M^{-\frac{4\kappa+2n+1}{4}} \left( \prod_{i=1}^{N_f} M_i{}^{\frac{1}{2}} \right)
w^{-n-1}\!\left( w^{2n+1}
+ \sum_{k=1}^{2n} U_k w^k - M^{\frac{2n+1}{2}} \right)\!.\quad 
\end{align}

\subsubsection{\texorpdfstring{$\mathrm{SU}(N)_\kappa+1\mathbf{AS}$}{SU(N)k+1AS} and 4d limit}

For simplicity, we consider $N_f = 0$. In this case, the Seiberg-Witten curve can be written as 
\begin{align}
&w^{\alpha} t^3
+ \left( - w^{\alpha} - \frac{(w+1)^2}{2w} + \frac{(-1)^{\kappa}(w-1)^2}{2w} + (w-w^{-1})^2 \hat{p}_2(w)
\right) t^2
\\
&~~ - \left( - w^{-\alpha} - \frac{(w+1)^2}{2w} + \frac{(-1)^{\kappa}(w-1)^2}{2w} + (w-w^{-1})^2 \hat{p}_2(w^{-1})
\right) t
- w^{-\alpha} = 0, \nonumber
\end{align}
with
\begin{align}
\hat{p}_2(w)= q^{-1} M^{-\frac{\kappa}{2}-\frac{N}{4}} w^{\alpha'} \prod_{i=1}^{N} (w-A_i) \ , 
\end{align}
where $\alpha=-\kappa$, $\alpha' = -\frac{N}{2}$ for even $N$, while $\alpha=-\kappa - \frac{3}{2}$, $\alpha' = -\frac{N+1}{2}$ for odd $N$.

Parameterizing $w=e^{-\beta v}$, $A_i=e^{-\beta a_i}$, $q = 4 (-1)^{N-1} (\beta \Lambda)^{N+2}$ and also taking $\beta \to 0$, we find that the curve above reduces to the 4d curve
\begin{align}
 t^3
& - \left( 3 + \Lambda^{-N-2} v^2 \prod_{i=1}^{N} (v-a_i)
\right) t^2
+ \left( 3 + (-1)^N \Lambda^{-N-2} v^2 \prod_{i=1}^{N} (v+a_i)
\right) t
- 1 = 0 \ , 
\end{align}
which agrees with the result in \cite{Landsteiner:1997ei} after changing the variable as $t = -\Lambda^{-N-2} y$.

\subsubsection{Decoupling of 
AS from \texorpdfstring{SU($2)_{\pi}+1\mathbf{AS}$}{SU(2)+1AS}}\label{sec:SU2anti-decouples}

Especially when $N=2$, the gauge group is SU(2), and the antisymmetric tensor is a singlet, which should not affect the low energy dynamics.
Considering the case $N_f=0$, $\kappa=0$ for simplicity, the Seiberg-Witten curve in this case reduces to 
\begin{align}\label{SW-SU2-singlet}
t^3 + \Big( - 3 + (w-w^{-1})^2 \hat{p}_2(w)
\Big) t^2
- \Big( - 3 + (w-w^{-1})^2 \hat{p}_2(w^{-1})
\Big) t
- 1  = 0\ , 	
\end{align}
where
\begin{align}
\hat{p}_2(w) = q^{-1} ( M^{-\frac{1}{2}} w + U + M^{\frac{1}{2}} w^{-1} ) \ .
\end{align}
In the following, we see that it is equivalent to the Seiberg-Witten curve for $\mathrm{SU}(2)_{\pi}$ gauge theory without a singlet, which gives a non-trivial consistency check.

For the massless case $M=1$, the Seiberg-Witten curve for SU(2) with the singlet is further simplified as
\begin{align}
\hat{p}_2(w)=q^{-1}(w+U+w^{-1}) = \hat{p}_2(w^{-1})\ .
\end{align}
In this case, the curve \eqref{SW-SU2-singlet} can be expressed as a factorized form, 
\begin{align}\label{eq:SU2+1AS-massless}
    (t-1)\Big[ (t-1)^2+q^{-1}(w-w^{-1})^2(w+U+w^{-1})t \Big]=0\ .
\end{align}
This computation is $N=1$ case of Higgsing from SU($2N$)+1$\bf AS$ to Sp($N$) discussed in the previous section. 
This clearly shows that the Seiberg-Witten curve \eqref{SW-SU2-singlet} for SU(2) with a singlet is factorized when the singlet is massless. The terms in the square bracket of \eqref{eq:SU2+1AS-massless} is the same Seiberg-Witten curve as Sp(1)$_\pi$ 
 given in \eqref{eq:p010}, \eqref{eq:SpNkp01},  and \eqref{eq:SpNkp1h}.
This also agrees with the one obtained from a 5-brane web with an O5-plane. (See also Eq. (2.14) in  \cite{Hayashi:2017btw}.) Therefore, we can conclude that, at least in the massless case, the singlet of the SU(2) gauge theory indeed decouples. 

Next, we consider the case with a generic mass for the singlet.
By computing the 
discriminant of the left-hand side of \eqref{SW-SU2-singlet} as a polynomial in $t$,
dropping the factor $q^{-4} (w-w^{-1})^{6}$, 
rewriting it in terms of $x=w+w^{-1}$,
and again by computing the discriminant of it in terms of $x$,
we obtain the following ``double discriminant''
\begin{align}\label{eq:discSU2AS}
\Delta 
&= \Delta_{\text{phys}} \Delta_{\text{unphys}}
\cr
\Delta_{\text{phys}}
&= U^4 + qU^3 - 8 U^2 - 36 qU - 27q^2 + 16,
\cr
\Delta_{\text{unphys}}
&= 4096 q^2 (M^{\frac12}-M^{-\frac12})^4 \biggl[ (M^{\frac12}-M^{-\frac12})^2 ( (M^{\frac12}+M^{-\frac12})^2 - U^2 )^3
\cr
& \qquad \qquad 
- 27 (M^{\frac12}-M^{-\frac12})^2
(M^2+M+M^{-1}+M^{-2} -U^2 )q U 
\cr
& \qquad \qquad 
- 27(M-1+M^{-1})^3 q^2 
\biggr]^3\ .
\end{align}
More detailed computation is given in Appendix \ref{app:SU2+1AS}.

Here, we have split the discriminant into the physical part and the unphysical part due to the following criteria:
For $U$ satisfying $\Delta_{\text{phys}}=0$, a non-trivial cycle integral of the Seiberg-Witten 1-form, which is non-zero at a generic value of $U$, vanishes. This indicates a massless BPS particle and/or a tensionless BPS object appears at such points.
On the contrary, $\Delta_{\text{unphys}}=0$ does not indicate any massless BPS particles or tensionless BPS objects. Thus, we discard the unphysical part $\Delta_{\text{unphys}}$.
The discussion on the unphysical part of the discriminant of Seiberg-Witten curves has already been given in several papers in the past, including \cite{Landsteiner:1997ei}.
A detailed explanation of how to distinguish the physical and the unphysical parts in our case is also given in Appendix \ref{app:SU2+1AS}.

We find that the physical part $\Delta_{\text{phys}}$ of the discriminant in \eqref{eq:discSU2AS}
agrees with
the double discriminant for the Seiberg-Witten curve for SU(2) gauge theory with discrete theta angle $\pi$.
This indicates that Seiberg-Witten curve \eqref{SW-SU2-singlet} is equivalent to the one for Sp(1)$_{\pi}$=SU(2)$_{\pi}$ gauge theory even for a generic mass for the singlet, which gives a non-trivial consistency check.

\subsubsection{Equivalence between \texorpdfstring{SU($3)+1\mathbf{AS}$}{SU(3)+1AS} and \texorpdfstring{SU($3)+1\mathbf{F}$}{SU(3)+1F}}\label{sec:equiSU3+1AS}

We now consider the case $N=3$. When the gauge group is SU(3), the antisymmetric tensor representation is the (anti-) fundamental representation.
Considering the case $N_f=0$, $\kappa=-\frac{1}{2}$ for simplicity, the Seiberg-Witten curve, in this case, reduces to 
\begin{align}\label{eq:SW-SU3-AS}
w^{-1} t^3
& + \left( -w- 2w^{-1} + (w-w^{-1})^2 \hat{p}_2(w)
\right) t^2
\cr
& - \left( -2 w -w^{-1} + (w-w^{-1})^2 \hat{p}_2(w^{-1})
\right) t
- w = 0\ , 
\end{align}
with 
\begin{align}
\hat{p}_2(w) 
= \hat{C}_{1} w + \hat{C}_0 + \hat{C}_{-1} w^{-1} + \hat{C}_{-2} w^{-2}.
\end{align}
Here, we parameterize each coefficient as
\begin{align}
\hat{C}_{1} = q^{-1} M^{-\frac{1}{2}}, \quad
\hat{C}_0 = - q^{-1} U,  \quad
\hat{C}_{-1} = q^{-1} V M^{\frac{1}{2}},  \quad
\hat{C}_{-2} = - q^{-1} M.
\end{align}

As discussed in the preceding subsection \ref{sec:SU2anti-decouples}, one computes the discriminant of the Seiberg-Witten curve \eqref{eq:SW-SU3-AS} for SU(3)$_{-\frac{1}{2}}+1\mathbf{AS}$ obtained from 5-brane webs to compare it with that of SU(3)$_{-\frac{1}{2}}+1\mathbf{F}$. This computation requires double discriminant factoring out the physical part of the discriminant $\Delta_{\rm phys}$ from the unphysical one. As it is a lengthy computation, we put the details in Appendix \ref{app:SU3+1AS=1F}.
We find that the physical part of the discriminant $\Delta_{\text{phys}}$ agrees with the double discriminant of the Seiberg-Witten curve for SU(3)$_{-\frac{1}{2}}+1\mathbf{F}$, 
\begin{align}\label{eq:SW-SU3_1.2+1F}
w t^2 + (w^2 - U w + V - w^{-1}) t - q M^{-\frac{1}{2}}(w-M) = 0\ .
\end{align}
Here, $M$ is the mass of the fundamental and antisymmetric hypermultiplet, $U,V$ are two Coulomb branch parameters, and $q$ is the instant factor.

\bigskip


\section{Comparison between 
\texorpdfstring{$\textrm{O7}^+$ and $\textrm{O7}^-\!+8\, \textrm{D7's}$}{O7+ and O7-+8D7's} }\label{sec:comparison}
In this section, we discuss the relation between O7$^+$ and O7$^-$+8D7's when the masses of eight D7-branes are specially tuned so that they are frozen at the O7$^-$-plane, and we show that the Seiberg-Witten curves obtained from 5-brane webs with O7$^-$+8D7's are equivalent to those from 5-brane webs with an O7$^+$-plane.
 
\subsection{Equivalence between two Seiberg-Witten curves}
In proceeding sections, we computed the Seiberg-Witten curves for the theories involving an O7$^+$-plane in section \ref{sec:O7+construction} and  those involving an O7$^-$-plane in section \ref{sec:O7-construction}. Now, we consider the cases involving an O7$^-$-plane with more than 8 flavors and we tune eight masses of hypermultiplets in the fundamental representation so that they are stuck at an O7$^-$-plane. We then compare the curves from O7$^-$+8D7's and O7$^+$. We will show that the resulting curves are factorized and, in particular, the physically relevant part of the factorized curves coincides with the curves associated with an O7$^+$-plane.

\paragraph{Sp($N$)$+(N_f+8) \mathbf{F}$ to SO($2N$)$+N_f\mathbf{F}$.}
First let us consider Sp($N$) gauge theory with $N_f< 2N+4$ flavors, whose Seiberg-Witten curve \eqref{eq:p010} is discussed in section \ref{sec:Sp+Nf}. The structure of the curve is given as
\begin{align}
	p_2(w)t^2 + p_1(w)t + p_2(w^{-1})= 0  \ . 
\end{align}
We tune 8 masses out of $(N_f+8)$ flavor masses to vanish in such a way that half of them are tuned with the opposite phase. In other words, the corresponding K\"ahler parameters, $M_i$~($i=1,\cdots, N_f+8$), are chosen as four $1$'s and four $-1$'s,
\begin{align}\label{eq:8massesturning}
    M_1=M_2=M_3=M_4 = 1 \quad {\rm and} \quad M_5=M_6=M_7=M_8 = -1\ .
\end{align}
This special tuning yields that $p_0 (w)$ becomes the following factorized form
\begin{align}
p_2 (w) = (w-w^{-1})^4 \, \check{p}_2(w)\ . 
\end{align}
Since $p_1(w)$ is given as in \eqref{eq:solp1-Sp},
it is then straightforward to see that $p_1(w)$ also reduces to a factorized form
\begin{align}
p_1(w) = (w-w^{-1})^2 \, \hat{p}_1(w) \ .
\end{align}
We see therefore that with the special tuning of eight masses, given in \eqref{eq:8massesturning}, the Seiberg-Witten curve \eqref{eq:p010} for Sp($N)+(N_f+8)\mathbf{F}$ factorizes as
\begin{align}\label{eq:facotredSp}
(w-w^{-1})^2\Big[ (w-w^{-1})^2\check{p}_3(w) t^2 + \hat{p}_1(w)t+(w-w^{-1})^2 \check{p}_3(w) \Big] = 0\ . 
\end{align}
After ignoring the overall factor $(w-w^{-1})^2 $, one finds that the Seiberg-Witten curve inside the square bracket is the one for SO($2N$) + $N_f$ $\mathbf{F}$, as given in \eqref{eq:structure of SO(2N)+Nf}.

\vspace{.5cm}
\paragraph{SU($N$)$+1\mathbf{AS}+(N_f+8) \mathbf{F}$ to SU($N$)$+1\mathbf{Sym}+N_f\mathbf{F}$.}
As discussed in section \ref{sec:SU+1AS}, the Seiberg-Witten curve for SU($N$)$+1\mathbf{AS}+(N_f+8) \mathbf{F}$ is given in \eqref{eq:p0110} which structurally takes the following form, 
\begin{align}
p_3(w)t^3 + p_2(w)t^2 + p_2(w^{-1}) t + p_3(w^{-1}) = 0 \ , \label{eq:SW4SU+1AS}
\end{align}
where $p_2(w)$ and $p_3(w)$ are related as shown in \eqref{eq:solforp1}. As done in the case 
Sp($N$)$+(N_f+8) \mathbf{F}$ and SO($2N$)$+N_f\mathbf{F}$, we tune the eight flavor masses to be special values as given in \eqref{eq:8massesturning}. This yields that $p_3(w)$ is factorized as
\begin{align}
p_3 (w) = (w-w^{-1})^4 \,\check{p}_3(w)\ . 
\end{align}
Since a generic form of $p_2(w)$ has some
$p_3(w=\pm1)$ dependent terms as given in \eqref{eq:solforp1}, it follows that $p_2(w)$ reduces to a factorized form as well, 
\begin{align}
p_2(w) = (w-w^{-1})^2 \Big( - (w-w^{-1})^2 \check{p}_3(w) + \hat{p}_2(w) \Big)\ .
\end{align}
With the special tuning of eight masses of fundamental hypermultiplets, one finds therefore that the Seiberg-Witten curve \eqref{eq:p0110} for SU($N)_\kappa+1\mathbf{AS}+(N_f+8)\mathbf{F}$ factorizes as
\begin{align}\label{eq:facotredSU}
(w-w^{-1})^2 
\bigg[ &(w-w^{-1})^2 \check{p}_3(w) t^3 + \Big( - (w-w^{-1})^2 \check{p}_3(w) + \hat{p}_2(w) \Big) t^2
\\
 &-\! \Big(\! - (w-w^{-1})^2 \hat{p}_3(w^{-1}) + \hat{p}_2(w^{-1}) \Big) t - (w-w^{-1})^2 \check{p}_3(w^{-1}) \bigg] = 0\ . \nonumber
\end{align}
After ignoring the overall factor 
$(w-w^{-1})^2 $, one finds that the Seiberg-Witten curve inside the square bracket is the one for SU($N$)$_\kappa + 1 \mathbf{Sym}+ (N_f-8) \mathbf{F}$, discussed in \eqref{eq:O7p-po123}
by identifying $p_2(w) = - (w-w^{-1})^2 \check{p}_3(w^{-1}) + \hat{p}_2(w^{-1})$. 
This confirms the equivalence between
O7$^+$ and O7$^- + 8 $D7's with specially tuned masses at the level of the Seiberg-Witten curve.

\begin{figure}
    \centering
    \includegraphics[width=13cm]{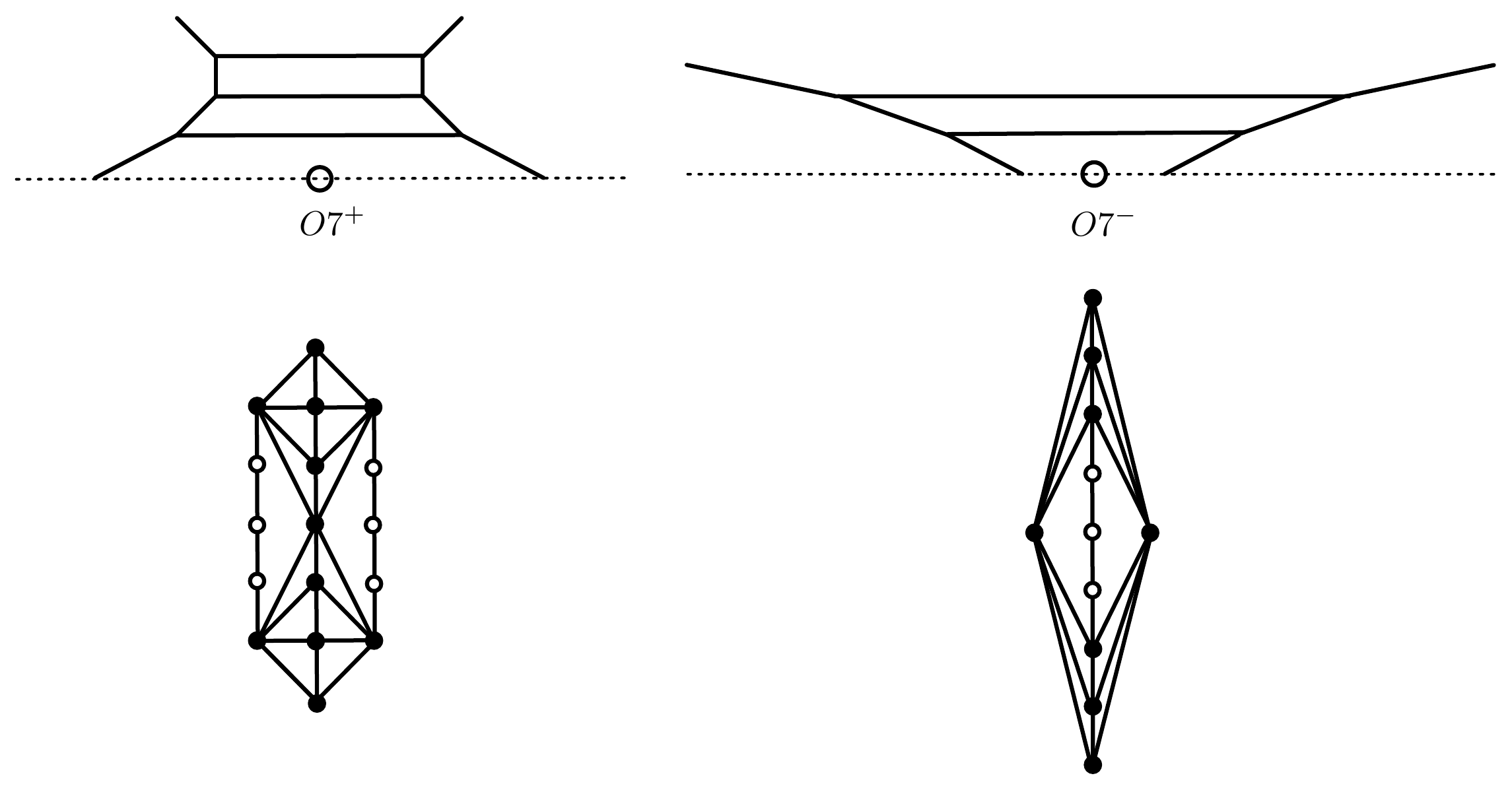}
    \caption{5-brane webs and their dual toric-like diagrams for pure SO($2N$) and Sp($N$) gauge theories involving an O7$^+$-plane and an O7$^-$-plane, respectively. As a representative example, SO($6$) gauge theory is on the left, and Sp($2$) gauge theory is on the right.}
    \label{fig:O7+-toric}
\end{figure}

\subsection{Equivalence from 5-brane webs}
\label{sec:equivalence}
We now discuss the equivalence relation between O7$^+$ and O7$^- + 8 $D7's with specially tuned masses from the perspective of 5-brane webs or their dual toric-like diagrams. 
\begin{figure}
    \centering
    \includegraphics[width=13cm]{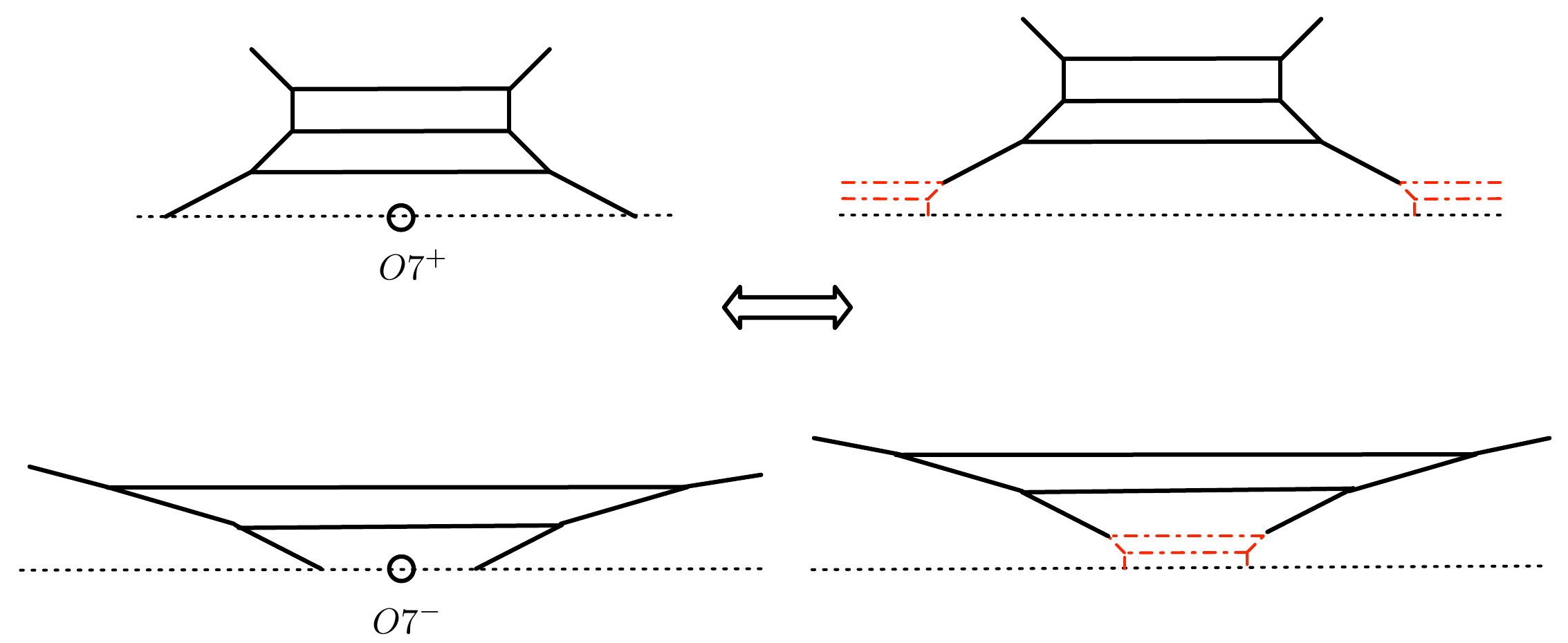}
    \caption{O7-planes and virtual 5-branes. The dash-dotted lines in red denote virtual 5-branes that are stuck at the position of O7-planes.}
    \label{fig:Virtual}
\end{figure}
In figure \ref{fig:O7+-toric}, we have a 5-brane web and its dual diagram for pure SO($6$) gauge theory which involves an O7$^+$-plane and a 5-brane web and dual diagram for pure Sp(2) gauge theory which involves an O7$^-$-plane. The dual diagrams have white dots but they are not the white dots representing multiple 5-branes bound to a single 7-brane. The white dots appearing in 5-brane webs of O7$^\pm$-planes (or 05$^\pm$-planes) represent the boundary conditions for O7-planes. To distinguish these dual diagrams from generalized toric diagrams, we call them toric-like diagrams. 

Though such white dots were introduced for boundary conditions of an O7-plane, one may regard the white dots as the presence of ``{\it virtual} 5-branes" which are frozen at the O7-plane such that these virtual 5-branes do not carry any physical degrees of freedom. In figure \ref{fig:Virtual}, 5-brane configurations are redrawn with virtual 5-branes which are denoted by dash-dotted lines. From dual toric-like diagrams for theories involving an O7$^+$-plane in the figure, there would be four {\it flavor} virtual 5-branes stuck at the O7$^+$-plane. We can also see the contribution of these virtual 5-branes from  the Seiberg-Witten curve, which appears in the $p_0(w), p_2(w)$ terms of the Seiberg-Witten curve for Sp($2N$) gauge theory such that the $(w-w^{-1})^2$ terms capture this virtual flavor 5-branes together with the $\mathbb{Z}_2$ reflected part due to the O7-plane. Similarly, in dual toric-like diagrams for theories involving an O7$^-$-plane in the figure \ref{fig:Virtual}, there are two {\it color} virtual 5-branes, which contribute to the Seiberg-Witten curves as $p_1(w) \propto (w-w^{-1})^2$. 

\begin{figure}
    \centering
    \includegraphics[width=14cm]{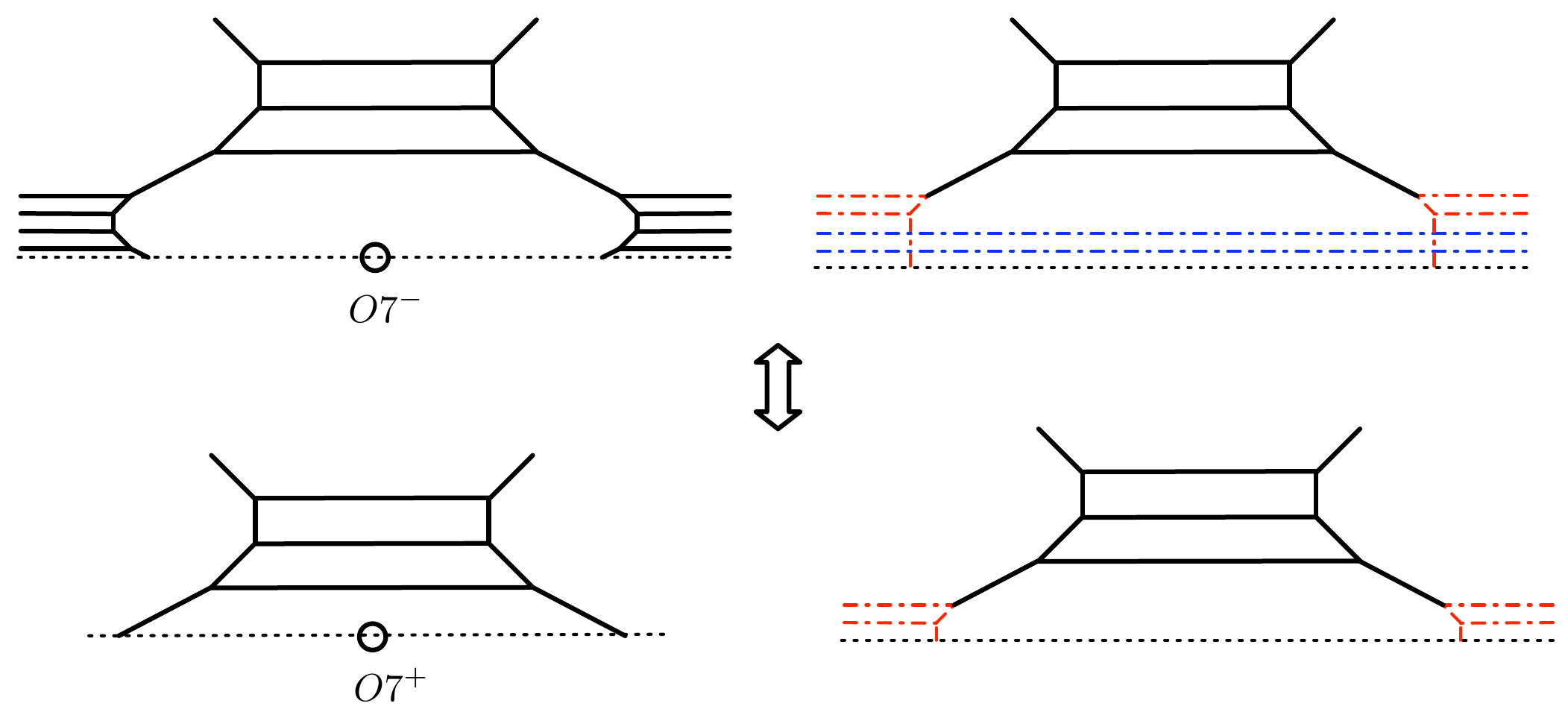}
    \caption{Removal 
    of virtual 5-branes in blue. The remaining virtual 5-branes (in red) account for an O7$^+$-plane. The upper two figures are a 5-brane web with an O7$^-+8D7$ with specially tuned masses, where the figure on the right is the corresponding brane configuration with virtual 5-branes. The lower two figures are a 5-brane with an O7$^+$ and its corresponding brane configuration with virtual 5-branes.
   }
    \label{fig:VirHiggsing}
\end{figure}
For Sp$(N)+(N_f+8)\mathbf{F}$ and SU$(N)+1\mathbf{AS}+(N_f+8)\mathbf{F}$ theories, if eight flavors whose masses are specially tuned such that the contribution from these eight flavors is expressed as $p_{0,2}(w)\propto (w-w^{-1})^4$, then they can be viewed as flavor virtual branes. Together with two color virtual already presented  in $p_1(w)$, the factor $(w-w^{-1})^2$ can be factored out so that the resulting Seiberg-Witten curve is rewritten as either \eqref{eq:facotredSp} or \eqref{eq:facotredSU}. This means that with the special tuning of their masses, eight flavors contribute as if they are 8 flavor virtual 5-branes, and four out of eight of them are aligned with two color virtual 5-branes such that two on the left and two on the right of the color virtual branes are recombined and removed.  
As a result, only four flavor virtual branes remain in the brane web and they are those existing for 5-brane configurations for O7$^+$-plane, as depicted in figure \ref{fig:VirHiggsing}.

\subsection{Non-Lagrangian theories involving an \texorpdfstring{O7$^+$}{O7+}-plane}

As non-trivial examples of theories whose brane realization involves an O7$^+$-plane, we consider non-Lagrangian theories of lower rank.

\subsubsection{Local \texorpdfstring{$\mathbb{P}^2 + 1\mathbf{Adj}$ from a web with O7$^+$-plane}{P2+1Adj}}\label{sec:LocalP2+1Adj}
\begin{figure}[t]
    \centering
    \includegraphics[width=13cm]{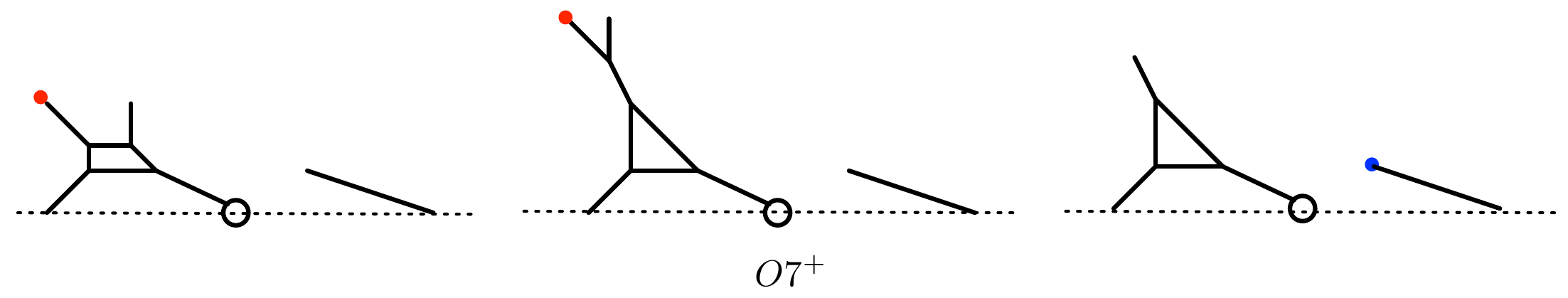}
    \caption{Left: SU$(2)+1\mathbf{Adj}$. Middle: Flop of instantonic hyper. Right: Local 
    $\mathbb{P}^2+1\mathbf{Adj}$.}
    \label{fig:web-P2+adj}
\end{figure}

\begin{figure}[t]
    \centering
    \includegraphics[width=14cm]{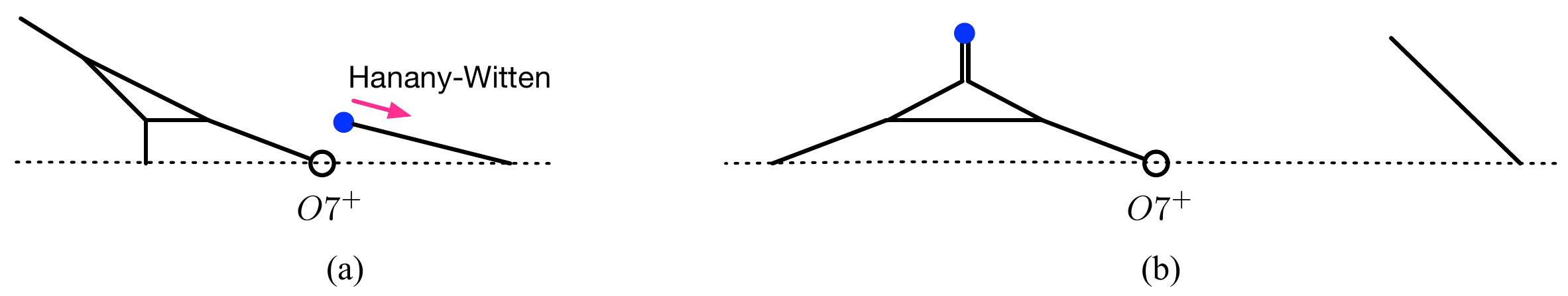}
    \caption{(a) A 5-brane web for local $\mathbb{P}^2+1\mathbf{Adj}$ that is $SL(2,\mathbb{Z})$ T-transformed from the last 5-brane web in figure \ref{fig:web-P2+adj}. (b) A 5-brane web where a Hanany-Witten move is performed on the configuration on the left.}
    \label{fig:web-P2+adj-different}
\end{figure}

Consider a local $\mathbb{P}^2$ with an adjoint ($\mathbb{P}^2+1\mathbf{Adj}$)  that is first discussed in \cite{Bhardwaj:2019jtr} as decoupling the instantonic hypermultiplet from the SU($2$)$_\pi$ gauge theory with an adjoint hypermultiplet (SU($2$)$_\pi+1\mathbf{Adj}-``1\mathbf{F}"$).
The corresponding 5-brane web is in given in \cite{Kim:2020hhh} which has an O7$^+$-plane, as given in figure \ref{fig:web-P2+adj}. We note that by an $SL(2,\mathbb{Z})$ T-transformation and Hanany-Witten transition, one can have another web diagram \cite{Kim:2020hhh} as given in figure \ref{fig:web-P2+adj-different}. 

\begin{figure}
    \centering
    \includegraphics[width=10cm]{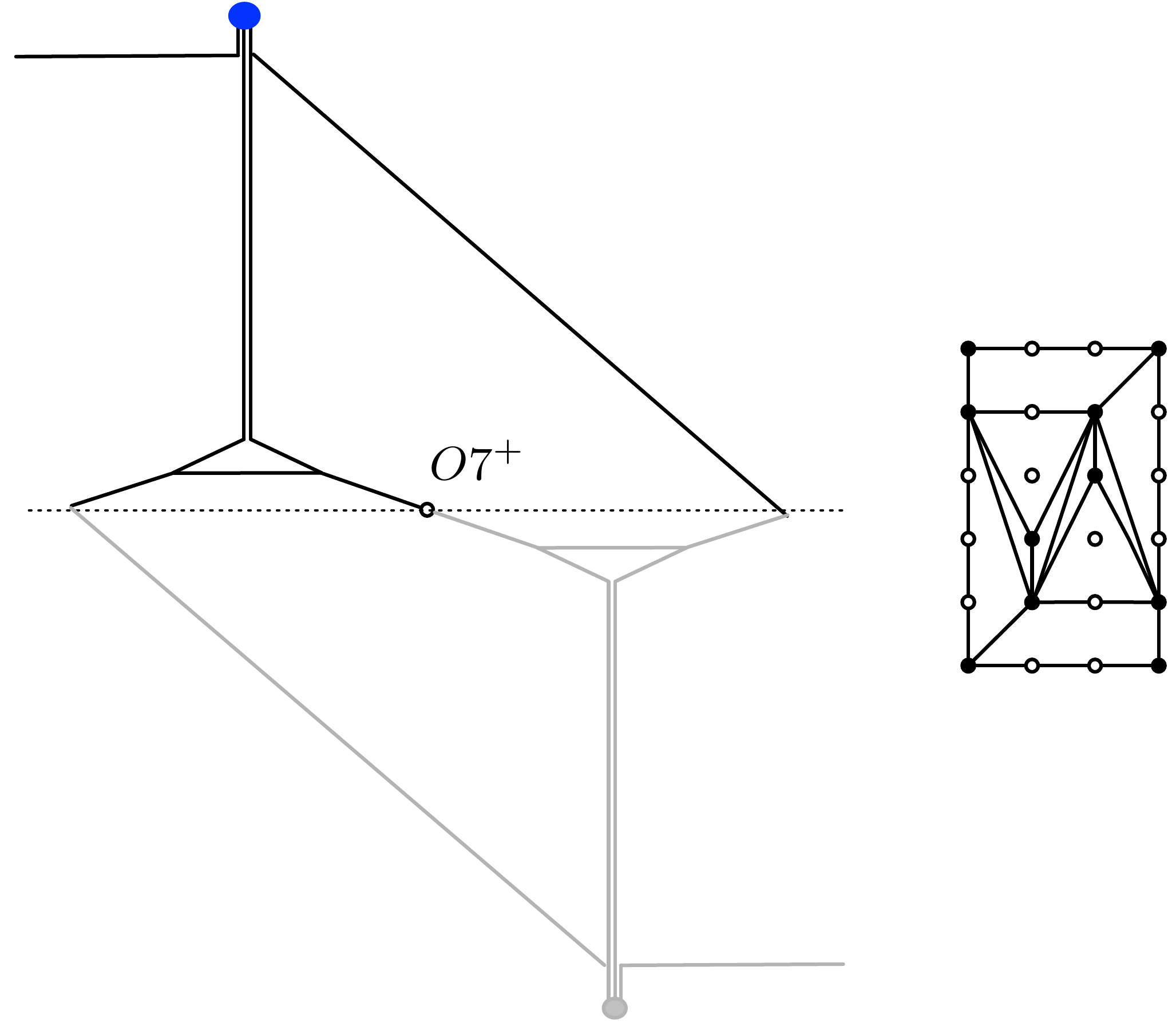}
    \caption{Left: Another 5-brane web for local $\mathbb{P}^2+1\mathbf{Adj}$ which includes the reflected images. Right: The corresponding toric-like diagram.}
    \label{fig:toric-P2+adj}
\end{figure}

We compute the Seiberg-Witten curve for $\mathbb{P}^2 + 1\mathbf{Adj}$ based on this 5-brane web with an O7$^+$-plane, given in figure \ref{fig:web-P2+adj-different}(b). To this end, we consider its covering space which includes the image due to an O7$^+$-plane. The corresponding dual toric diagram is shown in figure \ref{fig:toric-P2+adj}, which respects a point-wise reflection of an O7$^+$-plane, that is a $\mathbb{Z}_2$ rotation: $(t, w) \to (t^{-1}, w^{-1})$. We start with the following characteristic polynomial or ansatz, 
\begin{align}
    &\Big(  w^5 (t-a)^3 + (1-a t)^3\Big)\cr
    &+ b\Big( w^4 (t-a)^2(t+ b_1) +w (1-a t)^2(t+ b_1 t)\Big) \cr
    &+ c\Big( w^3 (t-a) (t^2+c_1t+c_2)+w^2(1-a t) (1+c_1 t+c_2 t^2)\Big) = 0  \ ,
\end{align}
where $a, b, b_1, c$, and $c_2$ are With proper boundary conditions discussed in the preceding subsections, one can compute the Seiberg-Witten curve for $\mathbb{P}^2 + 1\mathbf{Adj}$: 
\begin{align}\label{eq:SW for P2 with an adj}
&~~t^3 \left(w^2-1\right)^2 \left(w-M^3\right) \cr
&+t^2 \Big(-3 M w^5+ (2
   M^4 +M^{-2})w^4+U
   w^3-M U w^2-
   (M^5 +2M^{-1})w
+3 M^2\Big)\cr
&+t \Big(-3 M + (2
   M^4 +M^{-2})w+U
   w^2-M U w^3-
   (M^5 +2M^{-1})w^4
+3 M^2w^5\Big)\cr
&    +\left(1-w^2\right)^2 \left(1-M^3 w\right)=0\ .
\end{align}

\paragraph{local $\mathbb{P}^2$ limit. }
As a consistency condition, we consider the limit to a local $\mathbb{P}^2$ by decoupling the adjoint matter, {\it i.e.,} $M=e^{- \beta m_{\mathbf{Sym}}} \to 0$. To this end, we rescale the coordinates and parameters as  
\begin{align}\label{eq:locP2lim}
w \to M^{-1/3} w,\quad  t \to M t,\quad {\rm and}\quad U \to M^{-5/3} U\ .    
\end{align}
By multiplying an overall constant factor $M^\frac43$, one finds that this decoupling limit leads to 
the Seiberg-Witten curve for local $\mathbb{P}^2$, 
\begin{align}\label{eq:SWlocalP2}
 w^2 (t-1)^2+ U\, t+w^{-1} t =0\ . 
\end{align}
This curve can be re-expressed as
a more familiar form \eqref{eq:SWE0} by the coordinate transformations, first $t\to -t ,\, w\to -w, \, U\to -u$,  followed by $w\to (1+t^{-1})^{-1}w$ and then  by  $t\to  w^{-1} t$. 
The Weierstrass normal form for  local $\mathbb{P}^2$ is given by \cite{Kim:2014nqa}, 
\begin{align}\label{eq:WSSWlocalP2}
   y^2 = 4 x^3 - \Big(\frac{1}{12}\tilde{u}^4 -2\tilde{u} \Big)x- \frac{1}{216}\tilde{u}^6 +\frac16 \tilde{u}^3-1 \ ,
\end{align}
where $\tilde{u}$ is the Coulomb branch parameter which is related to $U$ by rescaling.


\subsubsection{Local \texorpdfstring{$\mathbb{P}^2 + 1\mathbf{Adj}$ from Sp(1)$+7\mathbf{F}$}{P2+1Adj from Sp(1)+7F}}

We now attempt to construct Seiberg-Witten curve for $\mathbb{P}^2 + 1\mathbf{Adj}$ from  
$\mathbb{P}^2 + 1\mathbf{AS}+8\mathbf{F}$. Because an 8-point blowup of $\mathbb{P}^2$ leads to SU(2)$+7\mathbf{F}$, and an antisymmetric hyper decouples for SU(2) theory. Local $\mathbb{P}^2 + 1\mathbf{AS}+8\mathbf{F}$ is hence equivalent to SU(2)$+7\mathbf{F}$, and, in turn, equivalent to Sp(1)$+7\mathbf{F}$, 
\begin{align}
    SU(2)_\pi+1\mathbf{Adj}-1\mathbf{F}\quad &\rightarrow\quad SU(2)_\pi+1\mathbf{AS}+8\mathbf{F}-\mathbf{F}\cr
    &=\quad Sp(1)+ 7\mathbf{F}\ .
\end{align}
We have obtained the Seiberg-Witten curve for 5d Sp($N$)$+(2N+5)\mathbf{F}$ 
in section \eqref{eq:SW-SpN-2N+5F}.
For $N=1$, the corresponding Seiberg-Witten curve is that for Sp(1)$+7\mathbf{F}$, which reads
\begin{align}\label{eq:Sp1+7FSW}
& t^2 w^{-4} \prod_{i=0}^7 (w-\mathsf{M}_i)
\cr
& 
+ \biggl[
- \frac{(w+1)^2}{2w}\prod_{i=0}^{7} (1 - \mathsf{M}_i) 
+ \frac{(w-1)^2}{2 w} \prod_{i=0}^{7} (1 + \mathsf{M}_i)
\cr
& ~~
+ (w-w^{-1})^2 
\biggl( 
- 2 (w^2+w^{-2})  \prod_{i=0}^{7} \mathsf{M}_i{}^{\frac{1}{2}} 
+ (w+w^{-1}) \sum_{i=0}^{7} (\mathsf{M}_i + \mathsf{M}_i^{-1}) \prod_{j=0}^{7} \mathsf{M}_j{}^{\frac{1}{2}}  + 2 \hat{C}_0
\biggr)
\biggr] t 
\cr
& + w^{4} \prod_{i=0}^7 (w^{-1}-\mathsf{M}_i)
  = 0 \ ,
\end{align}
where $\hat{C}_0$ is a constant that will be identified with the Coulomb branch parameter. We note here that this curve agrees with the SO($16$) manifest Seiberg-Witten curve obtained based on a 5-brane web with an O5-plane \cite{Hayashi:2017btw}.

Now, we specially tune the mass parameters $\mathsf{M}_i$ ($i=0,1,\cdots, 7$) as%
\footnote{One may wonder, among four mass parameters, $\mathsf{M}_{0,1,2,3}$, why one of them, $\mathsf{M}_0$ in this case, is tuned to $\widetilde{M}^{-1}$ rather than to $\widetilde{M}$. It is because if $\mathsf{M}_0$ is tuned to  $\widetilde{M}$, the resultant theory does not have the RG flow to local $\mathbb{P}^2$. These two different choices, $\widetilde{M}$ or $ \widetilde{M}^{-1}$, may be understood in a similar way as two discrete theta parameters of 5d SU(2)$_\theta$ theory with $\theta=0, \pi$ (mod $2\pi$), where SU(2)$_\pi$ theory can flow to local $\mathbb{P}^2$, while SU(2)$_0$ cannot.}
\begin{align}\label{eq:masstuning}
    \mathsf{M}_{0}=\widetilde{M}^{-1}, \qquad 
    \mathsf{M}_{1,2,3}=\widetilde{M},
    \qquad
    \mathsf{M}_{4,5, 6,7}=-\widetilde{M}\ .
\end{align}
This tuning reduces the curve \eqref{eq:Sp1+7FSW} to 
\begin{align}
& 
w^{-4} 
(w-\widetilde{M}^{-1}) (w+\widetilde{M})
(w^2 - \widetilde{M}^2)^3 t^2 
\cr
& + \Big[
\widetilde{M}^{-1} (1 - \widetilde{M}^2)^4 (w+w^{-1})
- 2 (w-w^{-1})^2 
\Big(\! 
\widetilde{M}^3 (w^2+w^{-2}) - \hat{C}_0
\Big)
\Big] t 
\cr
& + 
w^{4} (w^{-1}-\widetilde{M}^{-1}) (w^{-1}+\widetilde{M})(w^{-2}-\widetilde{M}^2)^3 = 0 .
\end{align}
We identify the tuned mass $\widetilde{M}$ as the mass of the adjoint hypermultiplet of local $\mathbb{P}^2+1\mathbf{Adj}$. 
By performing the coordinate transformation
\begin{align}
t \to \widetilde{M}^3 w^{4} 
(w-\widetilde{M}^{-1})^{-1} (w+\widetilde{M})^{-1} 
(w^2 - \widetilde{M}^2)^{-3}\, t\, ,
\end{align}
we can rewrite the curve as 
\begin{align}\label{eq:SU2+7F-O5}
&t^2 + \Big(
(\widetilde{M}-\widetilde{M}^{-1})^4 (w+w^{-1})
- 2 (w-w^{-1})^2 \big( w^2+w^{-2} - \widetilde{M}^{-3} \hat{C}_0 \big)
\Big)t\cr
&\quad 
+\widetilde{M}^{-8}(w^2-\widetilde{M}^2)^4(w^{-2}-\widetilde{M}^2)^4 =0\ .
\end{align}
This curve \eqref{eq:SU2+7F-O5}
can be simplified further by setting 
\begin{align}\label{eq:utoC0}
p = w+w^{-1},
\qquad
\chi_1 = \widetilde{M}^2+\widetilde{M}^{-2} ,
\qquad
u = 2 \widetilde{M}^{-3} \hat{C}_0 + 60,
\end{align}
as\footnote{Using the SO(16) manifest curve obtained from an O5-plane \cite{Hayashi:2017btw}, by the mass tuning \eqref{eq:masstuning}, one can obtain the same curve as \eqref{eq:SWquadratic-P6+Adj}  by identifying the instanton factor $q$ and the Coulomb branch parameter $U$ for Sp($1)+7\mathbf{F}$ with $q^{-2}= \mathsf{M}_0$ and $U=-\frac12 u+28$, respectively.}
\begin{align}\label{eq:SWquadratic-P6+Adj}
t^2- 2\Big( p^4\! - \!\frac12 (u-48)p^2 \!-\!\frac12(\chi_1-2)^2p +2 (u-56) \Big)t
    +(p^2-\chi_1-2)^4 =0\ .
\end{align}

To express this curve as the Weierstrass normal form, we perform the rescaling, 
$t \to {t}+ \big[p^4 - \frac12 (u-48)p^2 -\frac12(\chi_1-2)^2p +2 (u-56)\big]$ 
which leads to the following quartic curve
\begin{align}
 r^2 & +\big(u-4\chi_1-56\big) p^4 
 +\big(\chi_1-2\big)^2 p^3 
 +\Big(-\frac{u^2}{4}+24u+6\chi_1^2+8\chi_1-552\Big) p^2\cr
 &-\frac12(u-56)\big(\chi_1-2\big)^2p
 -\frac14 (\chi_1+2)^4
 +(u-56)^2=0\ ,
\end{align}
where $r^2= {t}^2/\big[(p+2)(p-2) \big].$
It follows from \cite{Huang:2013yta} that the Weierstrass normal form from this quartic curve is given by
\begin{align}\label{eq:WSnormalP2Adj}
    y^2 = 4 x^3 - g_2 x - g_3\ , 
\end{align}
where
\begin{align}
    g_2 & = \frac{1}{12}u^4-\frac{4}{3} \left(3 \chi_1^2+52\chi_1+1164\right)u^2\nonumber
\\
&\quad -
   \Big( 2 \chi_1^4+48 \chi_1^3-336 \chi_1^2-7488 \chi_1-115168\Big) u\nonumber\\
&\quad 
+16 \chi _1^5+288 \chi _1^4+3200 \chi _1^3-\frac{16640}{3} \chi_1^2-201472 \chi _1-2401792,\\
g_3 &= \frac{1}{216}u^6
-4u^5
-\frac{1}{9} \left(3 \chi_1^2-92 \chi_1-7764\right)u^4\nonumber\\
&\quad -\frac{1}{6} \left(\chi_1^4-72 \chi _1^3-744 \chi _1^2+12000 \chi_1+503824\right)u^3\nonumber\\
&\quad +\frac{4}{9} \left(3
   \chi _1^5-18 \chi_1^4-4776 \chi _1^3-33616 \chi_1^2+323184 \chi_1+9487968\right)u^2\nonumber\\
&\quad - \frac{4}{3} \left(15 \chi _1^6+236 \chi_1^5-1212 \chi_1^4-92256 \chi_1^3-553968 \chi_1^2+3339968 \chi_1+80170944\right)u \nonumber\\
 &\quad  
 +\chi _1^8
 +64 \chi _1^7
 +\frac{4912}{3}\chi_1^6
 +15488  \chi_1^5
 -\frac{95584 }{3}\chi_1^4
 -\frac{63023104 }{27}\chi_1^3\nonumber \\
&\quad 
   -\frac{39014656}{3} \chi_1^2
   +\frac{148920320}{3}\chi_1
   +1084823808
   .
\end{align}  
We note that this curve \eqref{eq:WSnormalP2Adj} coincides with the $E_8$ symmetry manifest Weierstrass normal form of the Seiberg-Witten curve for Sp$(1)+7\mathbf{F}$ \cite{Huang:2013yta,Kim:2014nqa} with all the masses being tuned as \eqref{eq:masstuning}. In other words, the mass tuning \eqref{eq:masstuning} makes the characters of $E_8$ symmetry to be combinations of the character $\chi_1$ of the fundamental representation of SU($2$) symmetry coming from the adjoint hypermultiplet of $\mathbb{P}^2+1\mathbf{Adj}$.

\paragraph{Local $\mathbb{P}^2$ limit.}
As a consistent condition, we can get the Seiberg-Witten curve for local $\mathbb{P}^2$ from the Seiberg-Witten curve  
\eqref{eq:SWquadratic-P6+Adj} of $\mathbb{P}^2+1\mathbf{Adj}$ by taking the limit where the adjoint hypermultiplet is decoupled. To this end, we take the mass of $\mathbf{Adj}$ to infinity, $\widetilde{M}=e^{-\beta m_{\mathbf{Adj}}}\to 0$, which leads to the following scaling, $\chi_1 \to L$ with $L = \widetilde{M}^{-2}$. To get the local $\mathbb{P}^2$, one also takes suitable rescalings for the coordinates and the Coulomb branch parameter as follows:
\begin{align}\label{eq:declimP2}
    t \to L^\frac{8}{3} t, \quad p  \to L^\frac{2}{3}p, \quad u \to L^\frac{4}{3} u \quad {\rm as} \quad  \chi_1 \to L\ . 
\end{align}
It follows that the leading contribution of \eqref{eq:SWquadratic-P6+Adj} becomes 
\begin{align}\label{eq:P2fromSp1+7F}
t^2 - 2 \Big(p^4 -\frac12 u \,p^2 -\frac12 p \Big)t + p^8 = 0 \ .     
\end{align}
The shifting
$t\to p t + \big(p^4 -\frac12 u \,p^2 -\frac12 p \big)$ 
then leads to a quartic curve
\begin{align}
    t^2 +u p^4 + p^3-\frac{u^2}{4} p^2 -\frac{u}{2} p -\frac14 =0\ .
\end{align}
It follows from \cite{Huang:2013yta} that the Weierstrass normal form for this quartic curve yields 
\begin{align}
    y^2 =  4 x^3 -g_2 x + g_3 \ , 
\end{align}
where $g_2$ and $g_3$ are the same as those in \eqref{eq:WSSWlocalP2} with the relabeling of the Coulomb branch parameter $u\to \tilde{u}$,  whose $j$-invariant is the same as that of \eqref{eq:WSSWlocalP2},
\begin{align}
    j \equiv \frac{1728 g_2^3}{g_2^3-27 g_3^2}=\frac{(u^4-24u)^3}{u^3-27}\ .
\end{align}
By the following rescaling  ${x}\to -{x}$, and ${y}^2\to -{y}^2$, one finds exactly the same Weierstrass form as the local $\mathbb{P}^2$ given in \eqref{eq:WSSWlocalP2}.
We note that one can also easily find the local $\mathbb{P}^2$ limit from the Weierstrass form by the following rescaling:
\begin{align}
    y \to L^4 y, \quad x  \to L^\frac{8}{3}x, \quad u \to L^\frac{4}{3} u, \quad {\rm and} \quad  \chi_1 \to L\ . 
\end{align}

\vspace{.5cm}
\subsubsection{Equivalence}
We have obtained two Seiberg-Witten curves for local $\mathbb{P}^2 + 1\mathbf{Adj}$ in two preceding sections. Here we discuss equivalence between two curves: one curve \eqref{eq:SW for P2 with an adj} is obtained from a 5-brane web with an O7$^+$-plane and the other curve \eqref{eq:SWquadratic-P6+Adj} is based on the freezing \eqref{eq:masstuning} of the Seiberg-Witten curve for Sp($1)+7\mathbf{F}.$

To begin with, we state that the curve based on an O7$^+$-plane \eqref{eq:SW for P2 with an adj} is cubic in $t$ and quintic in $w$, while the curve based on Sp$(1)+7\mathbf{F}$ is quadratic in $t$ and quartic in $p$. The corresponding Weierstrass normal form for \eqref{eq:SW for P2 with an adj} is not known but that for \eqref{eq:SWquadratic-P6+Adj} can be computed as given in \eqref{eq:WSnormalP2Adj}. 
So, it is not straightforward to compare them using Weierstrass forms.  We instead compare them based on consistency and some special case.

First, both curves have the proper limit to local $\mathbb{P}^2$ which is the limit where the adjoint hypermultiplet is decoupled. In this limit, the curve based on an O7$^+$-plane yields \eqref{eq:SWlocalP2} and the curve based on Sp$(1)+7\mathbf{F}$ becomes \eqref{eq:P2fromSp1+7F}, and their Weierstrass form perfectly agrees with that of local $\mathbb{P}^2$. Therefore, these two curves are the curve for a theory whose mass parameter can be decoupled giving rise to a local $\mathbb{P}^2$. We note here that both curves are inequivalent to the Sp($1)_\pi$ theory, as both theories have a $\mathbb{Z}_2$ symmetry from either $(t, w)\leftrightarrow (t^{-1}, w^{-1})$ or from $\widetilde{M}\leftrightarrow\widetilde{M}^{-1}$.

Next, we consider the massless case where the K\"ahler parameter for the mass of the adjoint hypermultiplet is set to 1. For this case, we start with the curve from $\mathrm{Sp}(1)+7\mathbf{F}$ given in \eqref{eq:SU2+7F-O5} with massless adjoint matter $\widetilde M=1$, 
\begin{align}\label{eq:SU2+7F-massless}
t^2+2\, t\,(w-w^{-1})^2\Big(U-1+(w+w^{-1})^2\Big) 
+(w-w^{-1})^8 =0\ ,
\end{align}
where $\widetilde{M}^{-3} \hat{C}_0$ 
is chosen according to $-U-1$. By rescaling $t\to t (w-w^{-1})^4$ and dropping out the overall factor, one readily reduces the curve \eqref{eq:SU2+7F-massless} to be
\begin{align} \label{eq:equivSp1+7F}
 t^2 ( w^2-1 )^2 - 2\, t\, (w^4 + (1 + U) w^2 + 1)+( w^2-1 )^2 = 0\ .
\end{align}
Now we consider the curve obtained from 5-brane web with an O7$^+$-plane. It follows from \eqref{eq:SW for P2 with an adj} that the Seiberg-Witten curve for local $\mathbb{P}^2+1\mathbf{Adj}$ with the massless case $M=1$, obtained from a 5-brane web is expressed as 
\begin{align}\label{eq:SWP2O7massless}
      (t-1) (w-1)\Big(t^2 (w^2-1 )^2 -\, t \,(2w^4  + (2 -U) w^2 +2 )+(w^2-1 )^2\Big)  = 0 \ . 
\end{align}
After dropping out the overall factor $(t-1) (w-1)$ and also rescaling $U\to -2 U$, one finds the curve  \eqref{eq:SWP2O7massless} exactly coincides with the curve \eqref{eq:equivSp1+7F} based on Sp($1)+7\mathbf{F}$.

Lastly, we consider the double discriminant $\Delta$ of two curves with non-zero generic mass. For the curve based on Sp$(1)+7\mathbf{F}$ with the frozen masses, it is convenient to use the Weierstrass form \eqref{eq:WSnormalP2Adj} as the double discriminant is given by
\begin{align} \label{eq:Deltafrozen}
\Delta_{\rm frozen}(u)   
   &= g_2^3-27g_3^2\cr
   &= \big(u^2-4  (\chi_1+30)u-\chi_1^3+6 \chi_1^2+244 \chi_1+3592\big)^4 \cr
&\quad~~\times\big(u^3+12 u^2 \chi_1-200 u^2-24 u \chi_1^2-1312 u \chi_1+12960 u\cr
&\quad \quad \quad -27\chi_1^4-8 \chi_1^3+1464 \chi_1^2+36064 \chi_1-274096\big)\ ,
\end{align}
where we took the double discriminant with respect to $y$ first and $x$ later, and we dropped an overall numerical factor.

Now consider the double discriminant of the curve \eqref{eq:SW for P2 with an adj} based on an O7$^+$-plane. The first discriminant $\mathcal{D}$ of \eqref{eq:SW for P2 with an adj} with respect to $t$ leads to a rather complicated expression with an overall factor $w^6$. As done in section \ref{sec:SU2anti-decouples} and Appendix \ref{app:SU3+1AS=1F}, we can properly rescale the first discriminant to use a new coordinate $x=w+w^{-1}$ and then we compute the discriminant of $\mathcal{D}$ with respect to $x$,
\begin{align}
    \Delta_{{\rm O7}^+} = {\rm Disc}\Big( \mathcal{D}(x)\Big)\ ,
\end{align}
which gives rise to a factorized form:
\begin{align}
    \Delta_{{\rm O7}^+}
    = \Delta_{\rm phys} \Delta_{\rm unphys}\ ,
\end{align}
where 
\begin{align}
    \Delta_{\rm phys}(U) = &~
    \bigg(\frac{U^2}{M^2}-2  \big(\chi_1+4\big)\frac{U}{M}-\chi_1^3+3 \chi_1^2+12 \chi_1+8\bigg)^5 \cr
&~\times\bigg(\frac{U^3}{M^3}+ \big(15\chi_1-32\big) \frac{U^2}{M^2}+ \big(3\chi_1^2-32\chi_1-32\big) \frac{U}{M}\cr
&\quad \quad -27\chi_1^4-19 \chi_1^3+120 \chi_1^2+192 \chi_1+80\bigg) ,\label{eq:DeltaphysO7+}\\
\Delta_{\rm unphys}(U) = &~  
256\, M^{36}  (\chi_1-2)^3
\cr
 &~\times \bigg(\frac{U^3}{M^3}-3  \big(\chi_1^2-4\chi_1+10\big)\frac{U^2}{M^2}+3\big(\chi_1^4-17 \chi_1^3+45\chi_1^2-8 \chi_1-8\big)\frac{U}{M}\cr 
 &\quad~~-\chi_1^6+12 \chi_1^5+57\chi_1^4-236\chi_1^3+30 \chi_1^2+120\chi_1+80\bigg)^3\ .
 \end{align}

We note that 
the solution for $\Delta_{\rm phys}(U)=0$ in \eqref{eq:DeltaphysO7+} is identical to the solution for $\Delta_{\rm frozen}(u)=0$ in \eqref{eq:Deltafrozen} up to rescaling of the Coulomb branch parameter. In other words,  under the rescaling $ U  \to M (u- 56-\chi_1)$, 
\begin{align}
   \Delta_{\rm phys}(U)=0 \quad \leftrightarrow \quad \Delta_{\rm frozen}(u)=0\ , 
\end{align}
where we set the mass parameters to be the same, $M=\widetilde{M}$, and hence the SU(2) character  $\chi_1=M^2+M^{-2}$ is equivalent to that in \eqref{eq:Deltafrozen}. 
Notice also that the double discriminant $\Delta_{{\rm O7}^+}$ is expressed as $\Delta_{\rm phys} \Delta_{\rm unphys}$, which appears to be a common structure that the double discriminant for the Seiberg-Witten curves obtained from a 5-brane with an O7$^+$-plane possesses the unphysical part $\Delta_{\rm unphys}$, as discussed in section \ref{sec:SU2anti-decouples} 
and in particular, in Appendix \ref{app:SU2+1AS},
for decoupling antisymmetric matter of SU(2) gauge theory as well as in section \ref{sec:equiSU3+1AS} and Appendix \ref{app:SU3+1AS=1F} for equivalence between SU($3)+1\mathbf{AS}$ and SU($3)+1\mathbf{F}$. 
We also give an interpretation of the physical part of the discriminant in this case in Appendix \ref{app:P2+1Adj}.

Three cases that we considered in this subsection, the local $\mathbb{P}^2$ limit, equivalence for the massless case, and the same physical part for the double discriminant $\Delta_{\rm phys}$  with generic mass, provide strong and convincing evidence that two seemingly different expressions of the Seiberg-Witten curve for local $\mathbb{P}^2+1\mathbf{Adj}$, \eqref{eq:SW for P2 with an adj} and \eqref{eq:SWquadratic-P6+Adj}, are indeed equivalent.


\bigskip
\section{Conclusion}\label{sec:concl}
In this paper, we proposed how to compute the Seiberg-Witten curves for 5d theories whose 5-brane configuration involves an O7-plane. The theories that we considered include: (i) SO($2N$) gauge theories with fundamental hypermultiplets and SU($2N$) gauge theories with a hypermultiplet in the symmetric representation and with hypermultiplets in the fundamental representation, which require an O7$^+$-plane in their 5-brane webs, and (ii) Sp($N$) gauge theories with hypermultiplets in the fundamental representation and SU($2N$) gauge theories with a hypermultiplet in the antisymmetric representation and with hypermultiplets in the fundamental representation which have an O7$^-$-plane in their 5-brane webs as well. We confirmed our proposal for the Seiberg-Witten curves based on an O7-plane by checking consistency conditions: the 4d limits of the theories we discussed  yielding the curves proposed by \cite{Landsteiner:1997ei}, and decoupling of an antisymmetric hypermultiplet of SU(2) gauge theory, and also equivalence between an antisymmetry hypermultiplet of SU(3) and a fundamental hypermultiplet which is discussed in detail in Appendix \ref{app:SU3+1AS=1F}.

We also proposed that an intriguing relation between the theories with an O7$^+$-plane and those with an O7$^-$-plane that a 5-brane web involving an O7$^+$-plane can be understood from the perspective of a 5-brane configuration with O7$^-$-plane and 8 D7-branes ($  \textrm{O7}^+\leftrightarrow\textrm{O7}^-\!+8\, \textrm{D7's} $) where the masses of 8 D7-branes are specially tuned (or ``frozen'') such that they are stuck on the O7$^-$-plane with half of them having the opposite phase factor. It is to tune the K\"ahler parameters of flavor masses such that four masses are assigned to $1$ while the other four masses are assigned to $-1$. We explicitly checked this relation by considering the Seiberg-Witten curves for the theories with an O7$^+$ and with ${O7}^-\!+8\, \textrm{D7's}$.  As a by-product, one can compute the Seiberg-Witten curves for 5d non-Lagrangian theories whose 5-brane configurations involve an O7$^+$-plane. As a representative example, we consider a local $\mathbb{P}^2$ with an adjoint matter ($\mathbb{P}^2+1\mathbf{Adj}$) where a local $\mathbb{P}^2$ can be obtained by decoupling of the instantonic hypermultiplet from pure Sp(1)$_\pi$ theory and an adjoint matter is a symmetric hypermultiplet from the point of view of Sp(1)$_\pi$ theory. In other words, local $\mathbb{P}^2+1\mathbf{Adj}$ can be obtained by decoupling the instantonic hypermultiplet from  Sp(1)$_\pi$ with a symmetric. We obtained the Seiberg-Witten curve for local $\mathbb{P}^2+1\mathbf{Adj}$ in two different perspectives: One is to directly compute it from a web diagram with an O7$^+$-plane, and the other is to obtain it via the special tuning of the mass parameters from Seiberg-Witten curve for Sp($1$) gauge theory with seven fundamentals (${\rm Sp}(1)+7\mathbf{F}$). We here remark that as an antisymmetric matter ($\mathbf{AS}$) of Sp(1) gauge theory decouples, ${\rm Sp}(1)+7\mathbf{F}$ can be understood as ${\rm Sp}(1)+7\mathbf{F}+1\mathbf{AS}$, and Sp($1$) gauge theory with seven fundamentals is understood as an 8 point blowups of local $\mathbb{P}^2$. In other words, ${\rm Sp}(1)+7\mathbf{F}$ can be viewed as local  $\mathbb{P}^2+8\mathbf{F}+1\mathbf{AS}$. As the curve obtained from a web with an O7$^+$-plane leads to a cubic expression while the curve obtained from ${\rm Sp}(1)+7\mathbf{F}$ is quadratic, we checked their equivalence from the double discriminants. We also computed the Weierstrass normal from for local $\mathbb{P}^2+1\mathbf{Adj}$ based on the quadratic curve obtained from ${\rm Sp}(1)+7\mathbf{F}$ or local $\mathbb{P}^2+8\mathbf{F}$ by ``freezing'' O7$^+$ and $8D7$s.

One immediate application of our proposal is to extend the construction of Seiberg-Witten curves for quiver gauge theories of production gauge groups involving an O7$^+$-plane as well as those involving an O7$^-$-plane, which we discussed in Appendix \ref{App:productgaugeroups}.

It would be an interesting direction to pursue how the relation between O7$^+$ and O7$^-+$ 8D7's that we proposed can be realized for other physical observables such as the supersymmetric partition function on $\mathbb{R}^4\times S^1$ in the Omega background \cite{KLNY:2023, Kim:2023qwh}, as well as 6d Seiberg-Witten curves and quantization of the Seiberg-Witten curves. These would shed some light on new perspectives of frozen singularities involving O7$^+$-planes.

\bigskip

\acknowledgments
We thank Songling He, Hee-Cheol Kim, Minsung Kim, Xiaobin Li, Yongchao L\"u, Satoshi Nawata, Xin Wang, and Gabi Zafrir for useful discussions and comments. SK thanks  the hospitality of Postech where part of this work was done and KIAS for hosting ``KIAS Autumn Symposium on String Theory 2022'' where this work was presented. The work of HH is supported in part by JSPS KAKENHI Grant Number JP18K13543 and JP23K03396. SK is supported by the NSFC grant No. 12250610188. KL is supported by KIAS Individual Grant PG006904 and by the National Research Foundation of Korea (NRF) Grant funded by the Korea government (MSIT) (No.2017R1D1A1B06034369). FY is supported by the NSFC grant No. 11950410490 and by Start-up research grant YH1199911312101. 

\bigskip

\appendix

\section{Product gauge groups}\label{App:productgaugeroups}

It is possible to generalize the method for computing Seiberg-Witten curves from an O7$^+$-plane to 5-brane webs which realize quiver theories. 5-brane webs with an O7$^+$-plane can yield different quiver theories from those which are obtained from 5-brane webs with an O5-plane. We focus on specific examples and compute their Seiberg-Witten curves. We utilize the perspective of treating an O7$^+$-plane effectively as an O7$^0$-plane and four D7-branes for writing down Seiberg-Witten curves.

\begin{figure}
\begin{tikzpicture}
\draw (-6,0)--(-6+1/2,1/4);
\draw (6,0)--(6-1/2,1/4);
\draw (-6+1/2,1/4)--(6-1/2,1/4);
\draw (-6+1/2,1/4+1/40)--(6-1/2,1/4+1/40);
\draw (-6+1/2,1/4+1/40)--(-6+1/2,3);
\draw (6-1/2,1/4+1/40)--(6-1/2,3);
\draw (-6+1/2,0)--(-6+1/2+1,1/2);
\draw (6-1/2,0)--(6-1/2-1,1/2);
\draw (-6+1/2+1,1/2)--(6-1/2-1,1/2);
\draw (-6+1/2+1,1/2+1/40)--(6-1/2-1,1/2+1/40);
\draw (-6+1/2+1,1/2+1/40)--(-6+1/2+1,3);
\draw (6-1/2-1,1/2+1/40)--(6-1/2-1,3);
\draw (-6+1/2+3/2,0)--(-6+1/2+3/2+3,3/2);
\draw (6-1/2-3/2,0)--(6-1/2-3/2-3,3/2);
\draw (-6+1/2+3/2+3,3/2)--(6-1/2-3/2-3,3/2);
\draw (-6+1/2+3/2+3,3/2+1/40)--(6-1/2-3/2-3,3/2+1/40);
\draw (-6+1/2+3/2+3,3/2+1/40)--(-6+1/2+3/2+3,3);
\draw (6-1/2-3/2-3,3/2+1/40)--(6-1/2-3/2-3,3);
\node at (-3-1/4,3+1/4) {$\overbrace{\hspace{5cm}}^{N}$};
\node at (3+1/4,3+1/4) {$\overbrace{\hspace{5cm}}^N$};
\node at (1/4,3/4+1/8) {$\left.\begin{array}{c}\\\\\\\end{array}\right\}$};
\node at (1/4+1/2,3/4+1/15) {${}^{2N}$};
\node at (0,1+1/8) {$\vdots$};
\node at (-3+1/4,3-1/4) {$\cdots$};
\node at (3-1/4,3-1/4) {$\cdots$};
\draw [dotted] (-7,0)--(7,0);
\draw (0,0) circle[radius=0.1];
\node at (0,-1/5) [below] {O7$^+$};
\end{tikzpicture}
\caption{A 5-brane web diagram for the quiver theory $\mathrm{SO}(4N) - \mathrm{SU}(4N-4)_0 \cdots - \mathrm{SU}(4)_0$.}
\label{fig:quiver1}
\end{figure}
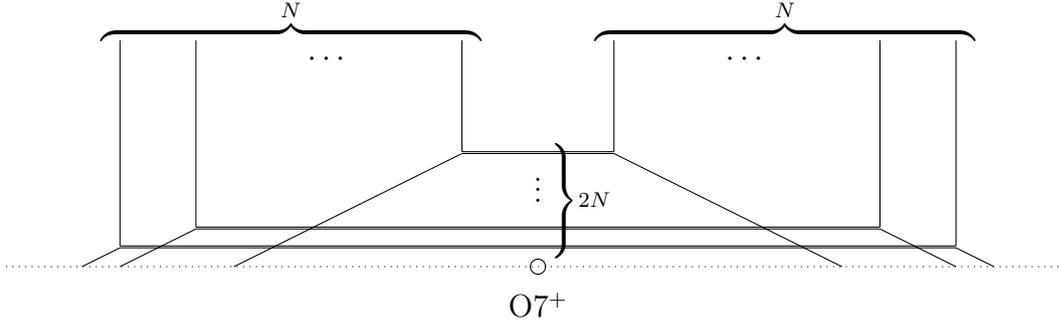
The first example is the quiver theory $\mathrm{SO}(4N) - \mathrm{SU}(4N-4)_0 \cdots - \mathrm{SU}(4)_0$, 
which has $N$ gauge nodes. The 5-brane web diagrams of the quiver theory are depicted in figure \ref{fig:quiver1}. The brane diagram 
 in the covering space contains $2N$ asymptotic NS5-branes and the general form of the Seiberg-Witten curve is 
\begin{equation}\label{SOSUcurve1}
F(t, w) = \sum_{m=0}^{2N}p_m(w)t^m = 0,
\end{equation}
where $p_m(w)$ is a polynomial of $w$. In the T-dual picture, the brane configuration contains an O6$^+$-plane at $t = 1, w = 1$ and another O6$^+$-plane at $t = 1, w = -1$. As far as writing the down Seiberg-Witten curves, the contribution of each O6$^+$-plane can be effectively thought of as that of a $\mathbb{Z}_2$ symmetry 
with $2$ D6-branes \cite{Landsteiner:1997ei}. Then the curve \eqref{SOSUcurve1} is invariant under the exchange of $(t, w) \leftrightarrow (t^{-1}, w^{-1})$ due to the effect of the O6$^0$-planes. The invariance gives a condition for the polynomial $p_m(w)$ given by
\begin{equation}\label{SOSUinvariance1}
p_{2N-m}(w) = p_m(w^{-1}),
\end{equation}
for $m=0, \cdots, N$. Hence it is enough to determine $p_m(w)$ for $m=0, \cdots, N$. The effect of the two effective D6-branes induces a bunch of virtual D4-branes at $w= \pm 1$. 
The polynomial $p_m(w)$ in \eqref{SOSUcurve1} can be written as \cite{Witten:1997sc,Landsteiner:1997vd,Brandhuber:1997cc}
\begin{equation}
p_m(w) = (w-w^{-1})^{2N-2m}\widetilde{p}_m(w),
\end{equation}
for $m=0, \cdots, N-1$. Then the condition \eqref{SOSUinvariance1} implies
\begin{equation}
p_{2N-m}(w) = (w-w^{-1})^{2N-2m}\widetilde{p}_{2N-m}(w)
\end{equation}
with
\begin{equation}\label{SOSUinvariance2}
\widetilde{p}_{2N-m}(w) = \widetilde{p}_m(w^{-1}).
\end{equation}
Since the location of color branes for the $\mathrm{SU}(4m)\; (m=1, \cdots, N-1)$ and the $\mathrm{SO}(4N)$ gauge groups are captured by the polynomials $\widetilde{p}_m(w)$ and $p_{N}(w)$ respectively, the polynomials can be written by 
\begin{equation}\label{pmforSU}
\widetilde{p}_m(w) = \sum_{n=-2m}^{2m}C_{m, n}w^n,
\end{equation}
for $m=1, \cdots, N-1$ and
\begin{equation}\label{pNforSO}
p_{N}(w) = \sum_{n=0}^{2N} C_{N, n}(w^n + w^{-n}).
\end{equation} 
For \eqref{pNforSO} we have taken into account the condition \eqref{SOSUinvariance1}. $\widetilde{p}_0(w)$ is constant and \eqref{pmforSU} also holds for $m=0$. Then the curve \eqref{SOSUinvariance1} becomes 
\begin{equation}\label{SOSUcurve2}
\begin{split}
&F(t, w) = \left(\sum_{m=0}^{N-1}\sum_{n=-2m}^{2m}(w-w^{-1})^{2N-2m}C_{m, n}w^nt^m\right) + \left(\sum_{n=0}^{2N}C_{N, n}(w^n + w^{-n})\right)\cr
&\hspace{5cm} +\left(\sum_{m=0}^{N-1}\sum_{n=-2m}^{2m}(w-w^{-1})^{2N-2m}C_{m, -n}w^nt^{2N-m}\right)=0.
\end{split} 
\end{equation}
\eqref{SOSUcurve2} gives the Seiberg-Witten curves of the quiver theory given by $\mathrm{SO}(4N) - \mathrm{SU}(4N-4)_0 - \cdots - \mathrm{SU}(4)_0$. 

The curve \eqref{SOSUcurve2} is parameterized by $C_{m, n}$'s which are related to Coulomb branch moduli, mass parameters of bifundamental hypermultiplets, and instanton fugacities. The mass parameter of the bifundamental hypermultiplet between $\mathrm{SU}(4m)$ and $\mathrm{SU}(4m+4)$ is given by the difference between the average of the color D5-branes for $\mathrm{SU}(4m)$ and that for $\mathrm{SU}(4m+4)$. Similarly, the mass parameter of the bifundamental hypermultiplets between $\mathrm{SU}(4N-4)$ and $\mathrm{SO}(4N)$ is given by the difference between the average of the color D5-branes for $\mathrm{SU}(4N-4)$ and that for $\mathrm{SO}(4N)$. Then the exponentiated mass parameters are written by
\begin{equation}\label{bifundamentalmass1}
M_m = \left(\frac{C_{m,2m}}{C_{m,-2m}}\right)^{\frac{1}{4m}}\left(\frac{C_{m+1,2(m+1)}}{C_{m+1,-2(m+1)}}\right)^{-\frac{1}{4(m+1)}},
\end{equation}
for $m=1, \cdots, N-2$ and
\begin{equation}\label{bifundamentalmass2}
M_{N-1} = \left(\frac{C_{N-1,2(N-1)}}{C_{N-1,-2(N-1)}}\right)^{\frac{1}{4(N-1)}}.
\end{equation}
For determining the instanton fugacities dependence we consider the behavior of the curve \eqref{SOSUcurve2} with $w$ large. When $w$ is large, the equation \eqref{SOSUcurve2} becomes
\begin{equation}\label{SOSUcurveforlargew}
F(t, w) \approx w^{2N}\left(\sum_{m=0}^{N}C_{m, 2m}t^m + \sum_{m=0}^{N-1}C_{m, -2m}t^{2N-m}\right) = 0.
\end{equation}
Let us denote the location of the NS5-branes at $w \to \infty$ by $t_1, \cdots, t_{2N}$ with $t_1 < t_2 < \cdots < t_{2N}$. Then \eqref{SOSUcurveforlargew} is given by 
\begin{equation}\label{instanton1}
\left(\sum_{m=0}^{N}C_{m, 2m}t^m + \sum_{m=0}^{N-1}C_{m, -2m}t^{2N-m}\right) = C_{0, 0}\prod_{m=1}^{2N}(t-t_m).
\end{equation}
Let $q_m$ and $q_{N}$ be the instanton fugacities for $\mathrm{SU}(4m)$ and $\mathrm{SO}(4N)$ respectively for $m=1, \cdots, N-1$. Then they are related to the location of the asymptotic NS5-branes by 
\begin{equation}\label{instanton2}
q_m = \sqrt{t_mt_{m+1}^{-1}t_{2N-m}t_{2N-m+1}^{-1}},
\end{equation}
for $m=1, \cdots, N$. Eqs. \eqref{bifundamentalmass1}, \eqref{bifundamentalmass2}, \eqref{instanton1}, and \eqref{instanton2} relate $C_{m, 2m}, C_{m,-2m}$ $\; (m=1, \cdots, N-1)$ and $C_{N, 2n}$ with the $N-1$ mass parameters of the bifundamental hypermultiplets and the $N$ instanton fugacities. We can set $C_{0, 0} = 1$ by the overall rescaling of the equation \eqref{SOSUcurve2}. The remaining parameters are related to the Coulomb branch moduli. More specifically, $C_{m, n}\; (n= -2m + 1, \cdots, 2m -1)$ for $m=1, \cdots, N-1$ are the Coulomb branch moduli of SU$(4m)$ and $C_{N, n}\; (n=0, \cdots, 2N-1)$ are the Coulomb branch moduli of SO$(4N)$. When $N=1$, the curve \eqref{SOSUcurve2} reduces to \eqref{pureSOcurve} with the instanton fugacity redefined.

\begin{figure}
\begin{tikzpicture}
\draw (-6,0)--(-6+1/2,1/4);
\draw (6,0)--(6-1/2,1/4);
\draw (-6+1/2,1/4)--(6-1/2,1/4);
\draw (-6+1/2,1/4+1/40)--(6-1/2,1/4+1/40);
\draw (-6+1/2,1/4+1/40)--(-6+1/2,3);
\draw (6-1/2,1/4+1/40)--(6-1/2,3);
\draw (-6+1/2,0)--(-6+1/2+1,1/2);
\draw (6-1/2,0)--(6-1/2-1,1/2);
\draw (-6+1/2+1,1/2)--(6-1/2-1,1/2);
\draw (-6+1/2+1,1/2+1/40)--(6-1/2-1,1/2+1/40);
\draw (-6+1/2+1,1/2+1/40)--(-6+1/2+1,3);
\draw (6-1/2-1,1/2+1/40)--(6-1/2-1,3);
\draw (-6+1/2+3/2,0)--(-6+1/2+3/2+3,3/2);
\draw (6-1/2-3/2,0)--(6-1/2-3/2-3,3/2);
\draw (-6+1/2+3/2+3,3/2)--(6-1/2-3/2-3,3/2);
\draw (-6+1/2+3/2+3,3/2+1/40)--(6-1/2-3/2-3,3/2+1/40);
\draw (-6+1/2+3/2+3,3/2+1/40)--(-6+1/2+3/2+3,3);
\draw (6-1/2-3/2-3,3/2+1/40)--(6-1/2-3/2-3,3);
\node at (-3-1/4,3+1/4) {$\overbrace{\hspace{5cm}}^{N}$};
\node at (3+1/4,3+1/4) {$\overbrace{\hspace{5cm}}^N$};
\node at (1/4+1/4,3/4+1/8) {$\left.\begin{array}{c}\\\\\\\end{array}\right\}$};
\node at (1/4+1/2-1/8+1/4+1/8,3/4+1/15) {${}^{2N}$};
\node at (1/4,1+1/8) {$\vdots$};
\node at (-3+1/4,3-1/4) {$\cdots$};
\node at (3-1/4,3-1/4) {$\cdots$};
\draw [dotted] (-7,0)--(7,0);
\draw (0,0) circle[radius=0.1];
\node at (0,-1/5) [below] {O7$^+$};
\draw (0,0)--(0,3);
\end{tikzpicture}
\caption{A 5-brane web diagram for the quiver theory $[1\mathbf{Sym}] - \mathrm{SU}(4N)_0 - \mathrm{SU}(4N-4)_0 - \cdots - \mathrm{SU}(4)_0$.}
\label{fig:quiver2}
\end{figure}
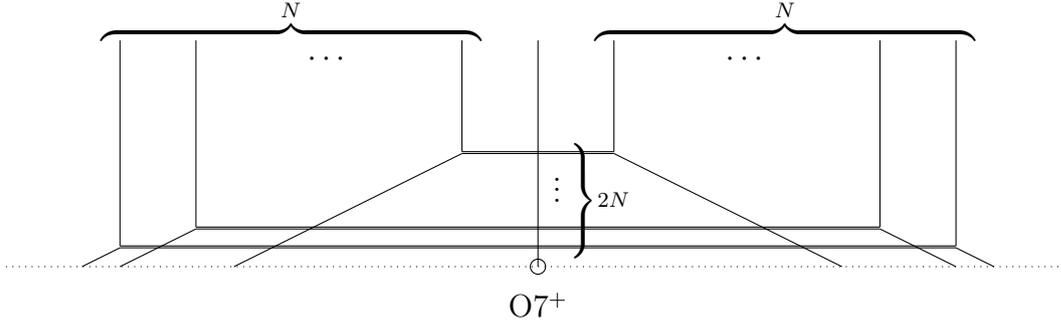
Next we consider the quiver theory $[1\mathbf{Sym}] - \mathrm{SU}(4N)_0 - \mathrm{SU}(4N-4)_0 - \cdots - \mathrm{SU}(4)_0$ where the number of the gauge nodes is $N$. The 5-brane web diagram is depicted in figure \ref{fig:quiver2}. The brane diagram in the covering space has $2N+1$ asymptotic NS5-branes and the Seiberg-Witten curve can be written by
\begin{equation}\label{SUsymSUcurve1}
F(t, w) = \sum_{m=0}^{2N+1}p_m(w)t^m = 0,
\end{equation}
where $p_m(w)$ is a polynomial of $w$. We first impose the invariance under the exchange of $(t, w) \leftrightarrow (t^{-1}, w^{-1})$ on \eqref{SUsymSUcurve1}. The invariance is realized by 
\begin{equation}\label{SUsymSUinvariance1}
p_{2N+1-m}(w) = - p_m(w^{-1}),
\end{equation}
for $m=0, \cdots, N$. The minus sign of \eqref{SUsymSUinvariance1} is chosen such that the O6$^+$-planes are placed at $(t, w) = (1, \pm 1)$. Next we consider the contribution of the effective virtual D6-branes and it can be incorporated by
\begin{equation}\label{SUsymSUfactored}
p_m(w) = (w - w^{-1})^{2N-2m}\widetilde{p}_m(w),
\end{equation}
for $m=0, \cdots, N$. Combining \eqref{SUsymSUinvariance1} with \eqref{SUsymSUfactored} gives
\begin{equation}
p_{2N+1-m}(w)  = -(w - w^{-1})^{2N-2m}\widetilde{p}_{2N+1-m}(w),
\end{equation}
with
\begin{equation}
\widetilde{p}_{2N+1-m}(w) = \widetilde{p}_m(w^{-1}). 
\end{equation}
The polynomial $\widetilde{p}_m(w)$ describes the color branes of $\mathrm{SU}(4m)$ and it can be written by
\begin{equation}\label{pmforSUsym}
\widetilde{p}_m(w) = \sum_{n=-2m}^{2m}C_{m, n}w^n,
\end{equation}
for $m=1, \cdots, N$. $\widetilde{p}_0(w)$ is constant and \eqref{pmforSUsym} also holds for $m=0$. Then \eqref{SUsymSUcurve1} becomes
\begin{equation} \label{SUsymSUcurve2}
\begin{split}
&F(t, w) = \sum_{m=0}^N\sum_{n=-2m}^{2m}(w - w^{-1})^{2N-2m}C_{m, n}w^nt^m\cr
&\hspace{4cm} - \sum_{m=0}^N\sum_{n=-2m}^{2m}(w - w^{-1})^{2N-2m}C_{m, -n}w^nt^{2N+1-m} = 0.
\end{split}
\end{equation}
\eqref{SOSUcurve2} gives the Seiberg-Witten curves of the quiver theory given by $[1\mathbf{Sym}] - \mathrm{SU}(4N)_0 - \mathrm{SU}(4N-4)_0 - \cdots - \mathrm{SU}(4)_0$. 

The parameters $C_{m, n}$'s of the curve \eqref{SUsymSUcurve2} are related to the parameters of the quiver theory. The mass parameter of the bifundamental hypermultiplet between $\mathrm{SU}(4m)$ and $\mathrm{SU}(4m+4)$ is given by the average of the color D5-branes for $\mathrm{SU}(4m)$ and that for $\mathrm{SU}(4m+4)$. Then the exponentiated mass parameter of the bifundamental hypermultiplet is
\begin{equation}\label{bifundamentalmass3}
M_m = \left(\frac{C_{m, 2m}}{C_{m, -2m}}\right)^{\frac{1}{4m}}\left(\frac{C_{m+1, 2(m+1)}}{C_{m+1, -2(m+1)}}\right)^{-\frac{1}{4(m+1)}},
\end{equation}
for $m=1, \cdots, N-1$. The mass of the symmetric hypermultiplet is given in a similar manner,
\begin{equation}\label{symmetricmass}
M_{\mathbf{Sym}} = \left(\frac{C_{N, 2N}}{C_{N, -2N}}\right)^{\frac{1}{4N}}\left(\frac{C_{N, -2N}}{C_{N, 2N}}\right)^{-\frac{1}{4N}}  = \left(\frac{C_{N, 2N}}{C_{N, -2N}}\right)^{\frac{1}{2N}}.
\end{equation}
In order to determine the instanton fugacities we consider the curve equation with large $w$. The equation \eqref{SUsymSUcurve2} around large $w$ becomes 
\begin{equation}\label{SUsymSUcurveforlargew}
F(t, w) = w^{2N}\left(\sum_{m=0}^NC_{m, 2m}t^m + \sum_{m=0}^NC_{m, -2m}t^{2N+1-m}\right) = 0.
\end{equation}
Let $t_m\;(m=1, \cdots, 2N+1)$ with $t_1 < \cdots < t_{2N+1}$ be the location of NS5-branes at $w \to \infty$. Then we can write \eqref{SUsymSUcurveforlargew} as
\begin{equation}\label{instanton3}
\sum_{m=0}^NC_{m, 2m}t^m + \sum_{m=0}^NC_{m, -2m}t^{2N+1-m} = C_{0,0}\prod_{m=1}^{2N+1}(t - t_m) = 0. 
\end{equation}
When we denote the instanton fugacities of $\mathrm{SU}(4m)$ by $q_m$ for $m=1, \cdots, N$, $q_m$ is given by
\begin{equation}\label{instanton4}
q_m = \sqrt{t_mt^{-1}_{m+1}t_{2N+1-m}t^{-1}_{2N+2-m}}.
\end{equation}
Then, \eqref{bifundamentalmass3}, \eqref{symmetricmass}, \eqref{instanton3}, and \eqref{instanton4} fix the parameters $C_{m, 2m}, C_{m, -2m}$ for $m=1, \cdots, N$. We can also set $C_{0, 0}=1$ by the overall rescaling of  \eqref{SUsymSUcurve2}. The remaining parameters $C_{m, n}\; (n=-2m+1, \cdots, 2m-1)$ correspond to the Coulomb branch moduli for the $\mathrm{SU}(4m)$ gauge theory for $m=1, \cdots, N$.

\section{Double Discriminant of Seiberg-Witten}
\label{app:Delta} 
In this appendix, we discuss double discriminant of Seiberg-Witten curves that we presented in connection with the physical part and the unphysical part.
\subsection{Double Discriminant of Seiberg-Witten curve for \texorpdfstring{$\mathrm{SU}(2)_{\pi}+1\mathbf{AS}$}{SU(2)+1AS}}
\label{app:SU2+1AS} 

First, we discuss the double discriminant of the Seiberg-Witten curves for $\mathrm{SU}(2)_{\pi}+1\mathbf{AS}$.
As in \eqref{SW-SU2-singlet}, the Seiberg-Witten curve is given by
\begin{align}\label{eq:SWSU2App}
t^3 + \Big( - 3 + (w-w^{-1})^2 \hat{p}_2(w)
\Big) t^2
- \Big( - 3 + (w-w^{-1})^2 \hat{p}_2(w^{-1})
\Big) t
- 1  = 0\ , 	
\end{align}
where
\begin{align}
\hat{p}_2(w) = q^{-1} ( M^{-\frac{1}{2}} w + U + M^{\frac{1}{2}} w^{-1} ) \ .
\end{align}

If we solve it in terms of $t$, we have three solutions $t=t_1(w), t_2(w), t_3(w)$. 
In order to find the branch points of these functions, we compute the discriminant of the left-hand side of \eqref{eq:SWSU2App} as a polynomial in $t$, which is given by
\begin{align}\label{eq:disc1SU2-app}
 q^{-4} (w-w^{-1})^{6} \Delta_1,
\end{align}
with
\begin{align}\label{eq:discSU2X}
\Delta_1 \equiv \sum_{n=0}^{6} c_n x^n,
\qquad 
x \equiv w+w^{-1}
\end{align}
and 
\begin{align}
&c_0 = - 4 ( M^{\frac12} - M^{-\frac12} )^4
- ( M^{\frac12} - M^{-\frac12} )^2 ( 27q^2 + 36 q U + 8 U^2 )
- 4 q U^3 - 4 U^4,
\cr
&c_1 = - 2(M^{\frac12}+M^{-\frac12}) \left[ (M^{\frac12} - M^{-\frac12})^2 ( 9 q + 4 U ) + 3 q U^2 + 4U^3
\right]
\cr
&c_2 = (M^{\frac12}-M^{-\frac12})^2 (M-10+M^{-1}) + 6 (M-4+M^{-1}) q U 
\cr
& \qquad -2(M+10+M^{-1}) U^2 + U^4,
\cr
&c_3 = 2 (M^{\frac12}+M^{-
\frac12}) \left[ (2M-5+2M^{-1}) q + (M-6+M^{-1})U + U^3 \right],
\cr
&c_4 = 2 (M - 4 + M^{-1}) + (M + 4 + M^{-1}) U^2, 
\cr
&c_5 = 2(M^{\frac12}+M^{-\frac12}) U, 
\cr
&c_6 = 1\ . 
\end{align}

The factor $(w-w^{-1})^6$ in \eqref{eq:disc1SU2-app} is originated from the condition \eqref{eq:constr1-AS} that ensures $t_1(\pm 1) = t_2(\pm 1) = t_3(\pm 1)$. 
Since this factor is not related to branch points corresponding to the non-trivial cycle of the Seiberg-Witten curve, we drop the factor $q^{-4} (w-w^{-1})^{6}$ from \eqref{eq:disc1SU2-app} and focus the remaining part $\Delta_1$.
Reflecting the symmetry $(w,t) \to (w^{-1},t^{-1})$ of the Seiberg-Witten curve, $\Delta_1$ is invariant under $w \to w^{-1}$ and can be rewritten 
as a polynomial of $x \equiv w+w^{-1}$. 
The six solutions $x=x_i$ ($i=1,\cdots, 6$) 
of $\Delta_1(x)=0$ gives the branch points of the functions $t_k(w)$ ($k=1,2,3$). 

Now, we would like to find the points in the Coulomb moduli that make at least two of the branch points $x_i$ coincide with each other.
Computing the discriminant of $\Delta_1$ as a polynomial of $x$, 
we obtain the ``double discriminant'' given in \eqref{eq:discSU2AS},
which we write down again for convenience:
\begin{align}
\Delta 
&= \Delta_{\text{phys}} \Delta_{\text{unphys}}
\cr
\Delta_{\text{phys}}
&= U^4 + qU^3 - 8 U^2 - 36 qU - 27q^2 + 16,
\cr
\Delta_{\text{unphys}}
&= 4096 q^2 (M^{\frac12}-M^{-\frac12})^4 \biggl[ (M^{\frac12}-M^{-\frac12})^2 ( (M^{\frac12}+M^{-\frac12})^2 - U^2 )^3
\cr
& \qquad \qquad 
- 27 (M^{\frac12}-M^{-\frac12})^2
(M^2+M+M^{-1}+M^{-2} -U^2 )q U 
\cr
& \qquad \qquad 
- 27(M-1+M^{-1})^3 q^2 
\biggr]^3\ .
\end{align}

In the following, we discuss how we have distinguished the physical part and the unphysical part of the discriminant in more detail. 

\subsubsection*{Physical part}
First, we consider the four solutions of $\Delta_{\text{phys}}(U)=0$. Expanding them in terms of small $q$, they are given by
\begin{align}
&U_1 = 2 + 2 q^{\frac12} - \frac14 q 
+ \frac{1}{16} q^{\frac32} 
- \frac{1}{64} q^2
+ \frac{3}{1024}q^{\frac52} 
+ \mathcal{O} (q^{\frac72})
\cr
&U_2 = - 2 - 2i q^{\frac12} - \frac14 q + \frac{i}{16} q^{\frac32} + \frac{1}{64} q^2 
- \frac{3i}{1024} q^{\frac52} 
+ \mathcal{O} (q^{\frac72})
\cr
&U_3 = 2 - 2 q^{\frac12} - \frac14 q - \frac{1}{16} q^{\frac32} - \frac{1}{64} q^2
- \frac{3}{1024}q^{\frac52} 
+ \mathcal{O} (q^{\frac72})
\cr
&U_4 = - 2 + 2i q^{\frac12} - \frac14 q - \frac{i}{16} q^{\frac32} + \frac{1}{64} q^2 
+ \frac{3i}{1024} q^{\frac52} 
+ \mathcal{O} (q^{\frac72})
\end{align}
We observe that $U_2, U_3, U_4$ are obtained from $U_1$ by transforming 
$q \to e^{\pi i} q$, $U \to e^{\pi i} U$ sequentially. This property is expected at all orders of $q$ because $\Delta_{\text{phys}}$ is invariant under this transformation.
Thus, what happens at these points is analogous to each other due to this symmetry. Therefore, it would be enough to study only one of them.

At $U=U_1$, the six solutions $x=x_i$ of $\Delta_1(x) =0$ are given by
\begin{align}
x_1 &= 2 + \frac{q}{(M^{\frac14} + M^{-\frac14})^2} + \mathcal{O}(q^{\frac32})
\cr
x_2 &= - 2 + \frac{q}{(M^{\frac14} - M^{-\frac14})^2} + \mathcal{O}(q^{\frac32})
\cr
x_3 &= - ( M^{\frac12} + M^{-\frac12} ) 
 + 2 (M^{\frac12}-M^{-\frac12}) q^{\frac14} 
 + \mathcal{O} (q^{\frac12})
\cr
x_4 &= - ( M^{\frac12} + M^{-\frac12} ) 
 - 2 (M^{\frac12}-M^{-\frac12}) q^{\frac14} 
 + \mathcal{O} (q^{\frac12})
\cr
x_5 &= x_6 =
- ( M^{\frac12} + M^{-\frac12} ) 
- \frac12 (M^{\frac12} + M^{-\frac12}) q^{\frac12}
+ \mathcal{O}(q^{\frac32} )
\equiv x^{(0)}.
\end{align}
We find that $x_5$ and $x_6$ coincide with each other, which is the source of the vanishing of $\Delta_{\text{phys}}$. 
Going back to the original variable $w$, the values $x=x^{(0)}$ corresponds to $w=w^{(0)}$ and $w =(w^{(0)})^{-1}$ with 
\begin{align}
w^{(0)} &\equiv  - \frac{1}{M^{\frac12}} - \frac{1+M}{2M^{\frac12} (1-M)}q^{\frac12} 
+ \frac{M^{\frac12} (1 + M)^2}{4 (1 - M)^3} q
+ \mathcal{O}(q^{\frac32}).
\end{align}
The three solutions $t_k(w)$ of the Seiberg-Witten curve \eqref{SW-SU2-singlet} at $w=w^{(0)}$ are given by
\begin{align}
t_1(w^{(0)}) &= (M^{\frac12}-M^{-\frac12})^4 q^{-1}
+ \frac12 (M^{\frac12}-M^{-\frac12})^2 (3 M^{-1} + 2 + 3M) q^{-\frac12}
+ \mathcal{O}(1),
\cr
t_2(w^{(0)}) &= t_3(w^{(0)}) = -\frac{1}{(M^{\frac12}-M^{-\frac12})^2} q^{\frac12} + \frac{(3M^{-1}+2+3M)}{4(M^{\frac12}-M^{-\frac12})^4} q 
+ \mathcal{O}(q^{\frac32}) \equiv t^{(0)}.
\end{align}
This implies that two of the branch points corresponding to $x_5$ and $x_6$ connect the second and the third solutions $t_2(w)$ and $t_3(w)$ and that these two branch points collide when $U=U_1$.

In order to see that this indicates the shrinking of a non-trivial cycle in the Seiberg-Witten curve at $U=U_1$,
we consider the small deviation from $U=U_1$
\begin{align}\label{eq:U-dev-phys}
U = U_1 + \varepsilon \delta U,
\end{align}
where $|\varepsilon| \ll 1$.
Then, we find that the coincident branch points are resolved as 
\begin{align}
w^{(0)} \to w^{(0)}
\pm  \left( M^{-\frac12} 
+ \frac{3-10 M -5M^2}{M^{\frac12} (1-M)^2} q^{\frac12}
+ \mathcal{O}(q) \right) (\varepsilon \delta U)^{\frac12} + \mathcal{O}(\varepsilon).
\end{align}
Taking this into account, 
we focus on the small region around $w=w^{(0)}$ as
\begin{align}
w &= w^{(0)} + \varepsilon^{\frac12} \, \delta w.
\end{align}
In order to see the variation of the two solution $t_2$ and $t_3$ from the value $t^{(0)}$, we find that we need to parameterized it as
\begin{align}\label{eq:localSW}
t &= t^{(0)} + q^{\frac14} \varepsilon^{\frac12} \, \delta t.
\end{align}
Using these parametrizations, the Seiberg-Witten curve around the point $(w,t)=(w^{(0)},t^{(0)})$, is approximately given as
\begin{align}\label{eq:microSW}
(M^{\frac12}-M^{-\frac12})^4 (\delta t)^2 + M (\delta w)^2 - \delta U + \mathcal{O}(q^{\frac14}) \sim 0,
\end{align}
which is obtained at the order $\mathcal{O}(\varepsilon)$ of the Seiberg-Witten curve \eqref{eq:SWSU2App}.

The curve \eqref{eq:microSW} has the branch cut structure connecting the two branch points $\delta w \sim \pm \left( \delta U / M  \right)^{\frac12}$. This indicates that the 
path encircling these two points gives a non-trivial cycle.
The corresponding cycle integral of the Seiberg-Witten 1-form gives
\begin{align}
\oint \lambda_{SW} 
\propto \oint \log w \frac{dt}{t}
\sim 
\frac{2 M q^{\frac14} \varepsilon^{\frac12} \log(w^{(0)})}{ (M^{\frac12}-M^{-\frac12})^2 t^{(0)}}
\int^{ \left( \frac{\delta U}{M} \right)^{\frac12} }_{-\left( \frac{\delta U}{M} \right)^{\frac12}} 
\frac{\delta w}{\sqrt{\delta U - M (\delta w)^2}} d(\delta w)
\end{align}
approximately at the leading order of $\varepsilon$ and $q$. 
Thus, we find that the cycle integral vanishes in the limit $\varepsilon \to 0$ at least at the leading order approximation in terms of $q$. We claim that this property is true for all orders in $q$.

In summary, the integral of the Seiberg-Witten 1-form over one of the cycles of the Seiberg-Witten curve, which is non-zero at a generic value of $U$, vanishes when $U=U_1$. 
This is what is expected at the point in the Coulomb moduli satisfying $\Delta_{\text{phys}}=0$. 

\subsubsection*{Unphysical part}
Next, we go on to the unphysical part of the discriminant.
The six solutions of $\Delta_{\text{unphys}}(U) = 0$ are given by 
\begin{align}
U'_n
=& e^{-\pi i n} M^{-\frac12} 
- \frac32 e^{-\frac{2 \pi i n}{3}} q^{\frac13} M^{-\frac13} 
- \frac38 e^{-\frac{\pi i n}{3}} q^{\frac23} M^{-\frac16} 
- \frac{9}{128} e^{\frac{\pi i n}{3}} q^{\frac43} M^{\frac16}
\cr
& \quad
+ e^{\pi i n} \left( 1 -\frac{27}{1024} q^2 \right) M^{\frac12}
+ \mathcal{O}(M^{\frac23})
\qquad \qquad 
( n = 0,1,2,3,4,5 )
\end{align}
This time, we have expanded them in terms of small $M$ because the computations become slightly simpler than the expansion in terms of $q$. 
We observe that $U_n'$ ($n=1,2,3,4,5$) are obtained from $U_0'$ by acting the transformation $M \to e^{2 \pi i} M$ sequentially. 
Thus, again, we study only for $U=U_0'$.

At $U=U_0'$, the six solutions of $\Delta_1 =0$ are given by
\begin{align}
x_1' &= 
-M^{-1} 
+ \frac32 q^{\frac13} M^{-\frac56} 
+ \frac38 q^{\frac23} M^{-\frac23} 
+ \frac{9}{128} q^{\frac43} M^{-\frac13} 
+ 2 q^{\frac12} M^{-\frac14} 
+ \mathcal{O} ( M^{-\frac{1}{12}} )
\cr
x_2' &= 
-M^{-1} 
+ \frac32 q^{\frac13} M^{-\frac56} 
+ \frac38 q^{\frac23} M^{-\frac23} 
+ \frac{9}{128} q^{\frac43} M^{-\frac13} 
- 2 q^{\frac12} M^{-\frac14} 
+ \mathcal{O} ( M^{-\frac{1}{12}} )
\cr
x_3' & = 2 + \frac{1}{2} q M^{\frac12}
+ \mathcal{O}(M^{\frac23})
\cr
x_4' & = - 2 - 6 q^{\frac23} M^{\frac13} + \mathcal{O}(M^{\frac12})
\cr
x_5' & =x_6' = -2 + \frac34 q^{\frac23} M^{\frac13} 
+ \frac{9}{64} q^{\frac43} M^{\frac23}
+ \mathcal{O}(M)
\equiv x^{(0)}{}'
\end{align}
We find that $x_5$ and $x_6$ coincide with each other, which is the source of the vanishing of $\Delta_{\text{unphys}}$. 
Going back to the original variable $w$, the values $x=x^{(0)}{}'$ corresponds to $w=w^{(0)}{}'$ and $w =(w^{(0)}{}')^{-1}$ with 
\begin{align}
w^{(0)}{}' &\equiv  
-1 
+ \frac{\sqrt{3}}{2} i  q^{\frac13} M^{\frac16} 
+ \frac38 q^{\frac23} M^{\frac13} 
+ \frac{9}{128} q^{\frac43} M^{\frac23}
+ \mathcal{O}(M)
\end{align}
This time, three solutions $t_k(w)$ ($k=1,2,3$) of the Seiberg-Witten curve \eqref{SW-SU2-singlet} at $w=w^{(0)}{}'$, we find all coincide with each other at
\begin{align}
t_1(w^{(0)}{}') = t_2(w^{(0)}{}') = t_3(w^{(0)}{}')
= e^{\frac{2 \pi i }{3}} \equiv t^{(0)}.
\end{align}
Assuming that these three are indeed equal, we find 
from the Seiberg-Witten curve \eqref{SW-SU2-singlet} 
that this value is exact rather than the approximation under small $M$.

Considering the small deviation from the value $U=U_0'$
\begin{align}
U = U_0' + \varepsilon \delta U,
\end{align}
we find that the coincident branch points are resolved at the next-to-next leading order of the deviation as
\begin{align}\label{eq:unphys-dev-w}
w^{(0)}{}' \to 
& w^{(0)}{}'
+ w^{(1)}{}' \varepsilon \delta U
\pm  w^{(2)}{}' (\varepsilon \delta U) ^{\frac32} 
+ \mathcal{O} (\varepsilon^{2})
\end{align}
with
\begin{align}
w^{(1)}{}'
&\equiv 
\frac{i}{\sqrt{3}} M^{\frac12} 
+ \frac{3 + i\sqrt{3}}{6} q^{\frac13} M^{\frac23}
+ \frac{3 + i\sqrt{3}}{12} q^{\frac23} M^{\frac56}
+ \mathcal{O}(M) 
\cr
w^{(2)}{}'
& \equiv 
 - \frac{4 \sqrt{2}}{9} q^{-\frac16} M^{\frac23} 
 + \frac{- 5\sqrt{2} + 2 \sqrt{6}i }{9} q^{\frac16} M^{\frac56}
+ \mathcal{O}(M)
\end{align}
Taking this into account, we focus on the small parameter region around the point
$w = w^{(0)}{}' + w^{(1)}{}' \varepsilon \delta U$
in the Seiberg-Witten curve as 
\begin{align}
w &= (w^{(0)}{}' + w^{(1)}{}' \varepsilon \delta U)
+ M^{\frac23} \varepsilon^{\frac32} \delta w,
\cr
t &= t^{(0)}{}' + M^{\frac16} \varepsilon^{\frac12} \delta t.
\end{align}
Then, we find the local structure of the Seiberg-Witten curve is given approximately as
\begin{align}\label{eq:SW-local-unphys}
q^{\frac13} (\delta t)^3 
+ 3 (1+\sqrt{3}i)\delta U (\delta t )
- 9 (\delta w)
+ \mathcal{O}(M^{\frac16})
\sim 0.
\end{align}

If we solve it for $\delta t$ as a function of $\delta w$, we find that they have two branch points corresponding to the ones discussed at \eqref{eq:unphys-dev-w}. However, unlike the previous case for physical discriminant, we do not find any non-trivial cycles around these 
branch points.
The situation would become clearer if we solve \eqref{eq:SW-local-unphys} for $\delta w$ as a function of $\delta t$ instead. Since $\delta w$ is simply a degree three polynomial in $\delta t$, there are no branch cuts nor poles at finite places. 
Thus, $\Delta_{\text{unphys}}=0$ does not mean the vanishing of any non-trivial period integral, and thus, no massless BPS particle nor tensionless BPS object appears. Therefore, we claim that $\Delta_{\text{unphys}}$ is unphysical.

\subsection{Discriminant of Seiberg-Witten curve for \texorpdfstring{$\mathrm{SU}(3)+1\mathbf{AS}$}{SU(3)+1AS}}
\label{app:SU3+1AS=1F} 
Here, we discuss some details of the double discriminant of the Seiberg-Witten curves for $\mathrm{SU}(3)_\frac12+1\mathbf{AS}$. The curves, given in \eqref{eq:SW-SU3-AS}, reads
\begin{align}\label{SW-SU3-AS}
w^{-1} t^3
& + \left( - w - 2w^{-1} + (w-w^{-1})^2 \hat{p}_2(w)
\right) t^2
\\
& - \left( - 2w - w^{-1} + (w-w^{-1})^2 \hat{p}_2(w^{-1})
\right) t
- w = 0\ ,\nonumber
\end{align}
where we denote
\begin{align}\label{eq:p1forSU3}
\hat{p}_2(w) = A w + B + C w^{-1} + D w^{-2}.
\end{align}
The discriminant of the left-hand side of \eqref{SW-SU3-AS} as a polynomial in $t$ is written in the form
\begin{align}
\frac{(w-1)^6(w+1)^6}{w^{6}} \left( \sum_{n=1}^{8} c_n (w^{n}+w^{-n}) + c_0 \right)
\end{align}
with
\begin{footnotesize}
\begin{align}
c_{8} =
&~ D{}^2 A{}^2, 
\cr
c_7 =
&~ 2 C D A{}^2 + 2 D{}^2 AB,
\cr
c_6 =
& ~B^2 D^2 + 4 B C D A - 2 D^2 A + 2 C D^2 A + C^2 A^2 + 2 B D A^2
- 2 D^2 A^2\ ,
\cr
%
c_5 =
&~ 2 (B^2 C D - B D^2 + B C D^2 + 2 D^3 + B C^2 A + 2 B^2 D A  
\cr   
 & - 2 C D A + 2 C^2 D A - 2 B D^2 A + 
   D^3 A + B C A^2 
\cr
&  - 2 D A^2 - 2 C D A^2 + D A^3)\ ,
\cr
c_4 =
&~ B^2 C^2 + 2 B^3 D - 4 B C D + 4 B C^2 D + D^2 - 2 B^2 D^2 + 10 C D^2 + C^2 D^2 
\cr
& + 2 B D^3 + 4 B^2 C A - 2 C^2 A + 2 C^3 A - 12 B D A - 4 B C D A  
\cr
& - 4 D^2 A + B^2 A^2 - 4 C A^2 - 2 C^2 A^2 + D^2 A^2 + 4 A^3 + 2 C A^3
\cr
c_3 =
&
~2 B^3 C - 2 B C^2 + 2 B C^3 - 8 B^2 D + 2 C D + 8 C^2 D + 2 C^3 D + 8 B D^2 
\cr
& - 2 D^3 + 2 C D^3 + 
 2 B^3 A - 12 B C A + 8 D A - 6 B^2 D A 
\cr 
& - 16 C D A - 6 C^2 D A + 6 B D^2 A - 4 D^3 A + 8 B A^2 - 6 D A^2 
\cr
& + 6 C D A^2 + 2 B A^3 - 4 D A^3
\cr
c_2 =
&~ B^4 - 8 B^2 C + C^2 + 2 B^2 C^2 + 2 C^3 + C^4 + 
 10 B D - 4 B^3 D + 4 B C D 
\cr 
& - 4 B C^2 D - 18 D^2 + 5 B^2 D^2 - 8 C D^2 + 2 C^2 D^2 - 4 B D^3 + D^4 + 2 B^2 A 
\cr
& + 8 C A - 4 B^2 C A - 12 C^2 A - 4 C^3 A - 
 16 B D A + 4 B C D A 
\cr
& + 4 D^2 A - 6 C D^2 A - 
 8 A^2 + 2 B^2 A^2 + 6 C A^2 + 5 C^2 A^2 
\cr 
& - 6 B D A^2 + 4 D^2 A^2 - 4 A^3 - 4 C A^3 + A^4
\cr
c_1 =
& -2 B^3 + 10 B C - 2 B^3 C - 4 B C^2 - 2 B C^3 - 
 4 D + 2 B^2 D - 28 C D 
 \cr
& - 2 B^2 C D - 14 C^2 D 
- 2 C^3 D - 12 B D^2 - 2 B C D^2 - 4 D^3 - 2 C D^3 
\cr
& - 
 8 B A - 2 B^3 A - 2 B C^2 A + 56 D A + 2 B^2 D A + 
 8 C D A 
\cr 
& + 2 C^2 D A - 4 B D^2 A + 2 D^3 A - 
 14 B A^2 - 2 B C A^2 + 4 D A^2 
 \cr
& - 4 C D A^2 - 
 2 B A^3 + 2 D A^3
\cr
c_0 =
&~ 
B^2 - 2 B^4 - 4 C + 4 B^2 C - 10 C^2 - 6 B^2 C^2 - 
 8 C^3 - 2 C^4 + 18 B D
\cr 
&  + 4 B^3 D - 24 B C D - 37 D^2 - 
 8 B^2 D^2 - 16 C D^2 - 6 C^2 D^2 + 4 B D^3 
\cr 
& - 2 D^4 + 
 4 A - 16 B^2 A + 40 C A + 16 C^2 A + 4 C^3 A + 
 32 B D A 
\cr 
& - 8 B C D A - 8 D^2 A + 8 C D^2 A - 
 28 A^2 - 6 B^2 A^2 - 16 C A^2 
\cr 
& - 8 C^2 A^2 + 8 B D A^2 - 8 D^2 A^2 - 4 A^3 + 4 C A^3 - 2 A^4\ .
\end{align}
\end{footnotesize}

Dropping the factor $(w-1)^6(w+1)^6/w^{6}$, rewriting the remaining polynomial in terms of 
$
x \equiv w+w^{-1}
$
and computing its discriminant in terms of $x$, we obtain the following double discriminant:
\begin{align}
&\Delta =
\Delta_{\text{unphys}} \Delta_{\text{phys}} 
\cr
&
\Delta_{\text{unphys}}
= 65536 A^{3} D^{4} 
\left( \sum_{m+n \le 9} c_{m,n} B{}^m C{}^n \right)^3
\cr
&
\Delta_{\text{phys}} 
= \sum_{m+n \le 9} d_{m,n} B{}^m C{}^n 
\end{align}
where
\footnotesize
\begin{align}
c_{00} = &
    D^6 + 84 D^8 - 159 D^{10} + D^{12} + 39 D^6 A + 186 D^8 A - 96 D^{10} A + 123 D^6 A^2 
\cr     
&  + 414 D^8 A^2  - 6 D^{10} A^2   - 12 D^4 A^3 + 73 D^6 A^3 + 156 D^8 A^3 - 12 D^4 A^4 - 336 D^6 A^4 
\cr     
&  + 15 D^8 A^4 - 39 D^4 A^5 + 93 D^6 A^5 + 48 D^2 A^6 + 114 D^4 A^6 - 20 D^6 A^6 - 156 D^2 A^7 
\cr
& - 255 D^4 A^7 - 81 D^2 A^8 + 15 D^4 A^8 - 64 A^9 + 87 D^2 A^9 + 48 A^{10} - 6 D^2 A^{10} + 15 A^{11} + A^{12}, 
\cr
c_{01} =&
 -3 (45 D^8 + 18 D^{10} - 8 D^6 A + 209 D^8 A + D^{10} A - 4 D^4 A^2 - 144 D^6 A^2 + 94 D^8 A^2 
\cr    
&  - 78 D^4 A^3 - 364 D^6 A^3 - 5 D^8 A^3 - 46 D^4 A^4 - 151 D^6 A^4 + 32 D^2 A^5 + 189 D^4 A^5  
\cr
& + 10 D^6 A^5 - 34 D^2 A^6 - 48 D^4 A^6 - 122 D^2 A^7 - {10} D^4 A^7 - 64 A^8 + 83 D^2 A^8 + 88 A^9 
\cr       
& + 5 D^2 A^9 + 4 A^{10} - A^{11}), 
\cr
c_{02} =&
 -3 (D^6 - 7 D^8 + D^{10} + D^4 A + {10}0 D^6 A + 54 D^8 A + 27 D^4 A^2 + 372 D^6 A^2 - 4 D^8 A^2 
 \cr
&  - 34 D^4 A^3 + 146 D^6 A^3 - 24 D^2 A^4 - 447 D^4 A^4 + 6 D^6 A^4 - 92 D^2 A^5 - 195 D^4 A^5 
\cr
& + 238 D^2 A^6 - 4 D^4 A^6 + 80 A^7 - 21 D^2 A^7 - 156 A^8 + D^2 A^8 + 16 A^9), 
\cr       
c_{03} =& 
27 D^8 - 3 D^4 A + 135 D^6 A + 3 D^8 A - 150 D^4 A^2 - 207 D^6 A^2 - 24 D^2 A^3 - 1104 D^4 A^3 
\cr
& - D^6 A^3 - 2{10} D^2 A^4 - 447 D^4 A^4 + 846 D^2 A^5 - 15 D^4 A^5 + 160 A^6 + 354 D^2 A^6 
\cr
& - 390 A^7 + 21 D^2 A^7 + 30 A^8 - 8 A^9, 
\cr     
c_{04}=& 
3 (D^6 + D^8 + 2 D^4 A + 33 D^6 A + D^2 A^2 + 76 D^4 A^2 + D^6 A^2 + 7 D^2 A^3 - 63 D^4 A^3 
\cr
& - 178 D^2 A^4 - 7 D^4 A^4 - 20 A^5 - 67 D^2 A^5 + 64 A^6 + 7 D^2 A^6 + 16 A^7 - 2 A^8), 
\cr       
c_{05}=& 
3 (2 D^4 A + D^6 A + 2 D^2 A^2 + 39 D^4 A^2 + 44 D^2 A^3 - 4 D^4 A^3 + 4 A^4 - 33 D^2 A^4 
\cr
& - 24 A^5 + D^2 A^5 - 8 A^6 + 2 A^7), 
\cr       
c_{06} =&  
-D^6 - 3 D^4 A - 6 D^4 A^2 - A^3 + 51 D^2 A^3 + 21 A^4 - 9 D^2 A^4 - 12 A^5 + 8 A^6, 
\cr     
c_{07} =& -3 (D^4 A + 2 D^2 A^2 + A^3 + 3 D^2 A^3 - 2 A^4), 
\cr       
c_{08} = & 
 -3 (D^2 A^2 + A^3 + A^4), 
\cr       
c_{09}=&
 -A^3, 
\cr         
c_{10} = & 
3 (-44 D^7 + 173 D^9 + D^{11} - 4 D^5 A - 117 D^7 A + 31 D^9 A - 15 D^5 A^2 - 301 D^7 A^2 
\cr
&- 5 D^9 A^2 + 35 D^5 A^3 + 92 D^7 A^3 + 32 D^3 A^4 + 135 D^5 A^4 + {10} D^7 A^4 - 97 D^3 A^5 
\cr
& - 219 D^5 A^5 - 59 D^3 A^6 - {10} D^5 A^6 - 64 D A^7 + 38 D^3 A^7 + 52 D A^8 + 5 D^3 A^8 
\cr
&
+ 58 D A^9 - D A^{10}), 
\cr 
c_{11} =& 
3 (D^5 + 146 D^7 + 28 D^9 + 21 D^5 A + 500 D^7 A + 2 D^9 A - 246 D^5 A^2 + 99 D^7 A^2 
\cr
& - 40 D^3 A^3 - 570 D^5 A^3 - 8 D^7 A^3 + 64 D^3 A^4 + 102 D^5 A^4 + 342 D^3 A^5 + 12 D^5 A^5 
\cr
& + 144 D A^6 - 127 D^3 A^6 - 272 D A^7 - 8 D^3 A^7 -  102 D A^8 + 2 D A^9), 
\cr       
c_{12}=&
-3 (-D^5 + 81 D^7 + D^9 - 165 D^5 A - 49 D^7 A - 16 D^3 A^2 - 765 D^5 A^2 + 7 D^7 A^2 
\cr
&- D^3 A^3  - 138 D^5 A^3 + 735 D^3 A^4 - 27 D^5 A^4 + 128 D A^5 - 63 D^3 A^5 
\cr
& - 483 D A^6 + 29 D^3 A^6  + 7 D A^7 - {10} D A^8), 
\cr       
c_{13} =& 
 -3 (2 D^5 + 20 D^7 + 2 D^3 A + 229 D^5 A + 10 D^7 A + 2 D^3 A^2 - 57 D^5 A^2 - 607 D^3 A^3 
 \cr
& - 22 D^5 A^3 - 56 D A^4 - 7 D^3 A^4 + 392 D A^5 + 14 D^3 A^5 - 91 D A^6 - 2 D A^7), 
\cr       
c_{14}=& 
-3 (2 D^5 + D^7 + 4 D^3 A + 39 D^5 A + 174 D^3 A^2 - 6 D^5 A^2 + 12 D A^3 - 55 D^3 A^3 
\cr
& - 148 D A^4 - 11 D^3 A^4 + 60 D A^5 + 16 D A^6), 
\cr       
c_{15}=&  
3 (D^5 + 6 D^5 A + D A^2 - 8 D^3 A^2 - 27 D A^3 + 20 D^3 A^3 + 8 D A^4 - {10} D A^5), 
\cr       
c_{16} =&
3 (D^5 + 4 D^3 A + 3 D A^2 + 13 D^3 A^2 + 9 D A^3 + 6 D A^4), 
\cr       
c_{17}=& 
3 (2 D^3 A + 3 D A^2 + 6 D A^3), 
\cr   
c_{18} = & 3 D A^2, 
\cr   
c_{20} = & 
-3 (-17 D^6 + 199 D^8 - {10}1 D^6 A - 69 D^8 A - 15 D^4 A^2 - 175 D^6 A^2 + D^8 A^2 
\cr
&+ 99 D^4 A^3 + 215 D^6 A^3 + 39 D^4 A^4 - 4 D^6 A^4 + 56 D^2 A^5 + 12 D^4 A^5 
\cr
& - 79 D^2 A^6 + 6 D^4 A^6 - 150 D^2 A^7 + 16 A^8 - 4 D^2 A^8 - 8 A^9 + A^{10}), 
\cr       
c_{21} =&
3 (-142 D^6 + 29 D^8 - 7 D^4 A - 393 D^6 A + 10 D^8 A + 159 D^4 A^2 + 5 D^6 A^2 
\cr
&+ 417 D^4 A^3 - 29 D^6 A^3 + 80 D^2 A^4 - 42 D^4 A^4 - 423 D^2 A^5 + 27 D^4 A^5 
\cr
& - 209 D^2 A^6 + 48 A^7 - 7 D^2 A^7 - 26 A^8 - A^9), 
\cr       
c_{22}=& 
3 (D^4 + 176 D^6 + 3 D^8 - 48 D^4 A + 32 D^6 A - 636 D^4 A^2 - 42 D^2 A^3 + 78 D^4 A^3 
\cr
&+ 645 D^2 A^4 - 6 D^4 A^4 - 60 D^2 A^5 - 56 A^6 + 31 A^7 + 3 A^8), 
\cr              
c_{23} =&
-3 (-2 D^4 + 9 D^6 - 264 D^4 A + 5 D^6 A - 10 D^2 A^2 + 36 D^4 A^2 + 350 D^2 A^3 
\cr
&+ 27 D^4 A^3 - 153 D^2 A^4 - 32 A^5 - 29 D^2 A^5 + 2 A^6 - 3 A^7), 
\cr       
c_{24}=&
 -3 (2 D^6 + D^2 A + 58 D^4 A - 66 D^2 A^2 + 39 D^4 A^2 - 13 D^2 A^3 + 9 A^4 &
 \cr
 &- 4 D^2 A^4 + 32 A^5 + 3 A^6), 
\cr       
 c_{25}=& 
-3 (2 D^4 + 3 D^2 A + 12 D^4 A + 60 D^2 A^2 - A^3 + 21 D^2 A^3 - 28 A^4 + 3 A^5), 
\cr       
 c_{26}=& -3 (D^4 + 3 D^2 A + 8 D^2 A^2 + 7 A^3 - A^4),
\cr   
c_{27}=& -3 (D^2 A - A^3), 
\cr    
c_{30}=& 
-3 D^5 + 231 D^7 - 8 D^9 - 189 D^5 A - 198 D^7 A - 54 D^5 A^2 + 21 D^7 A^2 - 16 D^3 A^3 
\cr
&- 201 D^5 A^3 + 237 D^3 A^4 - 15 D^5 A^4 + 591 D^3 A^5 - 144 D A^6 - D^3 A^6 + 51 D A^7 
\cr
&+ 3 D A^8, 
\cr     
c_{31} =& 
3 (43 D^5 - 55 D^7 + 151 D^5 A + 2 D^7 A - 2 D^3 A^2 + 37 D^5 A^2 - 323 D^3 A^3 - 14 D^5 A^3 
\cr
&- 181 D^3 A^4 + 104 D A^5 + 22 D^3 A^5 - 44 D A^6 - 10 D A^7),        
\cr              
c_{32} =&
3 (-134 D^5 + 3 D^7 + 45 D^5 A + 297 D^3 A^2 + 29 D^5 A^2 - 118 D^3 A^3 - 80 D A^4 - 27 D^3 A^4 
\cr
& +  39 D A^5 - 5 D A^6), 
\cr       
c_{33}=& 
D^3 + 147 D^5 - 237 D^3 A + 60 D^5 A + 120 D^3 A^2 + 78 D A^3 + 40 D^3 A^3 + 69 D A^4 + 60 D A^5, 
\cr     
c_{34}=& 
3 (D^3 + 2 D^5 + 85 D^3 A - 3 D A^2 + 11 D^3 A^2 - 56 D A^3 + 7 D A^4), 
\cr       
c_{35} =& 3 (D^3 + 2 D^3 A + 21 D A^2 - {10} D A^3), 
\cr   
c_{36} =& D^3 - 9 D A^2, 
\cr   
c_{40} =& -3 (-8 D^6 + 2 D^8 - 8 D^4 A + 42 D^6 A - 29 D^4 A^2 - 7 D^6 A^2 - 139 D^4 A^3 + 47 D^2 A^4 
\cr
& + 7 D^4 A^4 - 15 D^2 A^5 - D^2 A^6 + 4 A^7 - A^8), 
\cr       
c_{41} =& -3 (2 D^4 + 4 D^6 + 64 D^4 A + 16 D^6 A + 33 D^4 A^2 - 62 D^2 A^3 - 11 D^4 A^3 + 39 D^2 A^4 
\cr
& - 6 D^2 A^5 - 12 A^6 + A^7), 
\cr       
c_{42}=& -3 (-33 D^4 + 3 D^6 + 83 D^4 A + 24 D^2 A^2 - 4 D^4 A^2 - 32 D^2 A^3 + 39 D^2 A^4 + 13 A^5 + 2 A^6), 
\cr       
c_{43} = &
    3 (-36 D^4 + 3 D^2 A + 7 D^4 A + 21 D^2 A^2 + 
       11 D^2 A^3 + 6 A^4 + 2 A^5), 
\cr       
c_{44} =& 3 (D^4 - 21 D^2 A + 18 D^2 A^2 - A^3 + A^4), 
\cr      
c_{45} = &3 (3 D^2 A - A^3), 
\cr   
c_{50}= &
    3 (-7 D^5 + 2 D^7 + 39 D^5 A - 14 D^3 A^2 + D^5 A^2 + 11 D^3 A^3 - 4 D^3 A^4 - 12 D A^5 + D A^6), 
\cr       
c_{51} =& -3 (-26 D^5 - 6 D^3 A + {10} D^5 A +  28 D^3 A^2 - 20 D^3 A^3 - 25 D A^4 - 6 D A^5), 
\cr       
c_{52} = &-3 (D^3 + 3 D^5 - 14 D^3 A + 21 D^3 A^2 + 16 D A^3 + 12 D A^4), 
\cr       
c_{53} = &-3 (-7 D^3 + {10} D^3 A - 3 D A^2 - 2 D A^3), 
\cr   
c_{54} =& -3 (D^3 - 3 D A^2), 
\cr 
c_{60} = &
    3W D^4 + 8 D^6 + 15 D^4 A - 9 D^4 A^2 - 36 D^2 A^3 - 
     6 D^2 A^4 - A^6, 
\cr     
c_{61} = &
    3 (-7 D^4 + 6 D^4 A + 14 D^2 A^2 + 13 D^2 A^3 + A^5), 
\cr    
c_{62} = &3 (D^4 - 3 D^2 A - 8 D^2 A^2 - A^4), 
\cr   
c_{63} = &-9 D^2 A + A^3, 
\cr   
c_{70} = &-3 (4 D^3 A + 3 D^3 A^2 + D A^4), 
\cr   
c_{71} =& 3 (D^3 + 6 D^3 A + 2 D A^3), 
\cr      
c_{72} =& 3 (D^3 - D A^2), 
\cr   
c_{80}= &-3 (D^4 + D^2 A^2), 
\cr   
c_{81} = &3 D^2 A, 
\cr
c_{90}= & -D^3
\cr
\cr
d_{0,0} =
&
729 D^6 A + 432 D^4 A^2 + 2187 D^6 A^2 + 864 D^4 A^3 + 2187 D^6 A^3 
\cr
&
- 1512 D^4 A^4 + 729 D^6 A^4 - 1024 D^2 A^5 - 1944 D^4 A^5 - 729 D^4 A^6, 
\cr
d_{0,1} =
&
 3 (216 D^4 A^2 + 1404 D^4 A^3 + 512 D^2 A^4 + 702 D^4 A^4 - 768 D^2 A^5 + 243 D^4 A^5), ~~~~~~~~~~~~~~~~~
\cr
d_{0,2} =
&
 3 (-243 D^4 A^2 - 160 D^2 A^3 + 486 D^4 A^3 + 1120 D^2 A^4 - 504 D^2 A^5), 
\cr
d_{0,3} =
&
 -27 D^4 A - 16 D^2 A^2 - 540 D^4 A^2 - 984 D^2 A^3 + 216 D^4 A^3  + 2088 D^2 A^4 
\cr
& 
- 216 D^2 A^5, 
\cr
d_{0,4} =
& 8 (-3 D^2 A^2 - 69 D^2 A^3 - 2 A^4 + 27 D^2 A^4), 
\cr
d_{0,5} =
& 8 (3 D^2 A^2 + 2 A^3 - 4 A^4), 
\cr
d_{0,6} =
&
16 (D^2 A^2 + 2 A^3 - A^4),
\cr
d_{0,7} =
& 16 A^3, 
\cr
d_{0,8} = 
&
0,
\cr
d_{0,9} = 
&
0,
\cr
d_{1,0} =
&
-9 (108 D^5 A + 540 D^5 A^2 + 128 D^3 A^3 + 513 D^5 A^3 - 256 D^3 A^4 + 81 D^5 A^4
\cr
&
 - 288 D^3 A^5),
\cr
d_{1,1} =
&
-9 (-135 D^5 A - 64 D^3 A^2 - 27 D^5 A^2 + 656 D^3 A^3 + 108 D^5 A^3 + 336 D^3 A^4 
\cr
&
- 108 D^3 A^5),
\cr
d_{1,2} =
&
-4 (-513 D^3 A^2 + 297 D^3 A^3 - 32 D A^4 + 243 D^3 A^4), 
\cr
d_{1,3} = 
&
-4 (-9 D^3 A - 90 D^3 A^2 + 32 D A^3 + 54 D^3 A^3 - 60 D A^4), 
\cr
d_{1,4} = 
&
-4 (9 D^3 A + 4 D A^2 + 36 D^3 A^2 + 52 D A^3 - 36 D A^4),
\cr
d_{1,5} = 
&
-16 (4 D A^2 + 9 D A^3), 
\cr
d_{1,6} = 
&
-16 D A^2, 
\cr
d_{1,7} = 
&
0
\cr
d_{1,8} =
& 
0, 
\cr
d_{2,0} =
& 
 6 (36 D^4 A + 540 D^4 A^2 + 423 D^4 A^3 - 128 D^2 A^4), 
\cr
d_{2,1} =
&
 -1566 D^4 A - 351 D^4 A^2 + 1024 D^2 A^3 + 
  972 D^4 A^3, 
\cr
d_{2,2} =
&
 513 D^4 A - 112 D^2 A^2 + 270 D^4 A^2 - 504 D^2 A^3 - 
  270 D^2 A^4, 
\cr
d_{2,3} =
&
 2 (-4 D^2 A + 365 D^2 A^2 + 135 D^2 A^3), 
\cr
d_{2,4} =
&
 2 (23 D^2 A + 72 D^2 A^2 - 4 A^3), 
\cr
d_{2,5} =
&
 -8 (D^2 A - A^2 - A^3), 
\cr
d_{2,6} =
&
-8 A^2, 
\cr
d_{2,7} =
&
0, 
\cr
d_{3,0} =
&
 -27 D^5 - 540 D^5 A - 832 D^3 A^2 + 216 D^5 A^2 + 576 D^3 A^3 - 216 D^3 A^4, 
\cr
d_{3,1} =
&
12 (24 D^3 A - 43 D^3 A^2 + 18 D^3 A^3), 
\cr
d_{3,2} =
&
-591 D^3 A - 270 D^3 A^2 + 64 D A^3, 
\cr
d_{3,3} =
&
D^3 + 68 D^3 A - 64 D A^2 - 68 D A^3, 
\cr
d_{3,4} =
&
4 (-2 D A + 17 D A^2), 
\cr
d_{3,5} =
&
8 D A, 
\cr
d_{3,6} =
&
0, 
\cr
d_{4,0} = 
&
-3 (-9 D^4 - 252 D^4 A + 72 D^4 A^2 + 64 D^2 A^3), 
\cr
d_{4,1} = 
&
-4 (9 D^4 + 36 D^4 A - 56 D^2 A^2 - 36 D^2 A^3), 
\cr
d_{4,2} = 
&
-2 (-17 D^2 A + 72 D^2 A^2), 
\cr
d_{4,3} = 
&
-D^2 - 68 D^2 A, 
\cr
d_{4,4} = 
&
D^2 - A^2, 
\cr
d_{4,5} = 
&
A, 
\cr
d_{5,0} = 
&
 -200 D^3 A, 
\cr
d_{5,1} =
&
36 (D^3 + 4 D^3 A), 
\cr
d_{5,2} =
&
-8 (D^3 - D A^2), 
\cr
d_{5,3} = 
&
-8 D A, 
\cr
d_{5,4} = 
&
-D, 
\cr
d_{6,0} = 
&
 16 (D^4 - D^2 A^2), 
\cr
d_{6,1} =
&
16 D^2 A, 
\cr
d_{6,2} =
&
8 D^2, 
\cr
d_{6,3} =
&
0, 
\cr
d_{7,0} =
&
-16 D^3, 
\cr
d_{7,1} =
&
d_{7,2} =
d_{8,0} =
d_{8,1} =
d_{9,0} =
0.
\end{align}
\normalsize

If we identify the coefficients in \eqref{eq:p1forSU3} as 
\begin{align}
A = M^{-\frac{1}{2}} q^{-1}, \quad
B = - U q^{-1},  \quad
C = V M^{\frac{1}{2}} q^{-1},  \quad
D = - M q^{-1}, 
\end{align}
then the physical part of the discriminant $\Delta_{\text{phys}}$ agrees with the double discriminant of the Seiberg-Witten curve for SU(3)$_{-\frac{1}{2}}+1\mathbf{F}$ given in \eqref{eq:SW-SU3_1.2+1F} up to an overall factor independent of $U$ and $V$.

\subsection{Discriminant of Seiberg-Witten curve for Local \texorpdfstring{$\mathbb{P}^2 + 1\mathbf{Adj}$}{P2+1Adj}}
\label{app:P2+1Adj} 

In this appendix, we make some comments on the discriminant of the Seiberg-Witten curve for Local $\mathbb{P}^2 + 1\mathbf{Adj}$. 
As given in \eqref{eq:DeltaphysO7+}, the physical part of the double discriminant consists of the two factors as
\begin{align}
\Delta_{\rm phys}(U) = \Delta_{\rm phys1}(U)^5 \Delta_{\rm phys2}(U)
\end{align}
where
\begin{align}
\Delta_{\rm phys1}(U) 
&= \frac{U^2}{M^2}-2  \big(\chi_1+4\big)\frac{U}{M}-\chi_1^3+3 \chi_1^2+12 \chi_1+8,
\cr
\Delta_{\rm phys2}(U) 
&= \frac{U^3}{M^3}+ \big(15\chi_1-32\big) \frac{U^2}{M^2}+ \big(3\chi_1^2-32\chi_1-32\big) \frac{U}{M}
\cr
&\quad \quad -27\chi_1^4-19 \chi_1^3+120 \chi_1^2+192 \chi_1+80.
\end{align}
In the following, we discuss the singularity structure corresponding to these two factors.

\subsubsection*{Singularity corresponding to \texorpdfstring{$\Delta_{\rm phys1}(U)=0$}{Deltaphys1}}
First, we focus on the factor $\Delta_{\rm phys1}(U)$ which has the two roots $U=U_1, U_2$ with
\begin{align}
\frac{U_1}{M} &\equiv -M^{-3} + M^{-2} + M^{-1} + 4 + M + M^2 - M^3,
\cr
\frac{U_2}{M} &\equiv M^{-3} + M^{-2} - M^{-1} + 4 - M + M^2 + M^3.
\end{align}
where we have used $\chi_1 = M^2 + M^{-2}$. 
Substituting $U=U_1$ to the Seiberg-Witten curve \eqref{eq:SW for P2 with an adj}, we find that the curve factorizes into $(w-1)$ and the remaining factor as
\begin{align}
(w-1) F_{\text{rem}}(t,w) = 0.
\end{align}
Analogously, for $U=U_2$, the Seiberg-Witten curve is again factorized into the factor $(w+1)$ and the remaining. 

\begin{figure}
\centering
    \includegraphics[width=14cm]{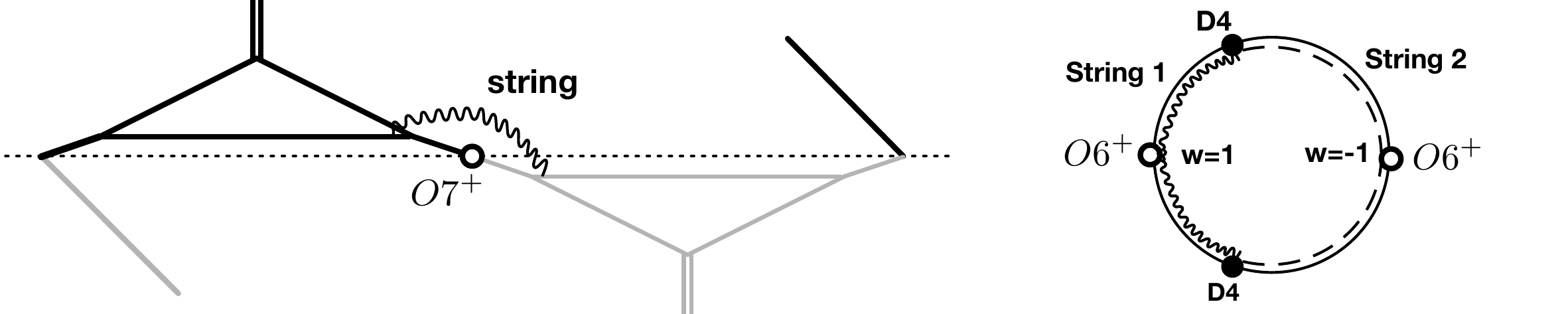}
\caption{Left: A string in the 5-brane web for local $\mathbb{P}^2+1\mathbf{Adj}$. Right: Two types of strings connecting a D4-brane and its mirror image in the T-dual picture.
}
\label{fig:P2BPS1}
\end{figure}

We interpret that these correspond to the cases where the string, which is depicted as a wavy line on the left of Figure \ref{fig:P2BPS1}, becomes massless. 
This string connects a D5-brane and its mirror image in the 5-brane web diagram for local $\mathbb{P}^2+1\mathbf{Adj}$.
When the D5-brane approaches the position of the O7$^+$-plane, the distance between this D5-brane and its mirror image also becomes zero, giving rise to the massless BPS particle.

The fact that there are two locations $w=\pm1$ where this string becomes massless can be naturally interpreted as follows. 
After T-duality, an O7$^+$-plane compactified on a circle becomes two O6$^+$-planes, which are located at two antipodal points of the circle, which are $w = \pm 1$ in this case. Also, the D5-brane wrapping the compactified circle becomes a D4-brane. 

In this setup, we can consider two types of strings connecting the D4-brane and its mirror image, as depicted on the right of Figure \ref{fig:P2BPS1}, which describes the compactified circle.
One of the strings is depicted as a wavy line and denoted as ``String 1'', while the other is depicted as a dashed line and denoted as ``String 2''. String 1 becomes massless at $U=U_1$, and String 2 becomes massless at $U=U_2$. 
The Seiberg-Witten curve is expected to capture the M-theory uplift of this situation.

In this way, we can understand that two singularities appear at the Coulomb moduli by adding an adjoint matter to the local $\mathbb{P}^2$ theory.

\subsubsection*{Singularity corresponding to \texorpdfstring{$\Delta_{\rm phys2}(U)$}{Deltaphys2}}
Next, we focus on the factor $\Delta_2(w)$. Instead of finding its roots explicitly, we consider the limit discussed in \eqref{eq:locP2lim}, or equivalently, \eqref{eq:declimP2}.
In this limit, the theory reduces to the local $\mathbb{P}^2$ theory without an adjoint matter. 
By rescaling $\chi_1$ and $U$ as 
\begin{align}
\frac{U}{M} \to U L^{\frac43},
\qquad
\chi_1 \to L
\end{align}
and taking the limit $L \to \infty$, 
we find that the discriminant simplifies as
\begin{align}
L^{-4} \Delta_{\rm phys2}(U) 
\to U^3 - 27.
\end{align}
This is exactly the discriminant of the local $\mathbb{P}^2$ theory without an adjoint matter. 

Therefore, we interpret that the three BPS particles that becomes massless at the roots of $\Delta_{\rm phys2}(U)$, respectively, is essentially identical to the 
ones that exist in the local $\mathbb{P}^2$ theory. 

\bibliographystyle{JHEP}
\bibliography{ref}
\end{document}